\documentclass[a4paper,12pt]{article}
\pdfoutput=1 

\usepackage{jheppub} 
\usepackage{fancyhdr}
\usepackage{graphicx}
\usepackage{epstopdf}
\usepackage{color}
\usepackage{hyperref}  
\usepackage{amsmath}
\usepackage{amsmath} 
\usepackage{enumerate} 
\usepackage{slashed}

\newcommand\ptwiddle[1]{\mathord{\mathop{#1}\limits^{\scriptscriptstyle(\sim)}}}
\newcommand{\nn}{\nonumber}

\title{\boldmath Characterising the 750~GeV diphoton excess}

\author[a]{J\'er\'emy Bernon,}
\author[b]{Andreas Goudelis,}
\author[a]{Sabine Kraml,}
\author[a,c]{Kentarou Mawatari,}
\author[a]{Dipan Sengupta}

\affiliation[a]{Laboratoire de Physique Subatomique et de Cosmologie, 
Universit\'e Grenoble-Alpes, \\ CNRS/IN2P3, 
53 Avenue des Martyrs, F-38026 Grenoble, France
}
\affiliation[b]{Institute of High Energy Physics, Austrian Academy of Sciences, Nikolsdorfergasse 18, \\
1050 Vienna, Austria
}
\affiliation[c]{Theoretische Natuurkunde and IIHE/ELEM,
 Vrije Universiteit Brussel,
 and International Solvay Institutes,
 Pleinlaan 2, B-1050 Brussels, Belgium
}
\emailAdd{bernon@lpsc.in2p3.fr}
\emailAdd{andreas.goudelis@oeaw.ac.at}
\emailAdd{sabine.kraml@lpsc.in2p3.fr}
\emailAdd{kentarou.mawatari@lpsc.in2p3.fr}
\emailAdd{dipan.sengupta@lpsc.in2p3.fr}

\abstract{
We study kinematic distributions that may help characterise the recently observed excess in diphoton events at 750 GeV at the LHC Run~2. Several scenarios are considered, including spin-0 and spin-2 750~GeV resonances that decay directly into photon pairs as well as heavier parent resonances that undergo three-body or cascade decays.
We find that combinations of the distributions of the diphoton system and the leading photon can distinguish  the topology and mass spectra of the different scenarios, while patterns of QCD radiation can help differentiate the production mechanisms.  
Moreover,  missing energy is a powerful discriminator for the heavy parent scenarios if they involve (effectively) invisible particles. 
While our study concentrates on the current excess at 750 GeV, the analysis is general and can also be useful for characterising other potential diphoton signals in the future. 
}

\begin{document}

\preprint{LPSC16048}
\maketitle

\section{Introduction}

The excess in the diphoton invariant mass spectrum around 750~GeV observed by ATLAS~\cite{ATLAS-CONF-2015-081} and CMS~\cite{CMS-PAS-EXO-15-004} in the first LHC Run~2 data led to a
sheer flood of theory papers trying to explain the alleged signal.\footnote{The ATLAS excess consists of 14 events in 3.2 fb$^{-1}$ of data; it has a local (global) significance of $3.6~\sigma$ ($2.0~\sigma$) and seems to favour a large width of about 45~GeV (see however \cite{Buckley:2016mbr}). The CMS excess consists of 10 events in 2.6 fb$^{-1}$ of data; it has a local (global) significance of $2.6~\sigma$ ($1.2~\sigma$) and is consistent with a narrow width.}
The interpretations put forward span a wide spectrum, including extra Higgs bosons, axions, sgoldstinos, radions, gravitons, hidden glueballs, hidden- or techni-pions and so on. Typically, the existence of additional new particles and/or new (strong) dynamics is invoked, in order to account for the increase in cross section from $\sqrt{s}=8$ to 13~TeV and to evade the often stringent bounds from null results in dijets, monojets and other search channels both at $\sqrt{s}=8$ and 13~TeV.  
The first theory papers discussing various ways to reproduce the observed diphoton rate as well as possibly a large width while avoiding existing constraints from Run~1 appeared on the arXiv already on the day after the announcement of the excess~\cite{Harigaya:2015ezk,Backovic:2015fnp,Nakai:2015ptz,Knapen:2015dap,Buttazzo:2015txu,Pilaftsis:2015ycr,Franceschini:2015kwy,DiChiara:2015vdm}.
More than 200 papers followed to date.

Whatever one may think of this ``ambulance chasing''~\cite{Backovic:2016xno}, 
an interesting question that arises is how to experimentally differentiate between this variety of possible interpretations. 
Needless to say this question will be of imminent importance should the observed excess turn into a discovery with the accumulation of more data.
One approach consists of observing the new state in different decay modes, as the predictions for the (ratios of) rates of specific final states vary between different concrete models. Another, complementary approach is to rely on the diphoton signal itself and attempt its detailed characterisation in terms of kinematic distributions. As a preparatory step in the latter direction, in this paper we study the expectations for differential distributions from various signal hypotheses and discuss ways to discriminate between them. 
We note in passing that both approaches---inclusive measurements in different final states and kinematic distributions---have been pursued successfully to scrutinise the 125~GeV Higgs signal in Run~1~\cite{Dittmaier:2011ti,Dittmaier:2012vm,Heinemeyer:2013tqa,Aad:2014lwa,Aad:2014tca,Khachatryan:2014kca,Aad:2015lha,Aad:2015mxa,Khachatryan:2015rxa,ATLAS-CONF-2015-044,ATLAS-CONF-2015-069,CMS-PAS-HIG-15-010,Khachatryan:2015yvw}.

Irrespective of the underlying model, the interpretations put forward generically fall in just a few classes.      
First, if we are dealing with a new particle with mass of 750~GeV which undergoes a two-body decay into two photons, the classification is by spin and production mechanism. 
The most straightforward option is a 750~GeV spin-0 (singlet scalar or pseudoscalar) particle produced in gluon fusion and decaying to photons e.g.\ via loops of new vector-like quarks. Bottom-quark ($b\bar b$) initiated production could also provide the necessary increase in cross section from $\sqrt{s}=8$ to 13~TeV of about a factor five~\cite{Franceschini:2015kwy}. 
If it has electroweak couplings, a scalar resonance can also be produced in vector boson fusion and vector-boson associated production.
Photon-initiated production has also been discussed~\cite{Fichet:2015vvy,Csaki:2015vek,Csaki:2016raa,Fichet:2016pvq}.

Another option is a spin-2 resonance, 
like the Kaluza--Klein (KK) graviton in Randall--Sundrum (RS)-type models~\cite{Randall:1999ee}, which might be produced from $gg$ or $q\bar q$ initial states. A spin-1 particle would not decay into photons,%
\footnote{See however~\cite{Chala:2015cev} for a scenario where a vector resonance decays to a photon and a light scalar, followed by a decay of the scalar into two highly collimated photons, $Z'\to \gamma + s\to 3\gamma$, which might appear as a diphoton final state.
A 750 GeV vector resonance is also considered in~\cite{Tsai:2016lfg}, where the resonance decays into a photon and a massive dark photon, $V(750)\to\gamma\gamma'$, followed by a displaced dark photon decay $\gamma'\to e^+e^-$ which can be misidentified as a photon.} 
and higher spins are not considered because they are disfavoured theoretically. 
In order to explain a large width, as seemingly favoured by ATLAS, the resonance should couple not only to gluons and photons (and perhaps quarks) but also to non-standard states such as dark matter or light hidden-valley particles. 
Invisible decays are, however, fairly constrained (although not excluded) by the 8~TeV mono-$X$ searches as discussed e.g.\ in~\cite{Barducci:2015gtd}.

Alternatively, the new particle can be (much) heavier than 750~GeV and undergo a three-body~\cite{Bernon:2015abk,An:2015cgp} or a cascade decay~\cite{Franceschini:2015kwy,Kim:2015ron,Huang:2015evq,Cho:2015nxy,Altmannshofer:2015xfo,Liu:2015yec,Han:2016bus} into two photons along with one or more light new particle(s). These light new particles would then need to be soft or invisible so as to avoid detection. Such a scenario could  ``naturally'' explain the apparent broadness of the diphoton invariant mass peak, as well as soften the tension with the 8~TeV data. Note that in this case the new states can in principle be scalars, vectors or fermions.   

Kinematic distributions for characterising the 750~GeV diphoton excess have been considered previously in the literature. 
For example, \cite{Gao:2015igz} discussed $gg$ versus $q\bar q$ initiated production of a spin-0 resonance, 
while \cite{Han:2015cty,Csaki:2016raa,Martini:2016ahj,Giddings:2016sfr} discussed kinematic distributions arising from  spin-2 resonances (in part comparing them to the spin-0 case). Reference~\cite{Csaki:2016raa} also discussed how to differentiate gluon- from photon-induced production for both spin-0 and spin-2 particles, while the authors of \cite{Harland-Lang:2016qjy} performed a spin-parity analysis for photoproduction including PDF uncertainties.  
For the case of a heavier parent resonance, 
\cite{Cho:2015nxy} considered $E_{T}^{\rm miss}$ and $E_\gamma$ distributions for various cascade-decay topologies. 

In this work we study the kinematic distributions arising from 750~GeV spin-0 and spin-2 resonances and compare them to those obtained from the production of heavier parent particle(s) that undergo two- or three-body decays. 
We go beyond the previous investigations listed above by analysing a consistent set of kinematic distributions 
for all these different cases. 
Moreover, we employ more realistic simulations including initial-state QCD radiation and parton shower matching.

We note that the results of our analysis hold regardless of the fate of the 750~GeV diphoton excess and should provide useful guidelines for the discrimination of other potential diphoton excesses that could appear during the 13~TeV LHC Run. 

The paper is organised as follows: in Section~\ref{sec:scenarios} we briefly present the scenarios we will consider and our choices for the parametrisation of the relevant interactions. Section~\ref{sec:simulations} describes the computational tools we employ for our analysis as well as some important technical features. 
Our main results on the kinematic distributions that could be used to discriminate among different explanations of the diphoton signal are presented in Section~\ref{sec:distributions}. We conclude in Section~\ref{sec:conclusions}. 
Appendices~\ref{app:s1special} and \ref{app:antler} contain supplementary considerations on two of the heavy parent scenarios discussed in the main part of the paper. 

\section{Scenarios for the 750~GeV diphoton excess}\label{sec:scenarios}

As briefly discussed in the Introduction, the various scenarios for the diphoton excess can be quite generically classified according to the number and the nature of the final state particles as well as according to the initial state producing them in the first place. A full survey of all types of models that have been proposed in the literature is not the scope of our study. Instead, our approach is mostly driven by the different types of topologies that could generate the excess at 750 GeV.

Even so, a choice is to be made for the parametrisation of the relevant interactions. It has already been argued that the relatively large, ${\cal{O}}(5\!-\!10)$ fb, cross section seemingly favoured by the ATLAS and CMS data~\cite{Falkowski:2015swt} would either require fairly light weakly coupled new physics, mostly likely in tension with observations, or might be pointing towards some type of strong dynamics or to some relatively singular threshold effect~\cite{Han:2016pab,Kats:2016kuz}. In this case, a complete description of the underlying physics is likely to involve some momentum-dependent form factors the form of which depends on the specific setup being invoked. 

A simpler approach, yet powerful enough to capture the main effects of interest for our work, is to parametrise the couplings of the new state(s) to the SM ones in terms of effective operators suppressed by appropriate powers of some effective field theory (EFT) scale $\Lambda$.%
\footnote{A discussion and motivation can be found, e.g.\ in the introduction of \cite{Artoisenet:2013puc}.}
Note that the scale $\Lambda$ does \textit{not} correspond to the cutoff scale of the theory: the two are rather typically \textit{related} to each other, but the relation depends on the specific UV-completion.
We also note that in everything that follows we will ignore Lagrangian terms that are not directly relevant to our analysis, whereas we will remain agnostic to whether the new states introduced in the effective Lagrangian description are fundamental or composite.

\subsection{750 GeV resonance}\label{sec:scenarios750}

The simplest way to accommodate the 750~GeV diphoton excess is by invoking a particle $X_s$ with a mass of 750 GeV that decays into a pair of photons.\footnote{For the reader who wants to dive into a plethora of realizations within concrete models, \cite{Staub:2016dxq} provides an extensive overview together with the actual model implementations.} 
The Landau-Yang theorem~\cite{Landau:1948kw,Yang:1950rg} then guarantees that $X_s$ can be a spin-0 particle $X_0$ or a spin-2 particle $X_2$. In either case it can be produced via gluon, $q\bar q$, photon or vector boson fusion, or in association with a vector boson or a pair of top quarks.

\begin{figure}
\center
 \includegraphics[width=.24\textwidth]{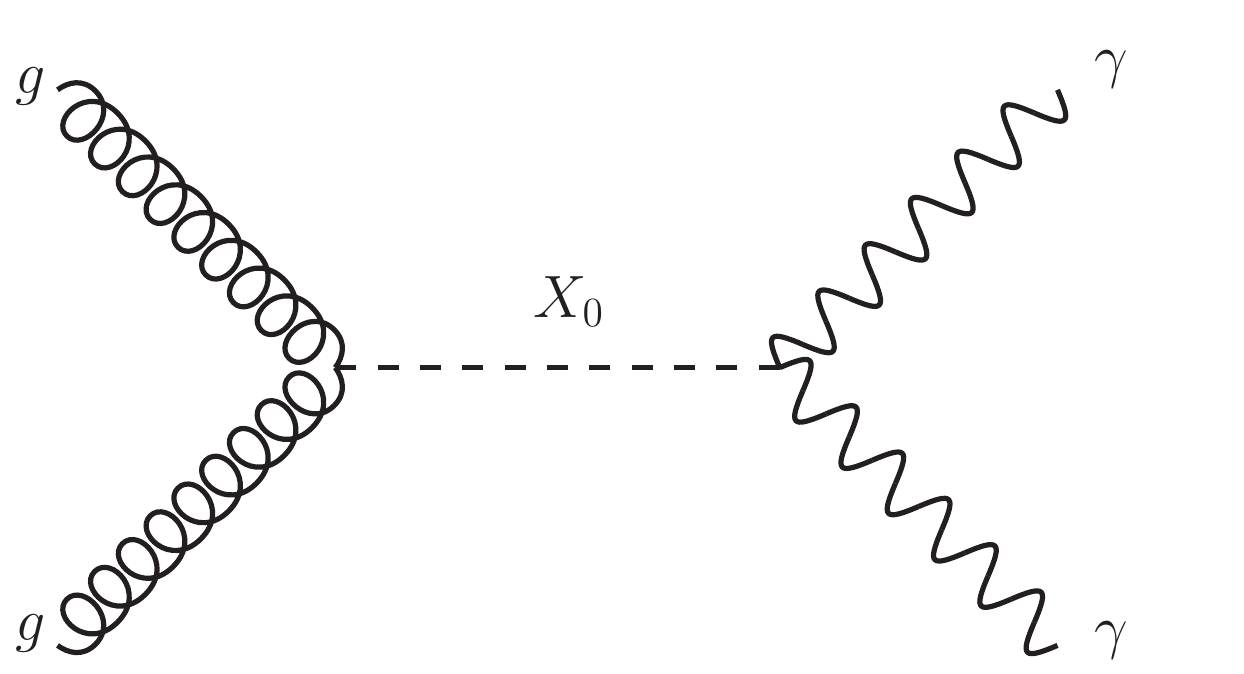}\!\!
 \includegraphics[width=.24\textwidth]{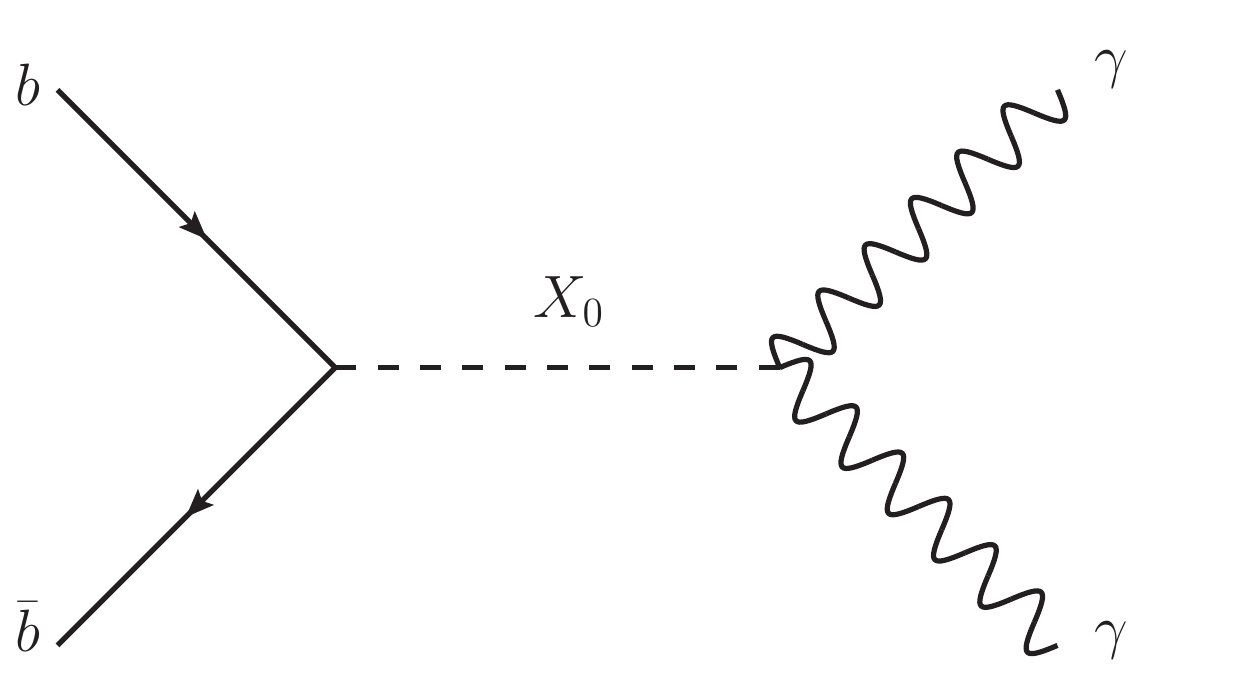}\quad
 \includegraphics[width=.24\textwidth]{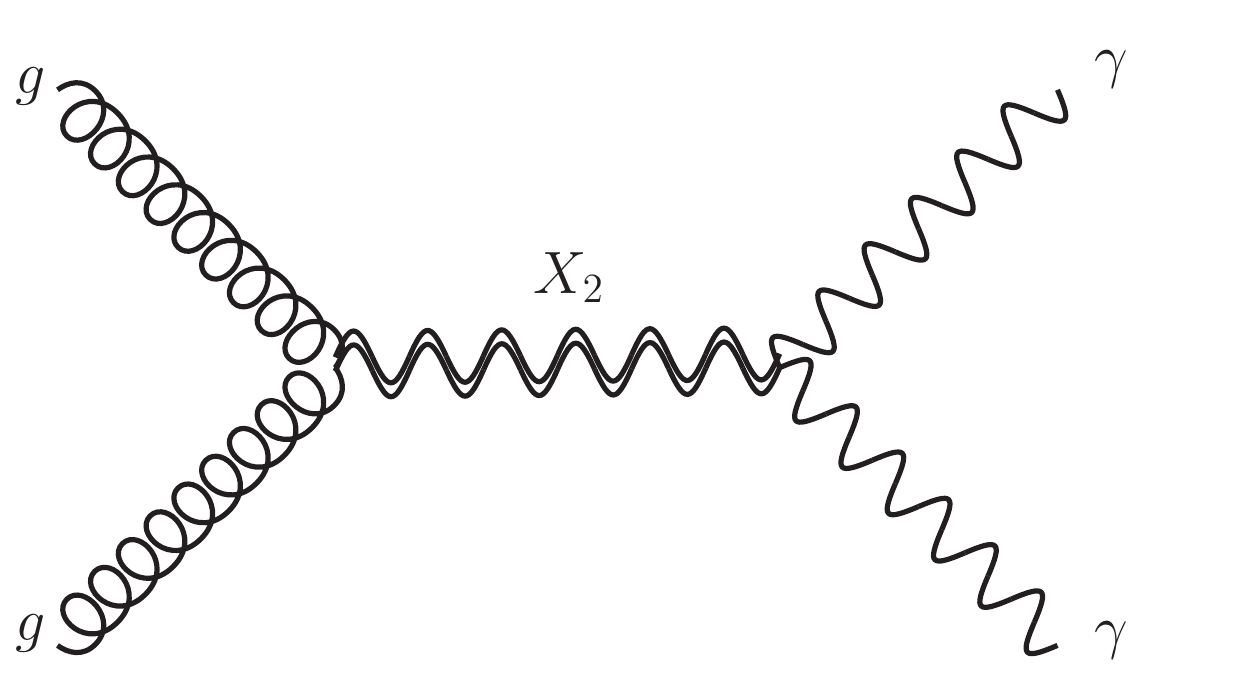}\!\! 
 \includegraphics[width=.24\textwidth]{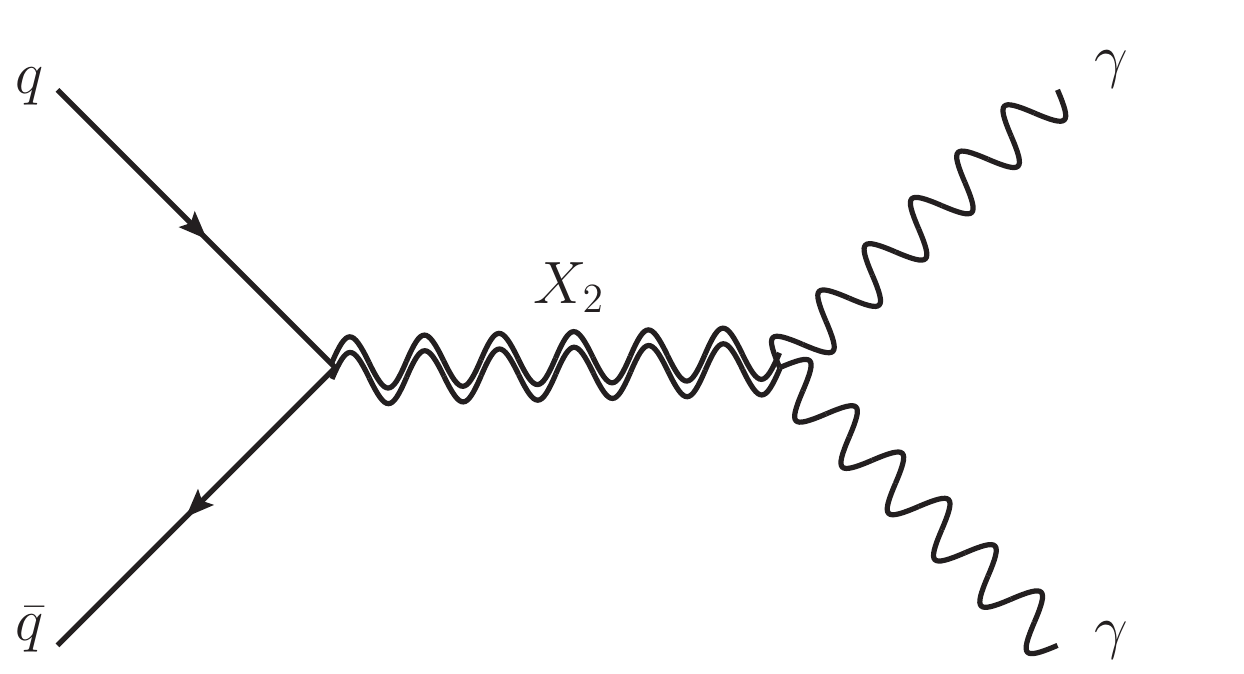}
 \caption{Diagrams of the 750~GeV resonance scenario for spin-0 (left) and spin-2 (right).}
\label{fig:diagrams1}
\end{figure}

Out of these cases, we consider $gg$ and $b\bar b$ production of a spin-0 resonance, and $gg$ and $q\bar q$ production of a spin-2 one, as depicted in Fig.~\ref{fig:diagrams1}. Note that $gg$ and $b\bar b$ production are particularly interesting because they yield the highest gain in parton luminosities when going from $\sqrt{s}=8$ to 13 TeV, being enhanced by factors of $\sim 4.7$ and $\sim 5.4$ respectively as opposed to valence quarks for which the corresponding factors are of the order of 2.5~\cite{Franceschini:2015kwy}. Similarly, in the case of vector boson fusion, naively assuming that the parton luminosities scale with the quark ones, one expects an enhancement of roughly a factor 2.7, well below the corresponding values for $b$ quarks and gluons. Besides, in the spin-0 case, if the new state is somehow involved in electroweak symmetry breaking, it is expected to couple to the SM fermions proportionally to their mass and should, hence, interact more with the third generation than with light-flavor quarks. 
For a spin-2 resonance, the prime example is the KK graviton, which has universal couplings to gluons and quarks. 
We thus consider $gg$ and $q\bar q$ production for the spin-2 case.

\subsubsection*{Spin-0}

In the spin-0 case, gluon fusion production followed by decay into a pair of photons can be parametrized by the effective Lagrangian 
\begin{align}\label{eq:LagSpin0ggf}
 {\cal L}_0^g = \frac{1}{4\Lambda}\big[
  {\ptwiddle\kappa}_{\!g}\, G_{\mu\nu}^a {\ptwiddle G}{}^{a,\mu\nu}
 +{\ptwiddle\kappa}_{\!\gamma}\, A_{\mu\nu}{\ptwiddle A}{}^{\mu\nu}
  \big] X_0\,,
\end{align}
where $G_{\mu\nu}^a$ and $A_{\mu\nu}$ are the $SU(3)_C$ and $U(1)_{\rm EM}$ field strength tensors respectively, 
$\tilde G^{a}_{\mu\nu}=\frac{1}{2}\epsilon_{\mu\nu\rho\sigma}G^{a,\rho\sigma}$ and 
$\tilde A_{\mu\nu}=\frac{1}{2}\epsilon_{\mu\nu\rho\sigma}A^{\rho\sigma}$ are their duals, and 
${\ptwiddle\kappa}_{\!g}$ and ${\ptwiddle\kappa}_{\!\gamma}$ are the 
CP-even (odd) couplings of $X_0$ to gluons and photons. 
This Lagrangian leads to $gg\to X_0\to\gamma\gamma/gg$ at the leading order (LO). 

To study the case of $X_0$ production from $b\bar b$ annihilation, we write an effective Lagragian as
\begin{align}\label{eq:LagSpin0bb}
 {\cal L}_0^{b} = {\ptwiddle\kappa}_{\!b}\, \bar b(i\gamma_5)b X_0
 +\frac{1}{4\Lambda}{\ptwiddle\kappa}_{\!\gamma}\, A_{\mu\nu}{\ptwiddle A}{}^{\mu\nu}
  X_0\,,
\end{align}
where ${\ptwiddle\kappa}_{\!b}$ parametrises the CP-even (odd) coupling of the $X_0$ to a pair of $b$-quarks. This Lagrangian leads to $b\bar b\to X_0\to\gamma\gamma/b\bar b$. 
The coefficients ${\ptwiddle\kappa}_{\!b}$ should generically be understood as ${\ptwiddle\kappa}_{\!b} \sim c_b m_b/\Lambda$ where the value of $c_b/\Lambda$ can vary from one model to another. As an example, if $X_0$ is taken to be the (heavy) pseudoscalar of a type-II two Higgs doublet model, one would expect $\Lambda \sim v = 246$ GeV and $c_b \sim \tan\beta$
(although in this case one should also include the corresponding
coupling to the top quark with $c_t \sim \cot\beta$, which would
contribute to gluon-induced production).

\subsubsection*{Spin-2}

As an alternative possibility, we also consider a massive spin-2 particle which couples to the SM gauge and matter fields through their energy--momentum tensors~\cite{Giudice:1998ck,Han:1998sg}.
As argued above, we only consider the interactions with
gluons, light quarks and photons~\cite{Martini:2016ahj}: 
\begin{align}
 {\cal L}_{2} = -\frac{1}{\Lambda}\big[
  \kappa_{g}\,T^{g}_{\mu\nu}
 +\kappa_{q}\,T^{q}_{\mu\nu}
 +\kappa_{\gamma}\,T^{\gamma}_{\mu\nu}
 \big]
 X_2^{\mu\nu}\,,
\label{lag}
\end{align}
where $X_2^{\mu\nu}$ is the spin-2 resonance and
$T^{g,q,\gamma}_{\mu\nu}$ are the energy--momentum tensors; see  
the explicit formulae, e.g., in~\cite{Han:1998sg,Hagiwara:2008jb}. 
While conventional graviton excitations have a universal coupling
strength $\Lambda^{-1}$, we adopt a more general parametrisation by introducing the coupling parameters $\kappa_{g}$,
$\kappa_{q}$ and $\kappa_{\gamma}$ without assuming any specific UV model~\cite{Ellis:2012jv,Englert:2012xt}. We consider three cases:
\begin{align}
	R \equiv \kappa_{q} / \kappa_{g} = \{0.1,\,1,\,10\}\,,
\end{align}
corresponding to the gluon-dominant, universal coupling, and quark-dominant scenarios, respectively.
These scenarios amount to 99\%, 87\% and 7\% gluon fusion contributions to the total 750~GeV spin-2 resonance production cross section at the 13~TeV LHC respectively~\cite{Martini:2016ahj}.

\subsection{Heavier parent resonance}\label{sec:scenariosParent}

Another way to induce a peak in the diphoton invariant mass distribution is by invoking more complicated decays or decay  chains of  a heavier parent particle, leading to three (or more)-body final states. In this case, one can envisage a number of different topologies. 
In our study, we consider the following possibilities:

\begin{figure}
\center
 \includegraphics[width=.24\textwidth]{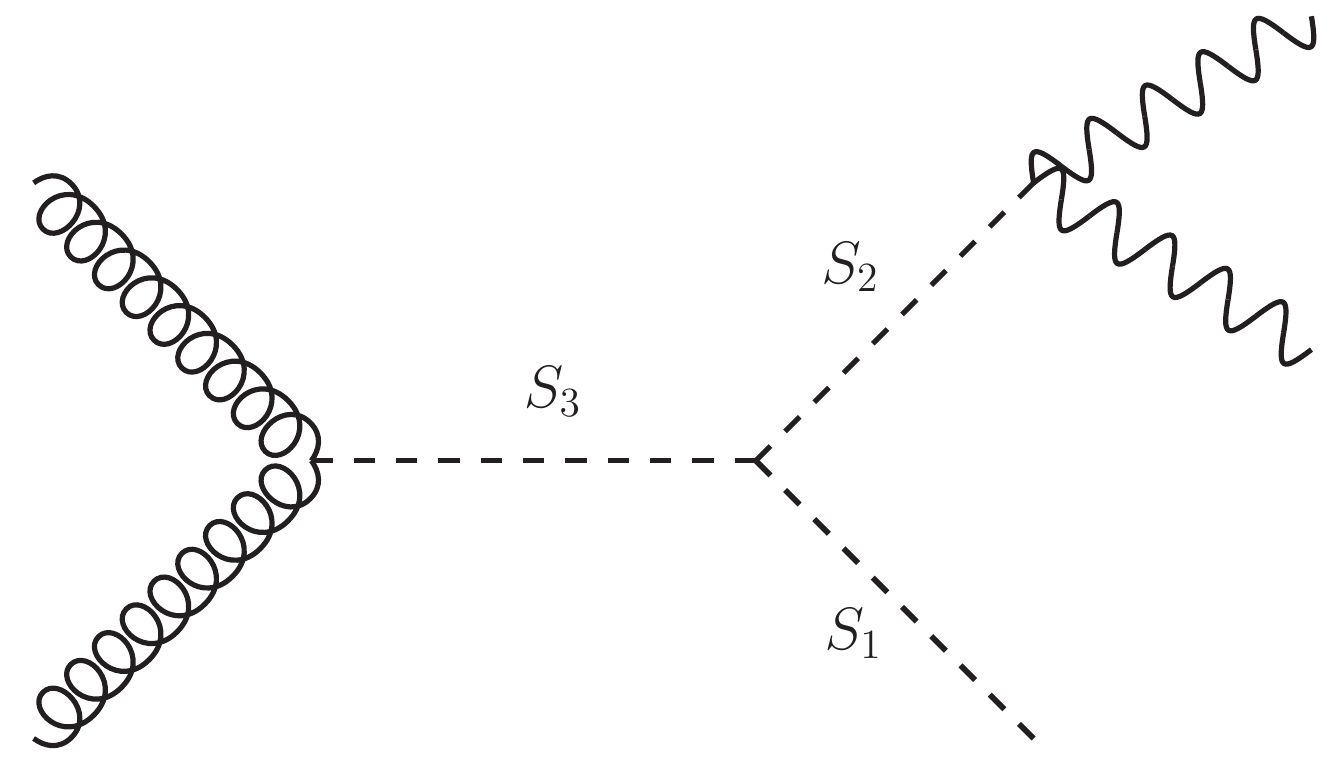}
 \includegraphics[width=.24\textwidth]{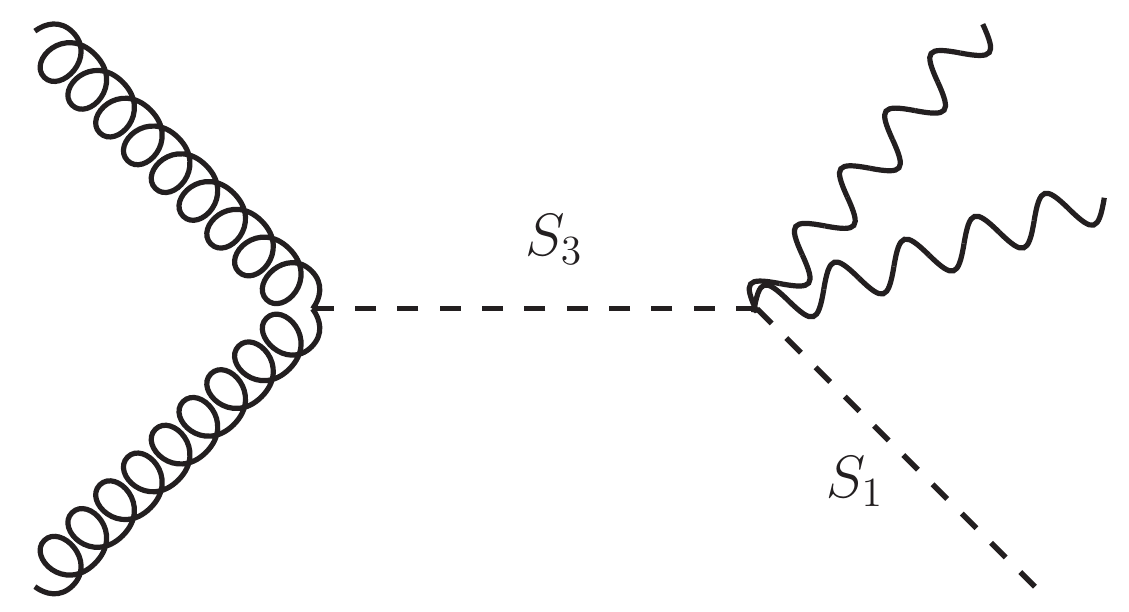}
 \includegraphics[width=.24\textwidth]{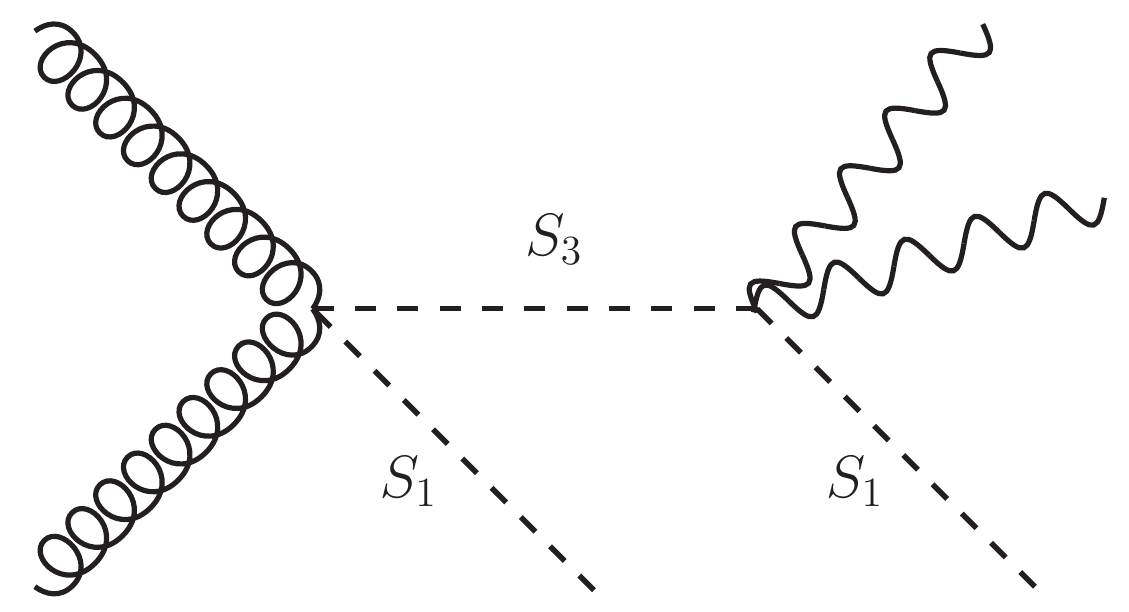}
 \includegraphics[width=.24\textwidth]{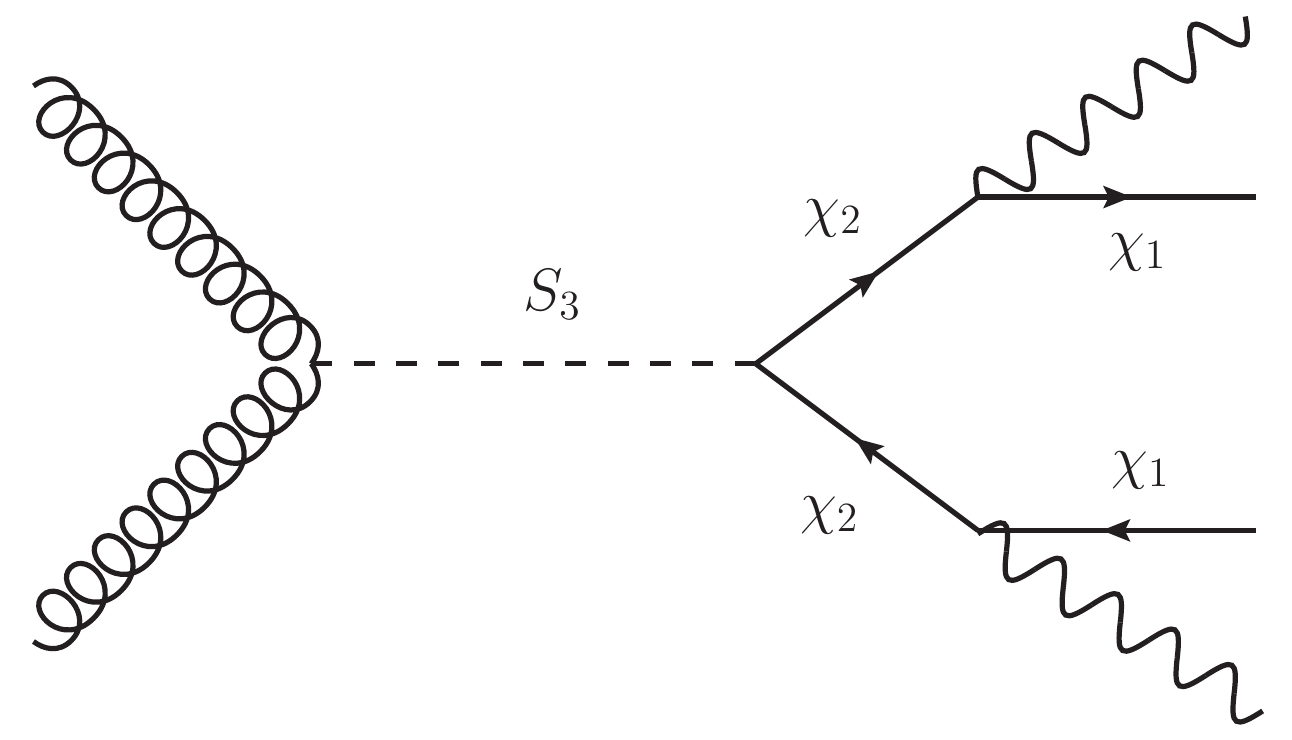}
 \caption{Diagrams for the heavier resonance scenarios I--IV.}
\label{fig:diagrams2}
\end{figure}

\begin{enumerate}[I)]
\item A process of the type $pp\to S_3 \to S_2+S_1$, $S_2\to \gamma\gamma$
with $S_1$ being invisible or leading to soft decay products. Such a scenario could e.g. be motivated by considering dark matter or ``hidden valley''~\cite{Strassler:2006im} models. 
In principle $S_3$ can be a fundamental scalar or vector~\cite{Franceschini:2015kwy} or a composite particle ($Q\bar Q$ bound state)~\cite{Harigaya:2015ezk}. 
\item A 3-body decay scenario with single production of the heavier resonance: $pp\to S_3\to S_1\gamma\gamma$~\cite{Bernon:2015abk}, where $S_1$ is again invisible or decays softly. This scenario is equivalent to the previous one in the limit that $S_2$ is heavy (virtual).
\item A 3-body decay scenario as above but with associated production of $S_3$ and $S_1$: $pp\to S_3 S_1$, $S_3\to S_1\gamma\gamma$~\cite{Bernon:2015abk}. 
Such a scenario has the advantage of allowing for the existence of a new conserved quantum number under which $S_1$ and $S_3$ are charged, and which would stabilize $S_1$, the lightest state of the new sector. 
\item The ``antler'' topology from a process of the type $pp\to S_3\to \chi_2\chi_2$, $\chi_2\to \chi_1+\gamma$ as proposed in~\cite{Cho:2015nxy}.%
\footnote{Note that in this scenario we take the $\chi_i$'s to be fermions, as the scalar case cannot be rendered gauge invariant at the leading operator order unless $\chi_{1,2}$ are mass degenerate.}
While it is rather difficult to envisage a realistic scenario, this topology is reminiscent of general gauge-mediated supersymmetry breaking scenarios \cite{Meade:2008wd} with a sufficiently short-lived neutralino NLSP, see e.g. \cite{Ruderman:2011vv}. 
\end{enumerate}

For concreteness,
we concentrate on gluon-initiated production of a CP-even spin-0 parent resonance. The relevant diagrams are shown in Fig.~\ref{fig:diagrams2}. 
These scenarios can be described by the Lagrangians 
\begin{align}\label{eq:heavyg}
 {\cal L}^g_H &= \sum_{i,j=1,2,3}
 \left[ \frac{1}{4\Lambda}
  \kappa_{i}^{ggS} G_{\mu\nu}^aG^{a,\mu\nu}S_i 
  +\frac{1}{4\Lambda^2}\kappa_{ij}^{ggSS} G_{\mu\nu}^aG^{a,\mu\nu}S_iS_j 
  \right] 
  \,,\\
\label{eq:heavysf}
 {\cal L}^{S,\chi}_H &= \sum_{i,j,k=1,2,3} \sum_{l,m=1,2}
  \left[
   \kappa_{ijk}^{SSS} m_3 S_i S_j S_k 
  +\kappa_{ilm}^{S\chi\chi} S_i\bar\chi_l\chi_m
  \right]
  \,,\\
\label{eq:heavygam}
 {\cal L}^{\gamma}_H &= \sum_{i,j=1,2,3} \sum_{l,m=1,2}
 \left[
   \frac{1}{4\Lambda}
   \kappa_{i}^{\gamma\gamma S} A_{\mu\nu}A^{\mu\nu}S_i 
  +\frac{1}{4\Lambda^2}\kappa_{ij}^{\gamma\gamma SS} A_{\mu\nu}A^{\mu\nu}S_iS_j \right. \nn\\
  & \hspace*{3cm} \left.\phantom{\frac{1}{4\Lambda}} + \frac{1}{\Lambda}\kappa_{ij}^{\gamma\chi\chi} \left( A_{\mu\nu}
   \bar\chi_l\sigma^{\mu\nu}\chi_m + h.c. \right)
  \right] 
  \,,
\end{align}
where in the last (magnetic-type) operator, $\sigma^{\mu\nu}=\frac{i}{2} \left[\gamma^\mu, \gamma^\nu \right]$. The couplings relevant for each of the scenarios I--IV are summarised in Table~\ref{tab:scenarios14}, together with the mass combinations that we consider as benchmarks. For simplicity, and regardless of naturalness arguments, all other couplings are taken to be zero throughout the subsequent analyses.

\begin{table}
\center
\begin{tabular}{lllccc}
\hline
 & scenarios & relevant couplings & $m_{3}$ & $m_{2}$ & $m_{1}$ \\
\hline
 I & Sequential resonance & $\kappa_{3}^{ggS}$, $\kappa_{321}^{SSS}$, $\kappa_{2}^{\gamma\gamma S}$ 
   & 1200/1200 & 750/750 & 440/40 \\
 II & 3-body decay (single) & $\kappa_{3}^{ggS}$, $\kappa_{31}^{\gamma\gamma SS}$ 
   & 900/1800 & heavy & 43/977 \\
 III & 3-body decay (assoc.) & $\kappa_{31}^{ggSS}$, $\kappa_{31}^{\gamma\gamma SS}$ 
   & 900/1800 & heavy & 43/977 \\
 IV & Antler & $\kappa_{3}^{ggS}$, $\kappa_{322}^{S\chi\chi}$, $\kappa_{21}^{\gamma\chi\chi}$ 
   & 1700/1600 & 849/798 & 175/10 \\
 \hline
\end{tabular}
\caption{\label{tab:scenarios14} Coupling assignments and benchmark mass combinations for scenarios I--IV; `single' and `assoc.' mean single and associated production as illustrated by the second and third diagram of Fig.~\ref{fig:diagrams2}, respectively.}
\end{table}

In \cite{Kim:2015ron} a scenario similar to our scenario I was considered, namely production of a heavier (pseudo)scalar resonance which decays into a pair of new pseudoscalars with mass of 750 GeV, which decay further into photons. This would correspond to scenario I with $m_2=m_1$ and is discussed in Appendix~\ref{app:s1special}.

For the antler topology, scenario IV, we note that one has to finely adjust the masses of $S_3$ and $\chi_2$, and even more so their decay widths, in order to obtain the desired diphoton invariant mass spectrum. For our simulations, we use $\Gamma(S_3)=0.1$ and $\Gamma(\chi_2)=0.05$~GeV for the antler topology. The sensitivity on the masses and widths will be commented upon in Appendix~\ref{app:antler}. For scenarios I--III, we use $\Gamma(S_3)=10$~GeV, which does not significantly affect the relevant distributions.  

We note that these cases do by no means exhaust all the possibilities for reproducing the 750 GeV excess. For instance, we do not consider scenarios with very light states decaying into pairs of highly boosted photons which would be misidentified as individual photons. Examples for this are $pp\to Z' \to \gamma s \to 3\gamma$~\cite{Chala:2015cev} or  $pp\to S \to aa$, $a \to \gamma\gamma$ \cite{Knapen:2015dap,Agrawal:2015dbf,Chang:2015sdy,Bi:2015lcf,Aparicio:2016iwr,Ellwanger:2016qax,Dasgupta:2016wxw}. Exhausting all these possibilities is beyond the scope of this work.

\section{Event simulations}\label{sec:simulations}

While we employ the Higgs
Characterisation (HC)~\cite{Artoisenet:2013puc} model
for the 750~GeV resonance scenarios,%
\footnote{The model file is publicly available at the
{\sc FeynRules} repository~\cite{FR-HC:Online}. Although the HC model is designed to study the spin--parity nature of the 125~GeV Higgs boson, one can easily change its mass $m_{X_{0,2}}$ as a parameter.}
we implemented the Lagrangians for heavier parent resonances in 
{\sc FeynRules}~\cite{Alloul:2013bka} to generate the model files which can be
interfaced~\cite{Degrande:2011ua,deAquino:2011ub}
to event generators. 

We generate inclusive signal samples 
by using the tree-level matrix-element plus parton-shower (ME+PS) merging procedure.
In practice, we make use of the shower-$k_T$ scheme~\cite{Alwall:2008qv}, implemeted in 
{\sc MadGraph5\_aMC@NLO}~\cite{Alwall:2014hca} with {\sc Pythia6}~\cite{Sjostrand:2006za}, and  
generate signal events with parton multiplicity from zero to two, e.g. $pp\to X_{0,2}+0,1,2$ partons. 
The merging separation parameter is set to $Q_{\rm cut}=200$~GeV for the 750~GeV resonance scenarios
and to  $Q_{\rm cut}=200,250,300$~GeV for the heavier parent scenarios with $m_3=900,1200,1600-1800$~GeV, 
respectively. 
Hadron-level events are analyzed in {\sc MadAnalysis5}~\cite{Conte:2012fm},
where we define jets by using the anti-$k_T$
algorithm~\cite{Cacciari:2008gp} as implemented in 
{\sc FastJet}~\cite{Cacciari:2011ma} with the jet cone radius $R=0.5$.

At the analysis level, we require for the transverse momentum $p_T$ and pseudorapidity $\eta$ of the photons and jets
\begin{align}
 &p_T(\gamma)>25~{\rm GeV}\,,\quad |\eta(\gamma)|<2.5\,,\\
 &p_T(j)>25~{\rm GeV}\,,\quad |\eta(j)|<5
\end{align}
respectively. Moreover, we assume 100\% reconstruction efficiency for photons and, when relevant, for $b$-jets. This is justified because we are only comparing shapes of distributions, not overall rates. The mild $p_T$ and $\eta$ dependence of the efficiencies can be neglected for our purpose.

\section{Results}\label{sec:distributions}

\subsection{750~GeV resonance}\label{sec:dis750}

\begin{figure}
\center
\includegraphics[width=0.45\textwidth]{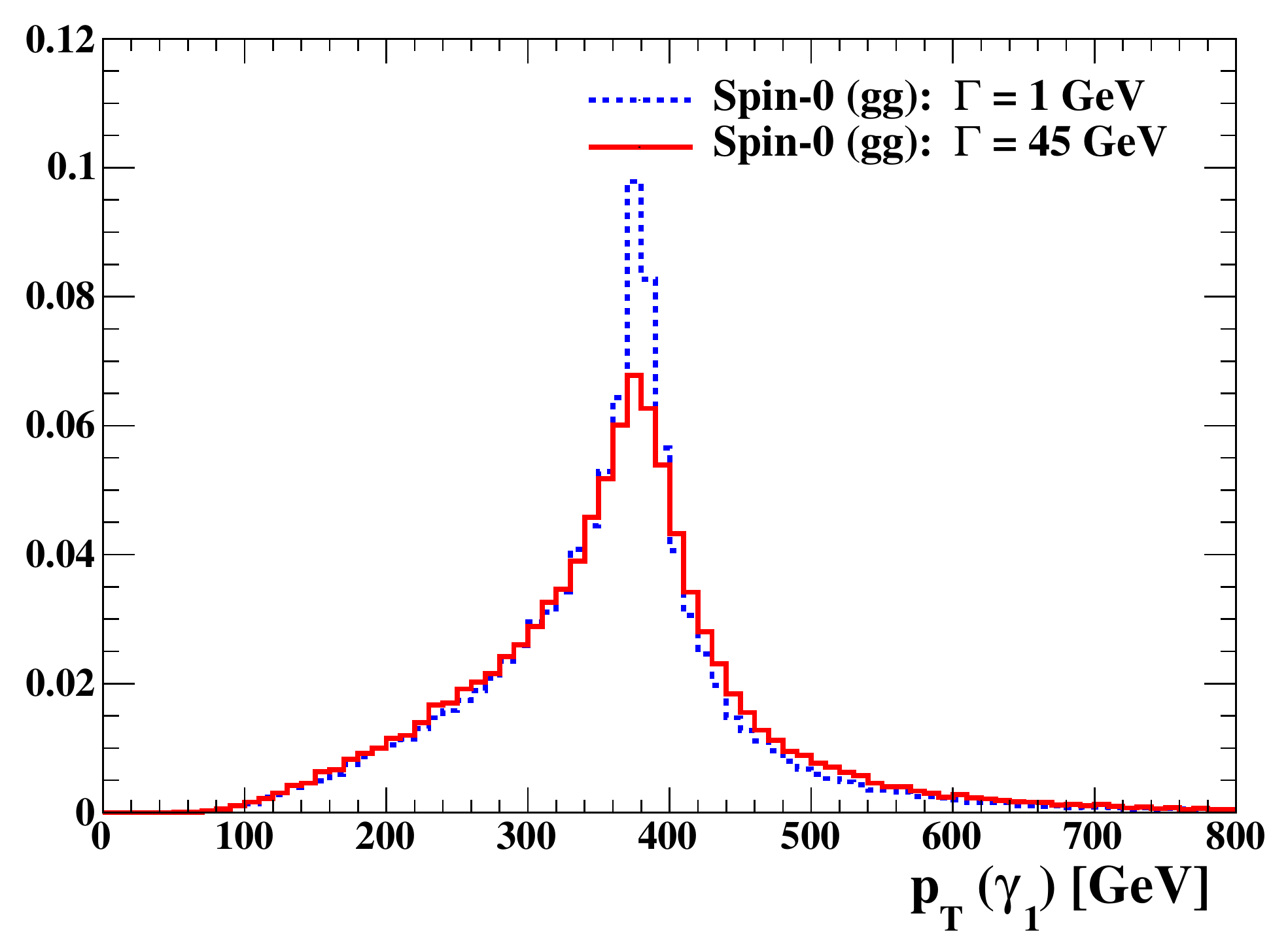}\quad
\includegraphics[width=0.45\textwidth]{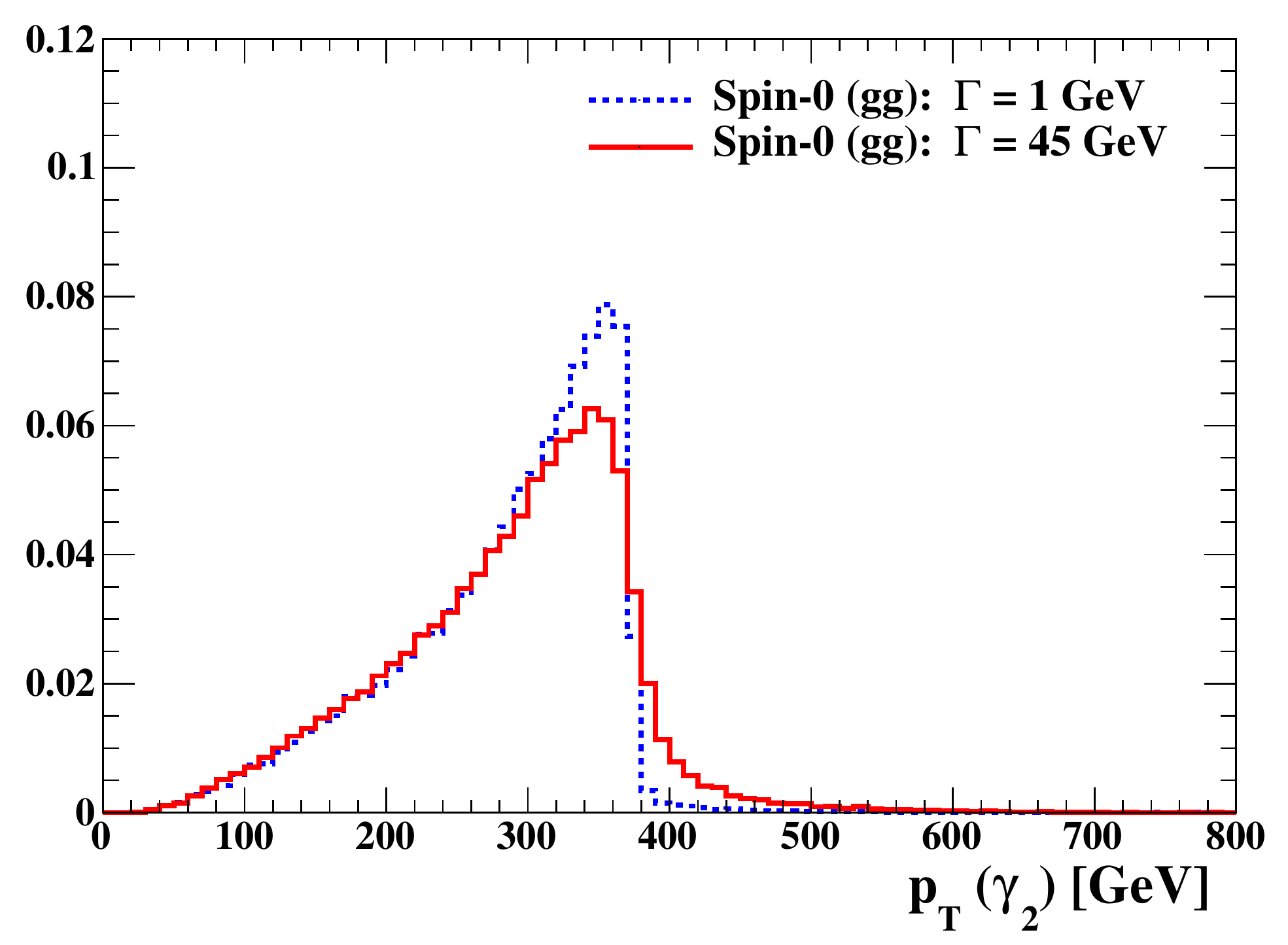}
\caption{ 
Normalised transverse momentum distributions of the leading and subleading photons, $p_T(\gamma_1)$ (left) and $p_T(\gamma_2)$ (right), for diphoton events produced from a gluon-induced 750~GeV spin-0 resonance at the 13~TeV LHC, comparing the narrow resonance ($\Gamma=1$~GeV) and the broad resonance ($\Gamma=45$~GeV) cases. }
\label{fig:width}
\end{figure}

We begin by considering the effect of a narrow vs.\ broad nature of a 750~GeV resonance on kinematic distributions. It turns out that there is very little sensitivity to the width, the largest effect occurring for the transverse momentum distributions of the photons  illustrated in Fig.~\ref{fig:width} for the case of a 750 GeV spin-0 resonance. In particular, for the leading photon  $p_T(\gamma_1)$ is more peaked for a smaller width. An analogous behaviour is observed for the second photon, where for a narrow width, the $p_T(\gamma_2)$ distribution has a sharp cut-off near $m_{X_{0}}/2=375$~GeV, while for a broad width there is a larger tail towards higher $p_T$ values. 
All other distributions that we will consider, including the transverse momentum of the diphoton system, $p_T(\gamma_1\gamma_2)$, show very little sensitivity to the width. 
The picture is essentially the same for a spin-2 resonance. In the following we will therefore consider only the $\Gamma=45$~GeV case for 750~GeV resonances and contrast it to the distributions obtained for heavier parents. 

\begin{figure}
\center
\includegraphics[width=.33\textwidth]{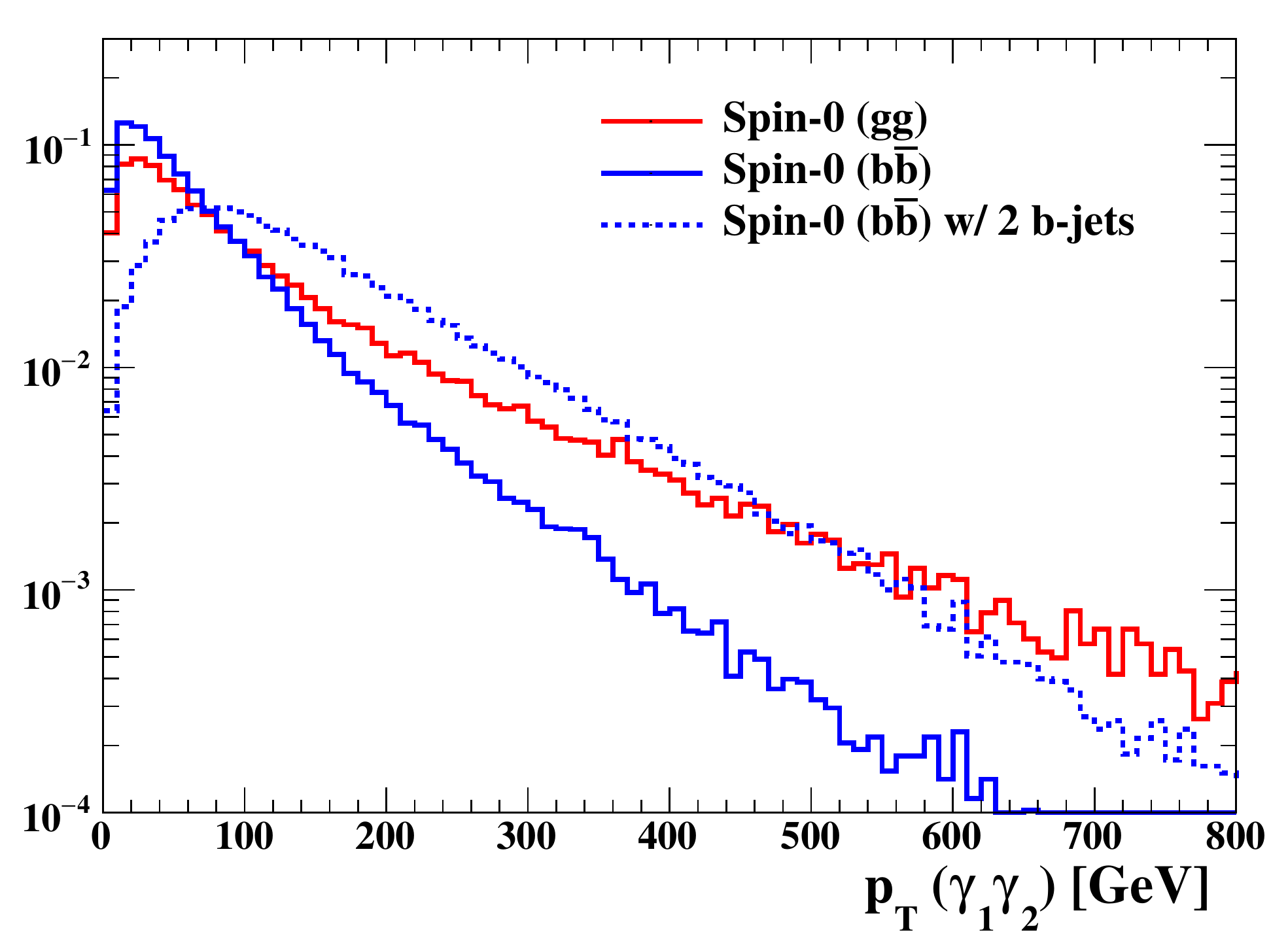}%
\includegraphics[width=.33\textwidth]{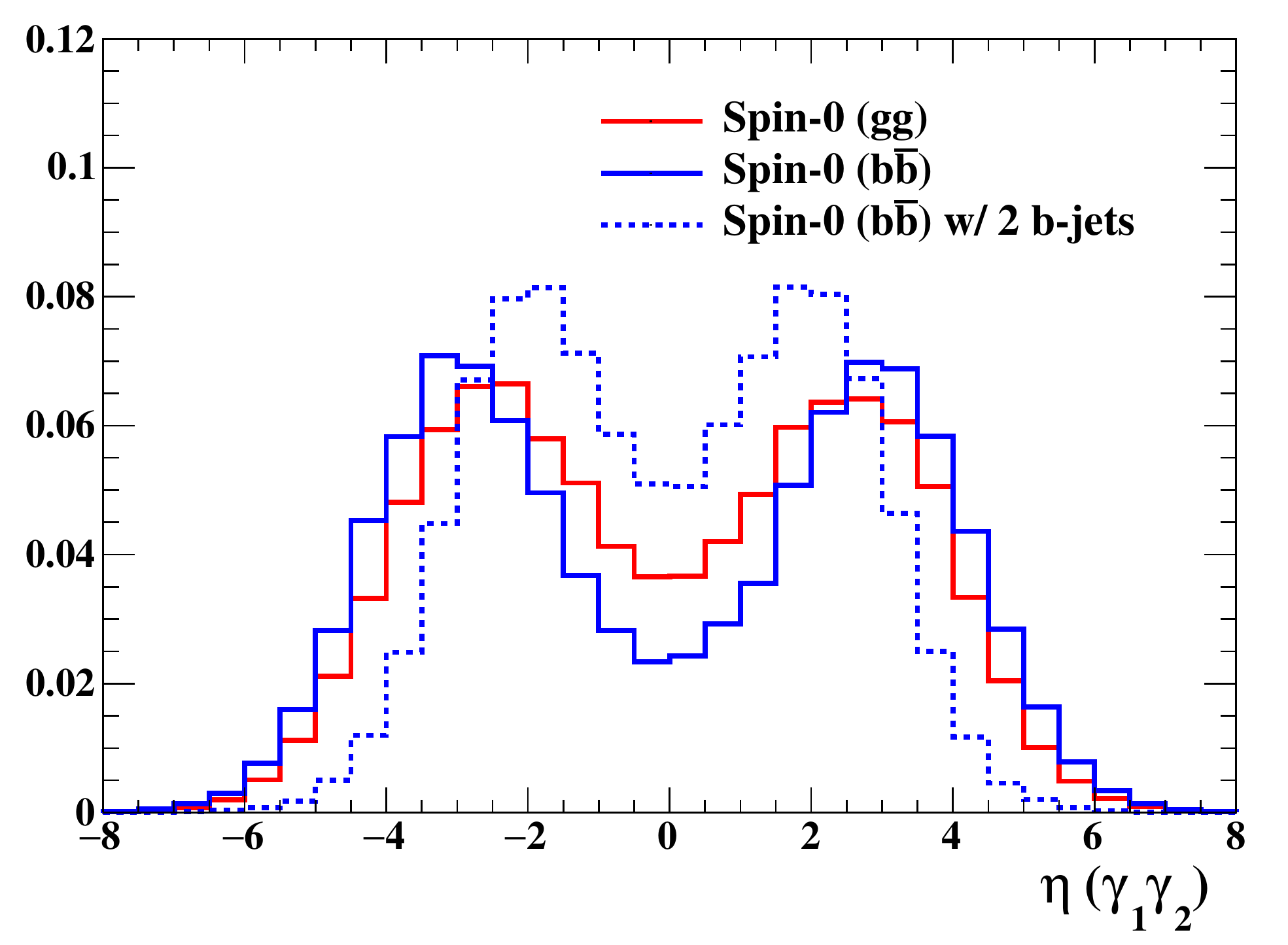}%
\includegraphics[width=.33\textwidth]{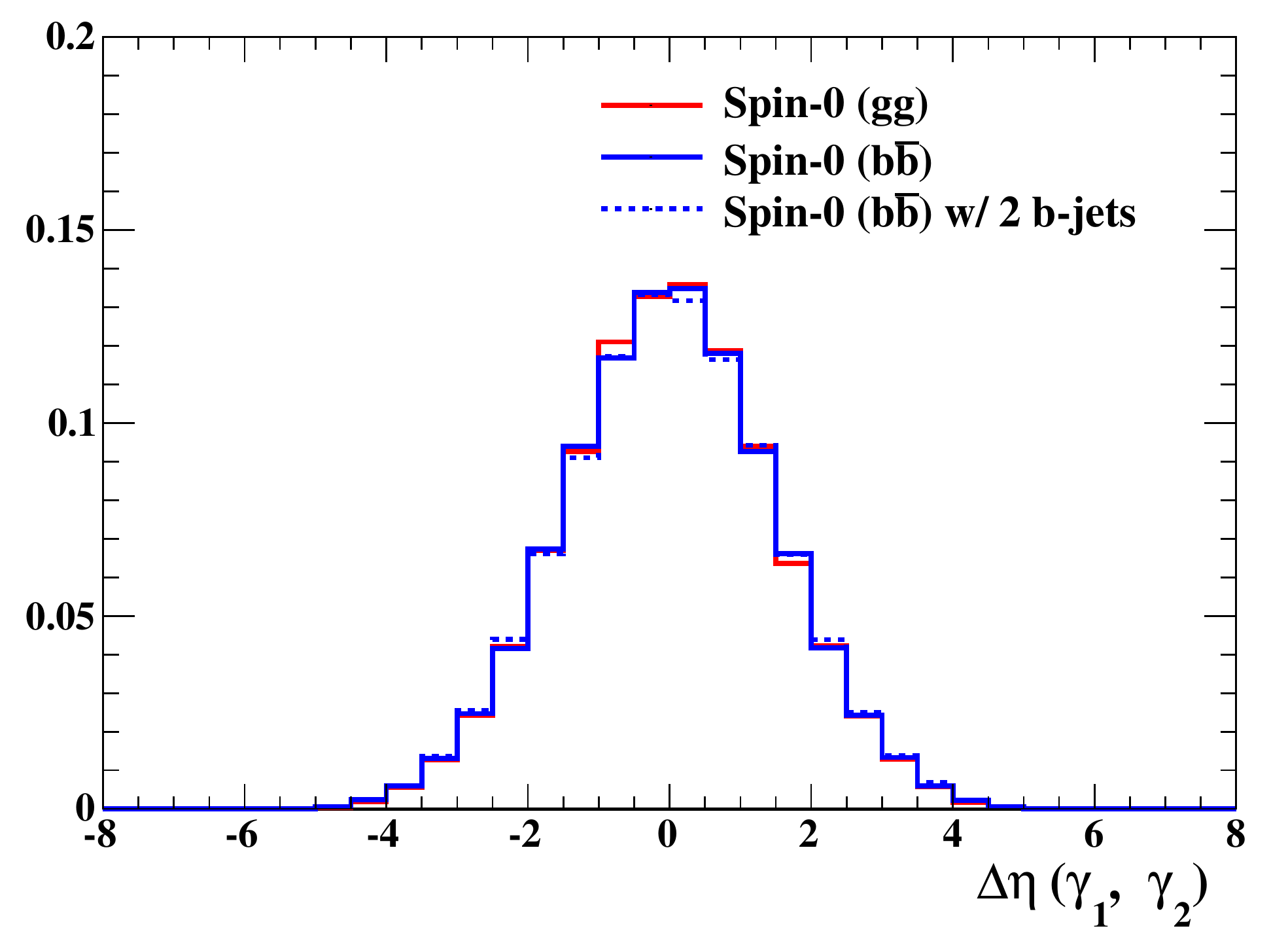}
\includegraphics[width=.33\textwidth]{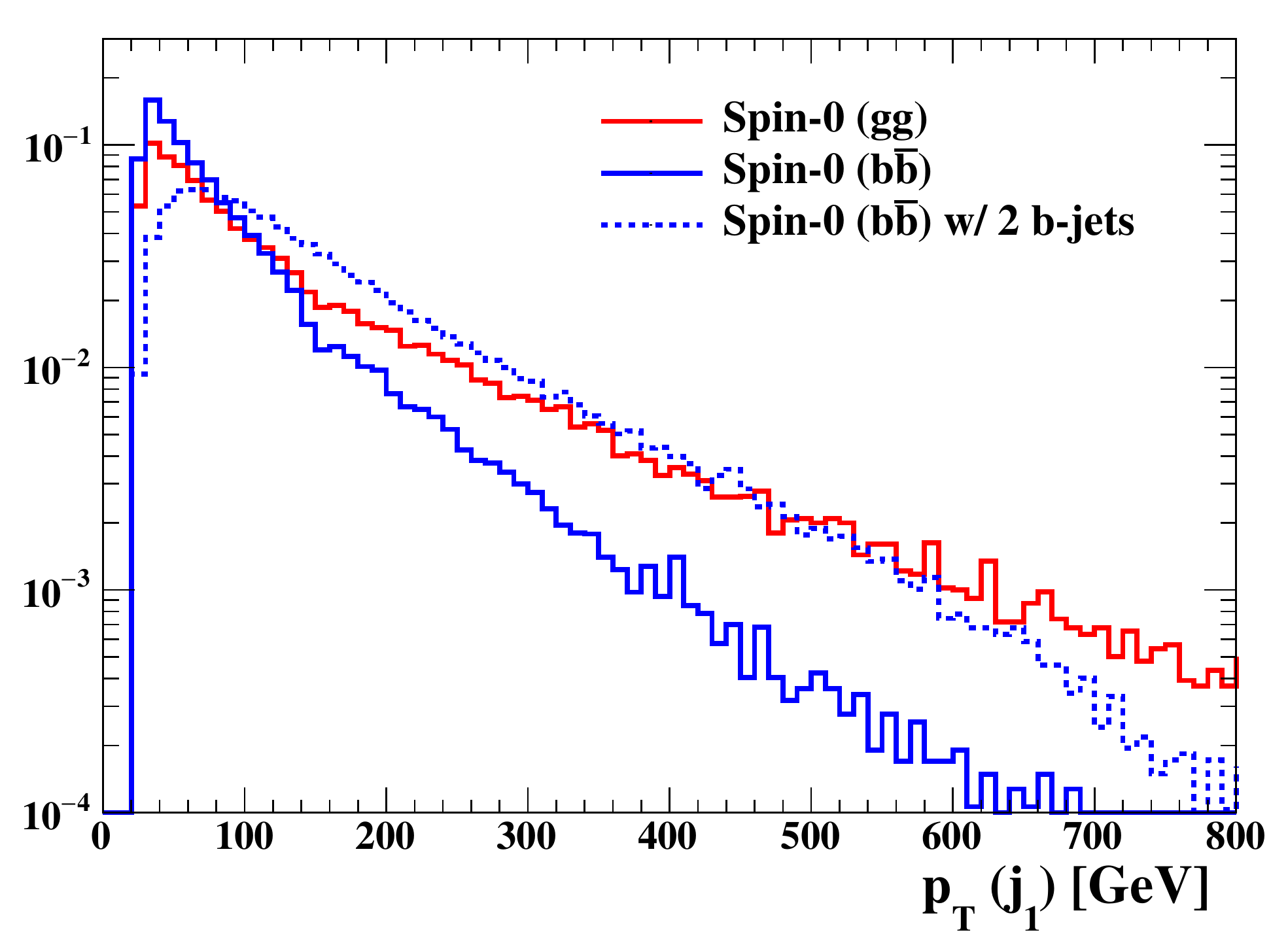}%
\includegraphics[width=.33\textwidth]{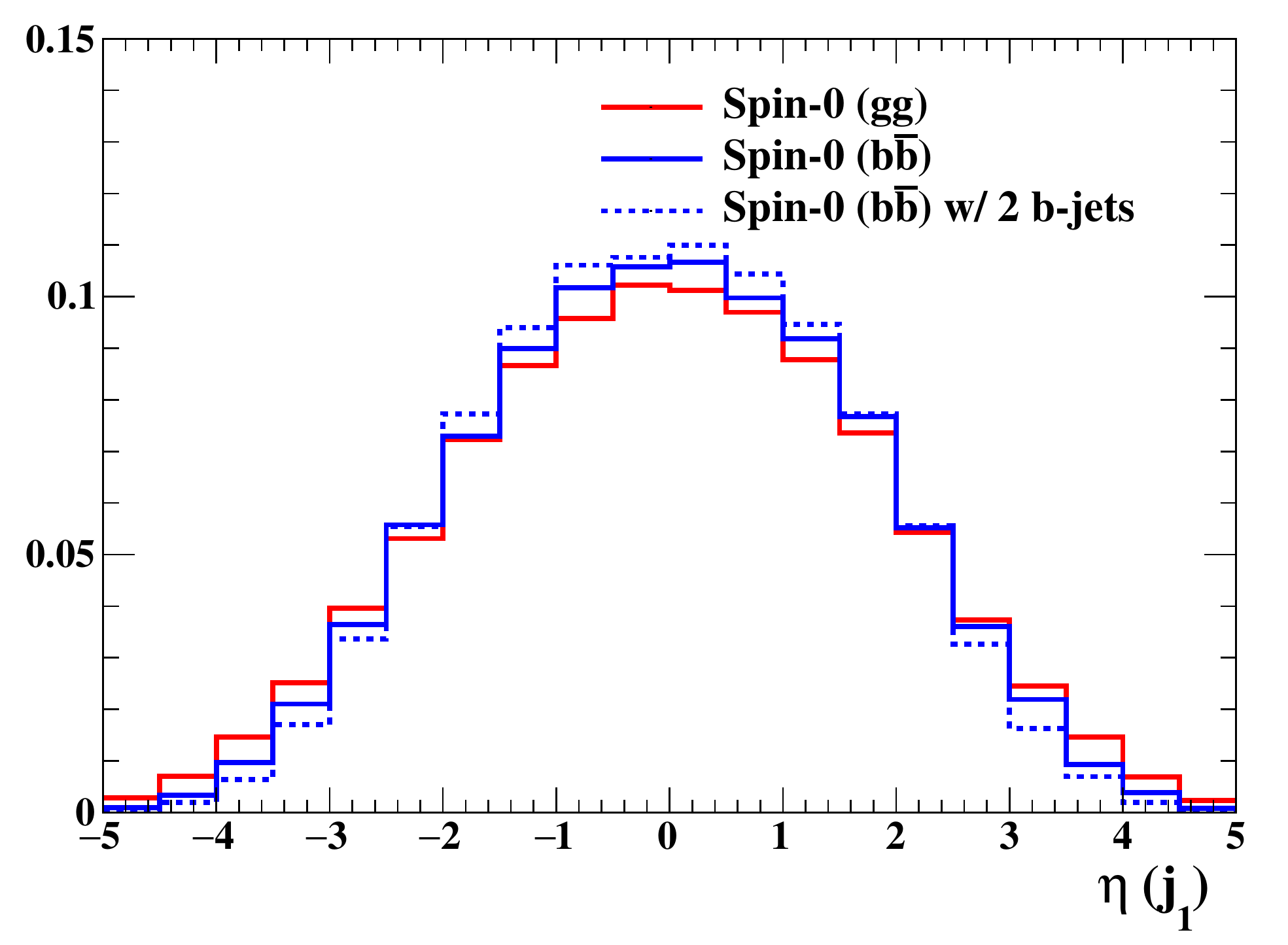}%
\includegraphics[width=.33\textwidth]{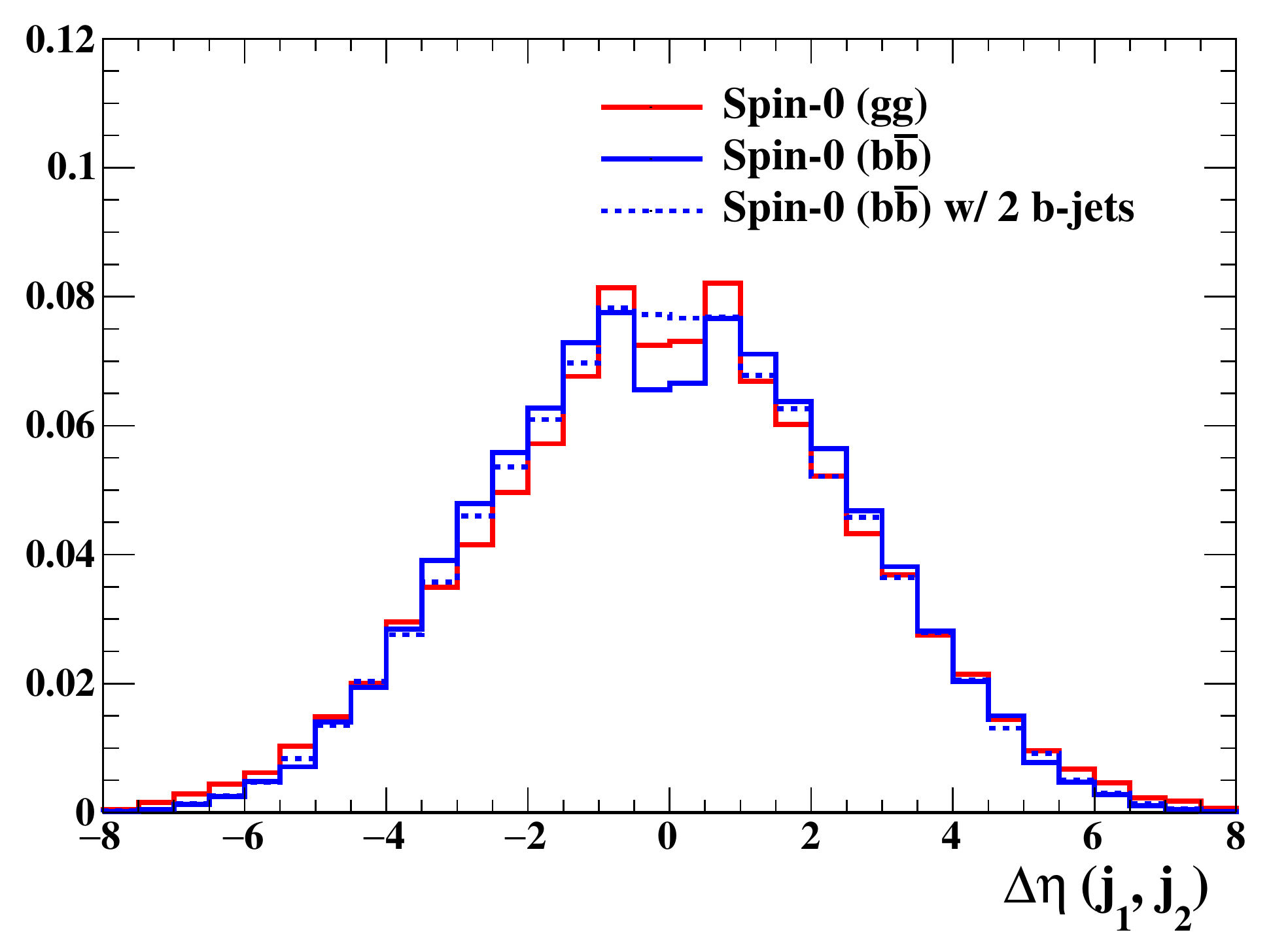}
\includegraphics[width=.33\textwidth]{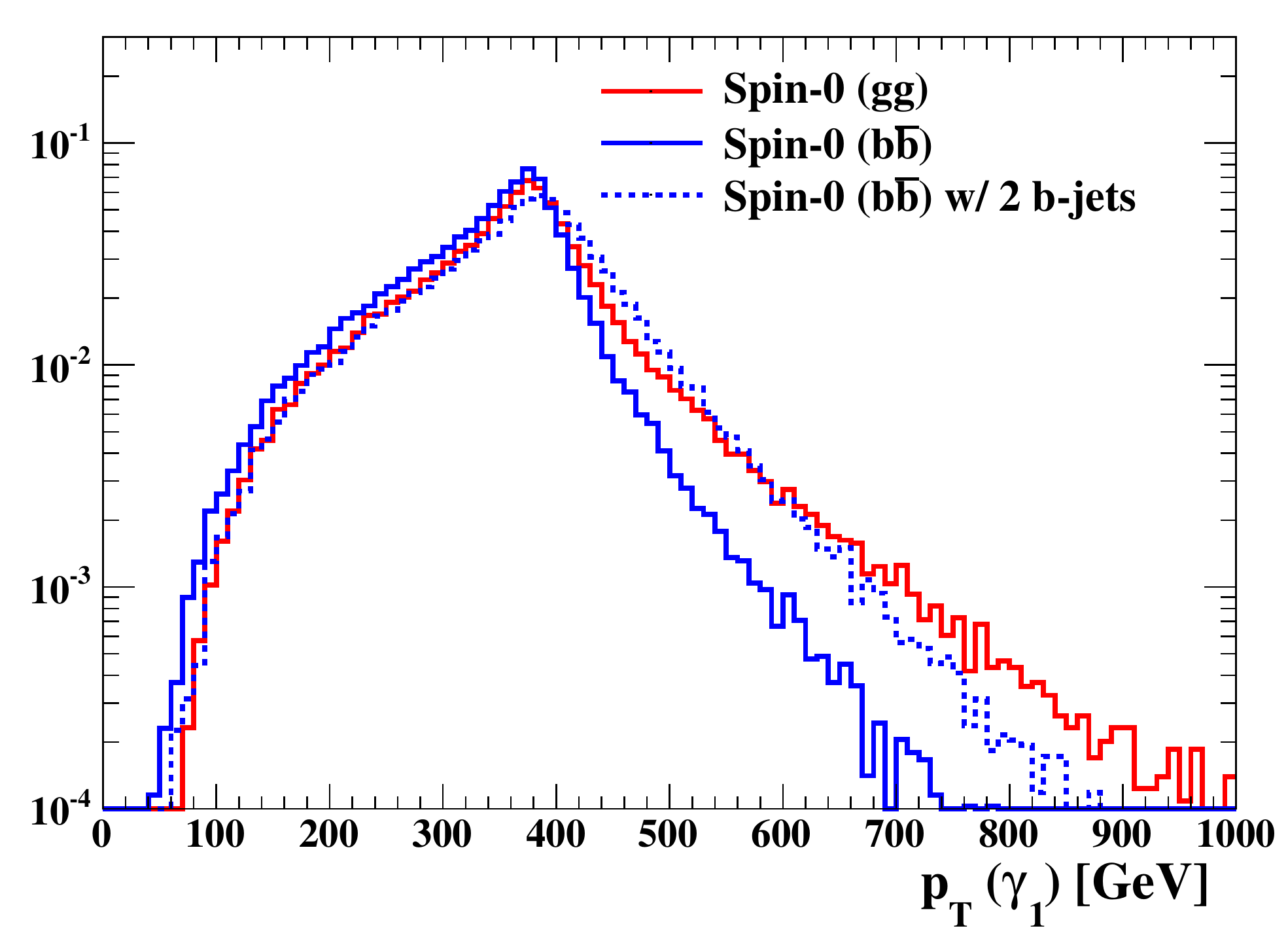}%
\includegraphics[width=.33\textwidth]{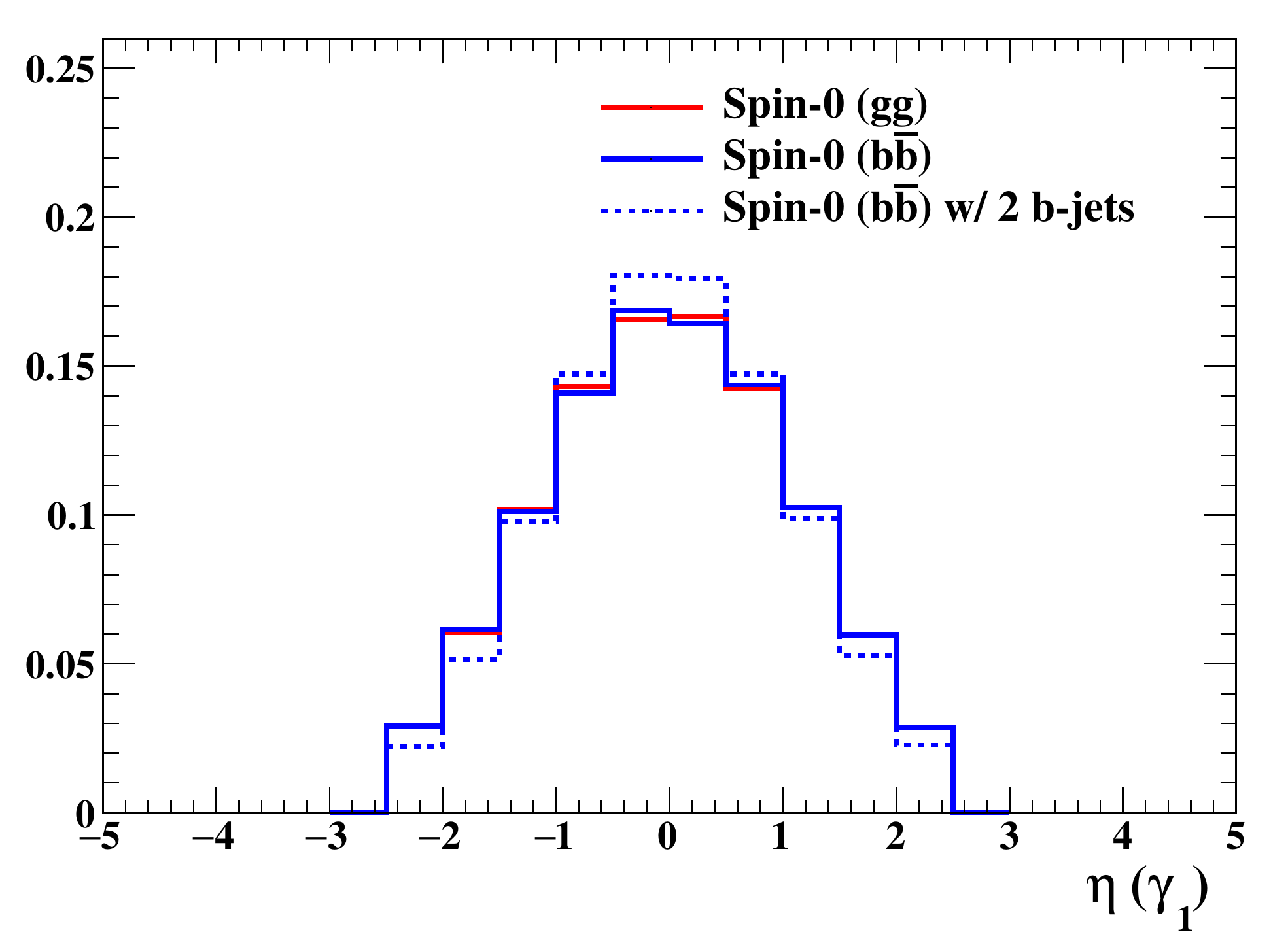}%
\includegraphics[width=.33\textwidth]{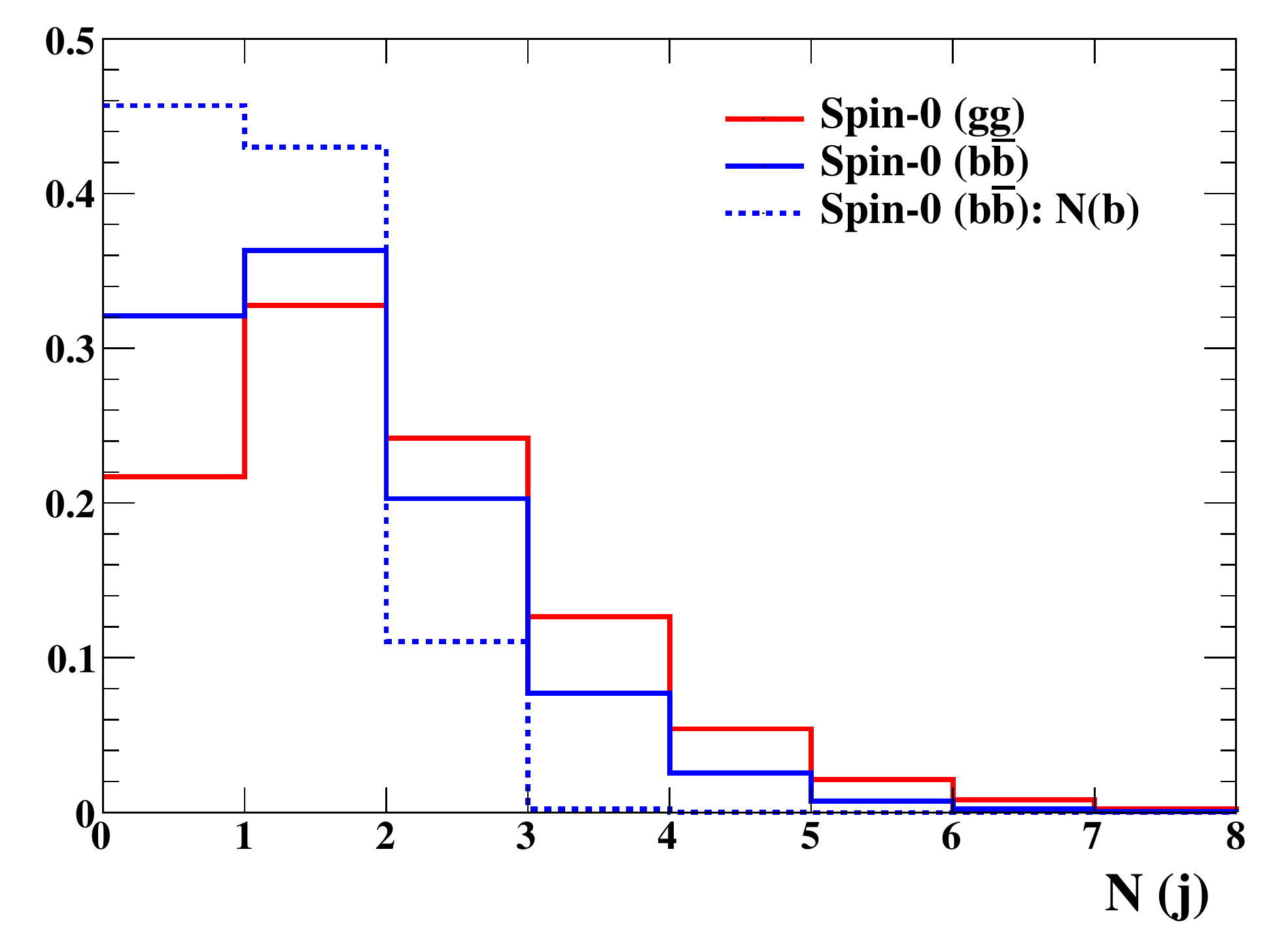}
\caption{
Normalised distributions in $p_T$ and $\eta$ of the diphoton system and the leading jet and photon as well as in $\Delta\eta$ of the photons and jets for the 750~GeV spin-0 scenario at the 13~TeV LHC.
The number of ($b$-)jets is also presented.
The gluon-induced and $b$-quark-induced cases are shown by red and blue solid lines, respectively.
The diphoton events with two $b$-jets for the $b$-induced case are also shown by blue dashed lines.
 }
\label{fig:spin0}
\end{figure}

Next, Fig.~\ref{fig:spin0} compares transverse momentum and pseudo-rapidity distributions of the diphoton system and the leading jet and photon as well as rapidity separation ($\Delta\eta$) distributions of jets and photons
for a spin-0 resonance produced in either $gg$ (red lines) or $b\bar b$ (blue lines) fusion. The blue dashed lines depict the case of associated production with a $b\bar b$ pair, $pp\to X_0b\bar b$ with 2 $b$-tagged jets. We also show the number of jets $N(j)$ (or $b$-jets $N(b)$).
We observe that QCD radiation leads to a rather hard $p_T$ spectrum of the diphoton system, which is zero at the LO, especially in the $gg$ fusion production case.
Moreover, a gluon-induced diphoton resonance tends to be produced more towards the central region than a $b\bar b$ one and involves a higher jet activity. 
The distinction between $gg$ and $b\bar b$ initiated production can be further enhanced by requiring 2 $b$-tagged jets, 
which changes the distributions for the $b\bar b$ case in a distinct way (for $gg$ production most events would be rejected). 
It is clear that the extra jet activity affects the higher tail of the $p_T$ of the leading photon.
The dependence of transverse momentum and rapidity distributions on the $gg$ and different flavour $q\bar q$ initial states was discussed in detail in~\cite{Gao:2015igz}.  

\begin{figure}
\center
\includegraphics[width=.5\textwidth]{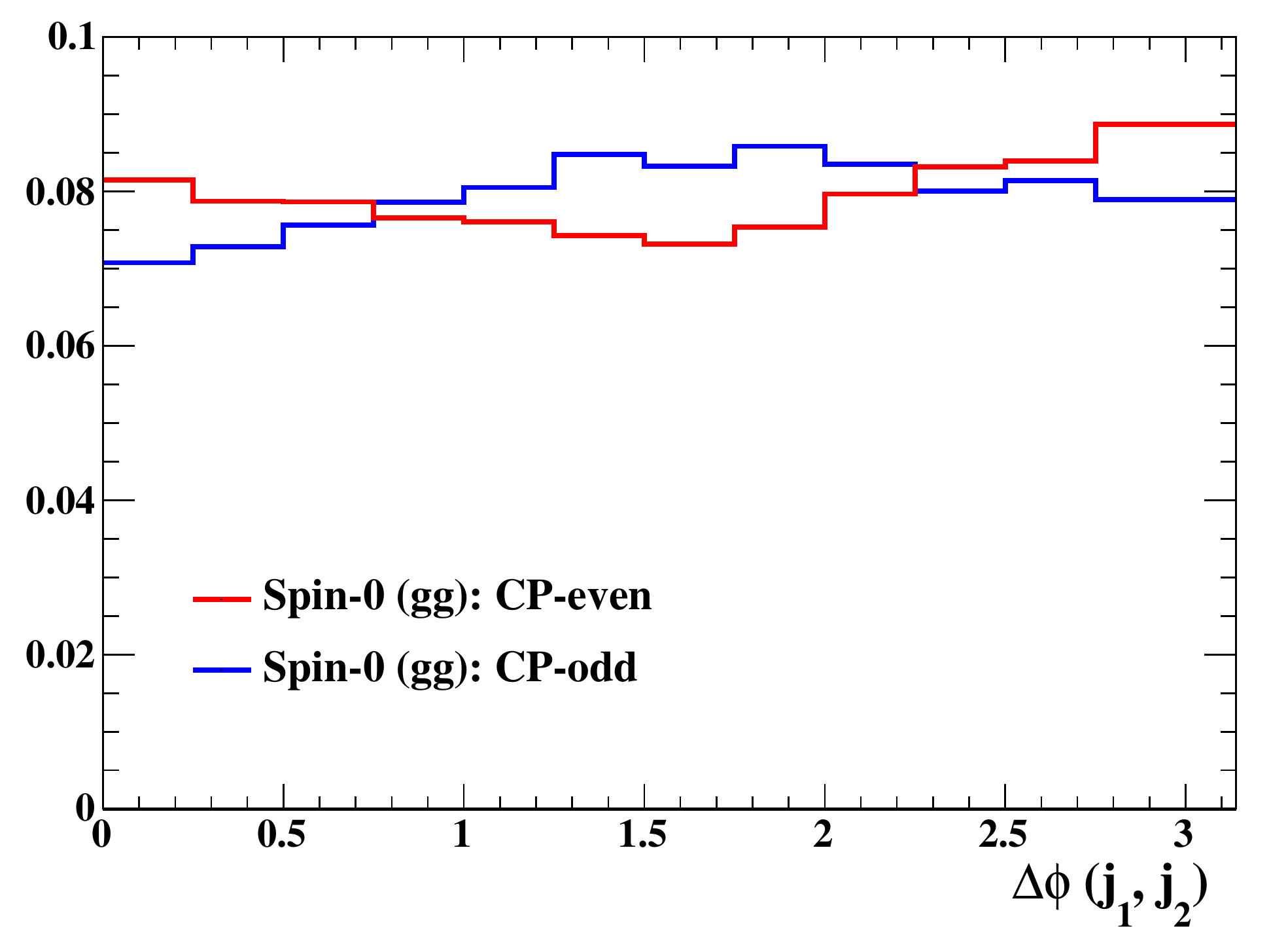}
\caption{
Normalised distribution of the azimuthal angle between the two tagging jets in $X_0+2\,{\rm jets}$ events for the CP-even (red) and -odd (blue) $X_0$ cases, requiring $|\Delta\eta(j_1,j_2)|>4$.
}
\label{fig:spin0cp}
\end{figure}

The distributions in Fig.~\ref{fig:spin0} are insensitive to the CP nature of $X_0$. 
To distinguish a CP-even from a CP-odd $X_0$ from the diphoton final state, one would need to exploit the azimuthal angle correlation of extra jets in $2\gamma+2\,{\rm jets}$ events with a vector boson fusion-like cut on $m(jj)$ and/or $\Delta\eta(j,j)$, analogous to Higgs studies~\cite{DelDuca:2001fn,Hankele:2006ja,Klamke:2007cu,Hagiwara:2009wt,Englert:2012xt}.
This is illustrated in Fig.~\ref{fig:spin0cp} but will require high luminosity to be potentially measurable.
We note that a heavier $X_0$ leads to a smaller $\Delta\phi_{jj}$ oscillation~\cite{Hagiwara:2009wt}, i.e. for the 750~GeV case it will be more difficult to measure CP effects than for the 125~GeV Higgs.

\begin{figure}
\center
\includegraphics[width=.33\textwidth]{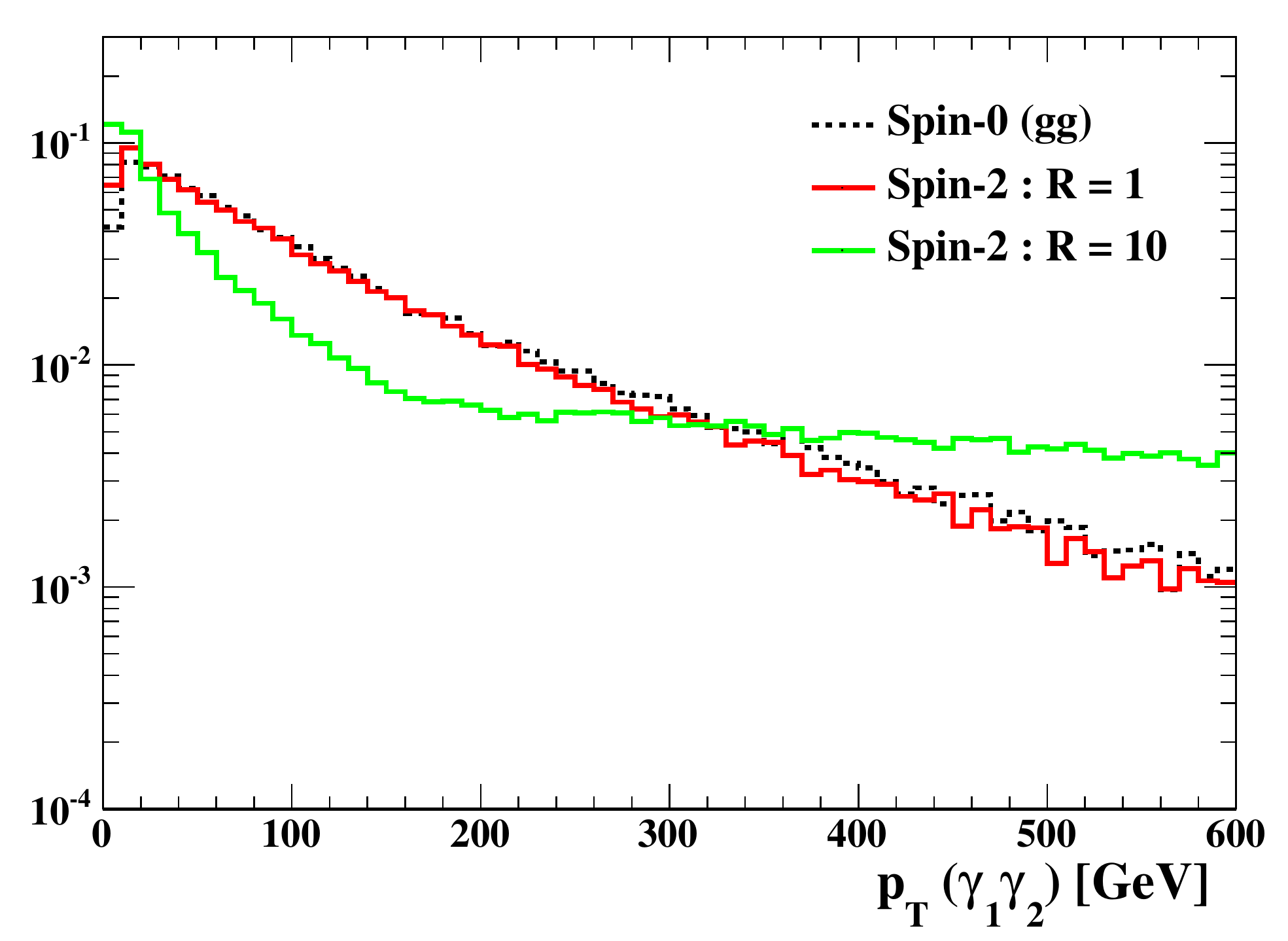}%
\includegraphics[width=.33\textwidth]{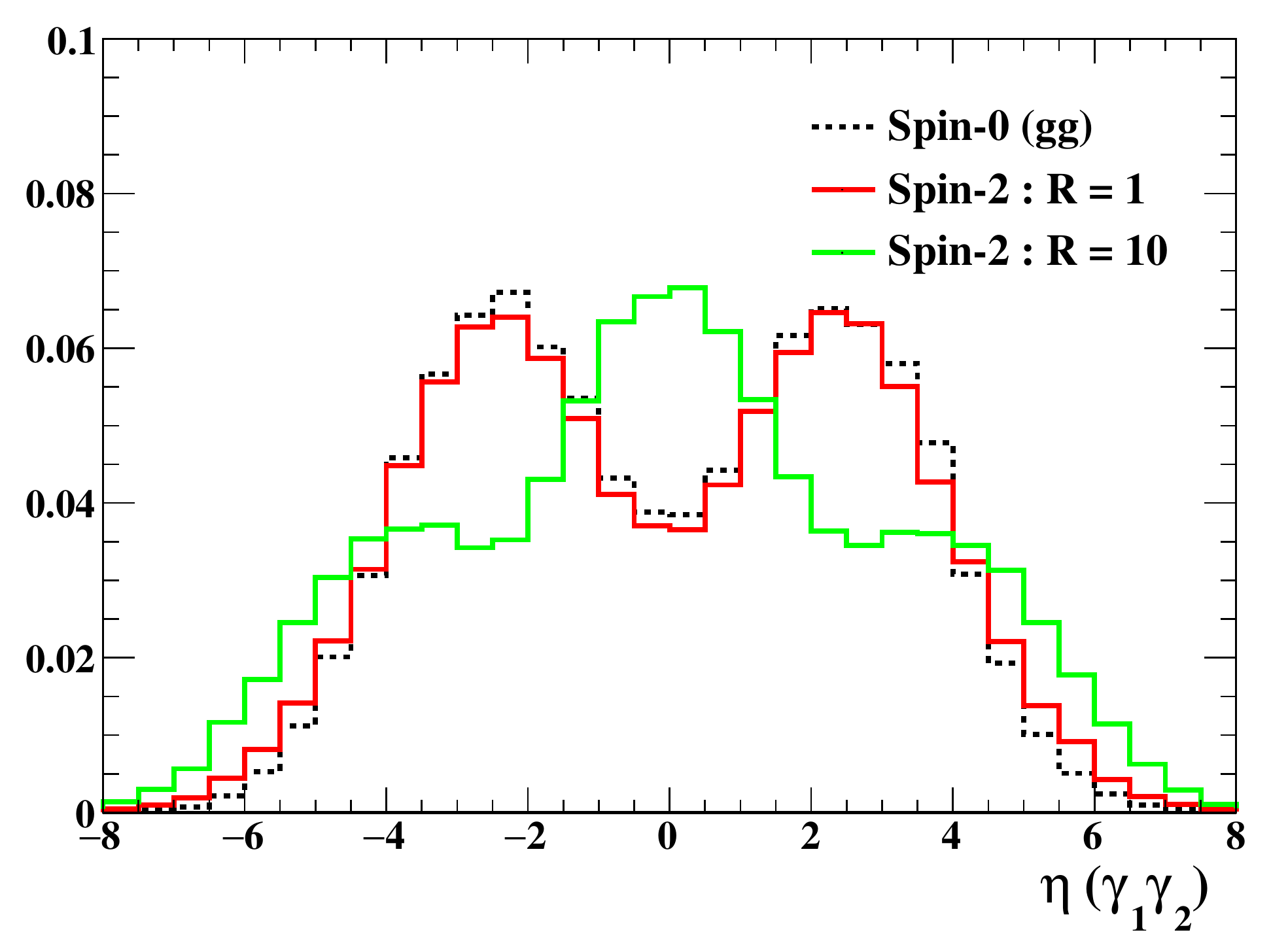}%
\includegraphics[width=.33\textwidth]{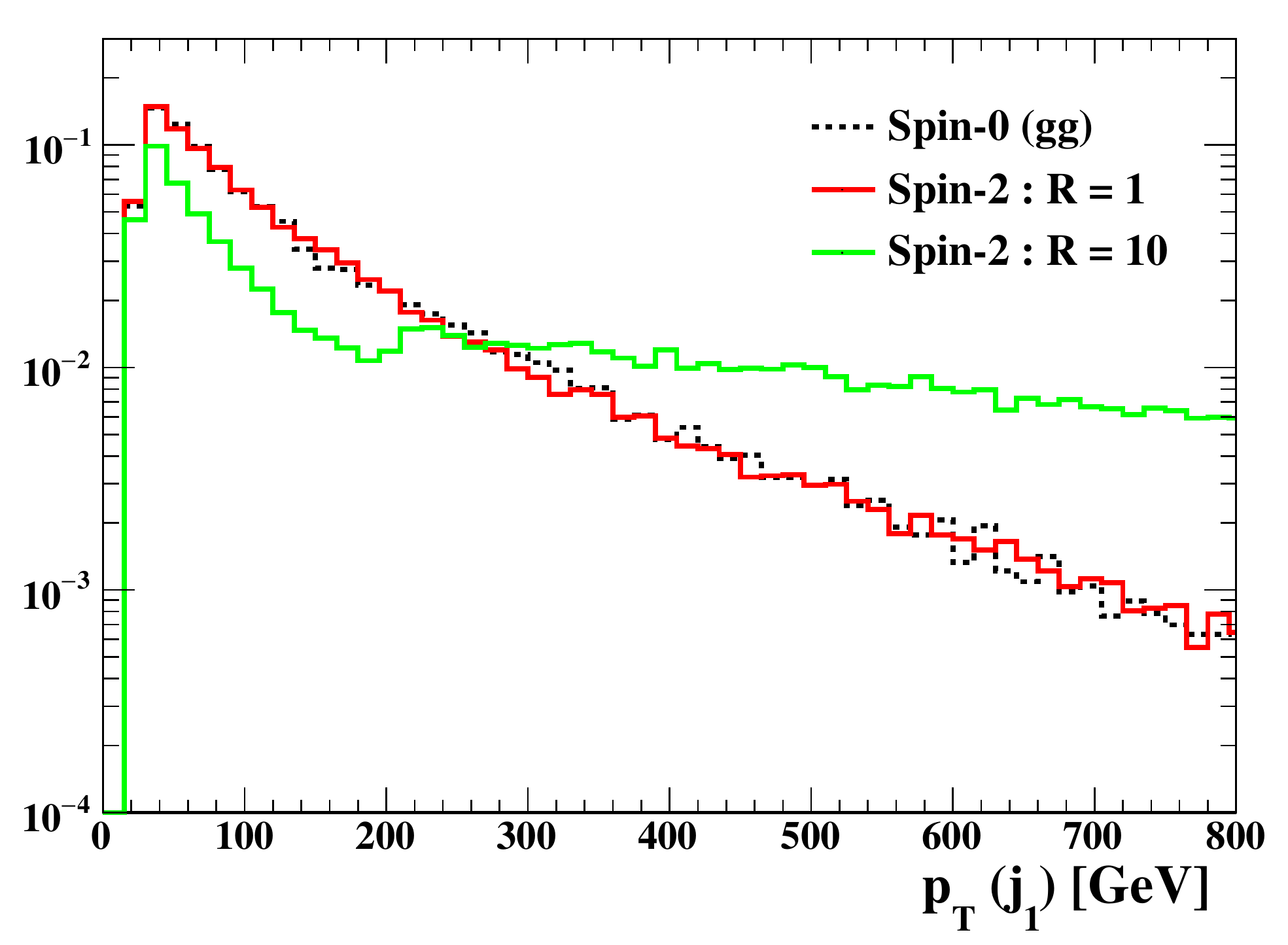}
\caption{
Normalised distributions in $p_T(\gamma_1\gamma_2)$, $\eta(\gamma_1\gamma_2)$ and $p_T(j_1)$ for the 750~GeV spin-2 scenario at the 13~TeV LHC. 
The red and green lines show the $R=1$ and $10$ cases, respectively. For reference, the gluon-initiated spin-0 case is shown as black dashed line.
Note the unitarity-violating behaviour for the $R=10$ case. 
}
\label{fig:spin2_nocut}
\end{figure}

\begin{figure}
\center
\includegraphics[width=.33\textwidth]{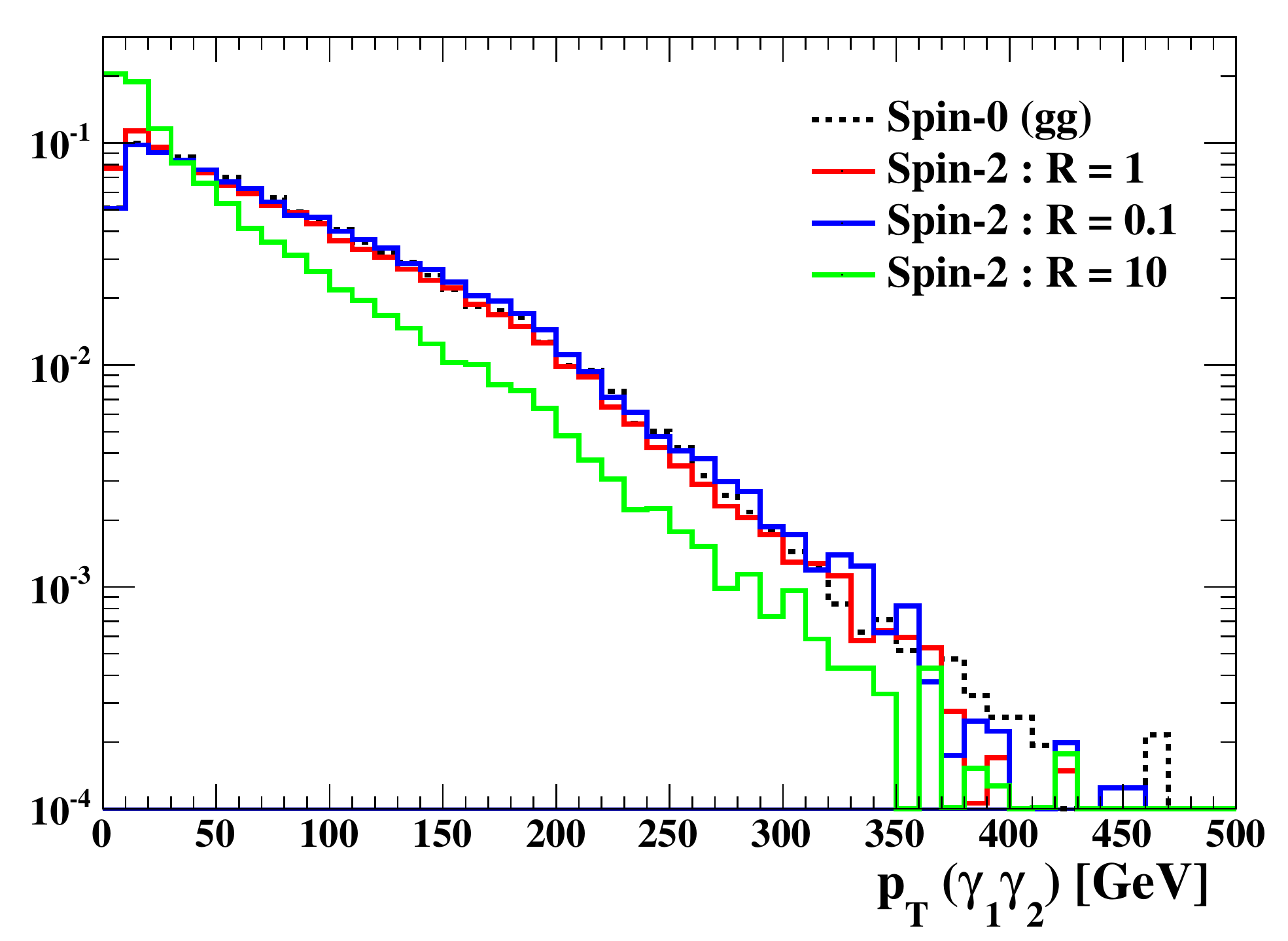}%
\includegraphics[width=.33\textwidth]{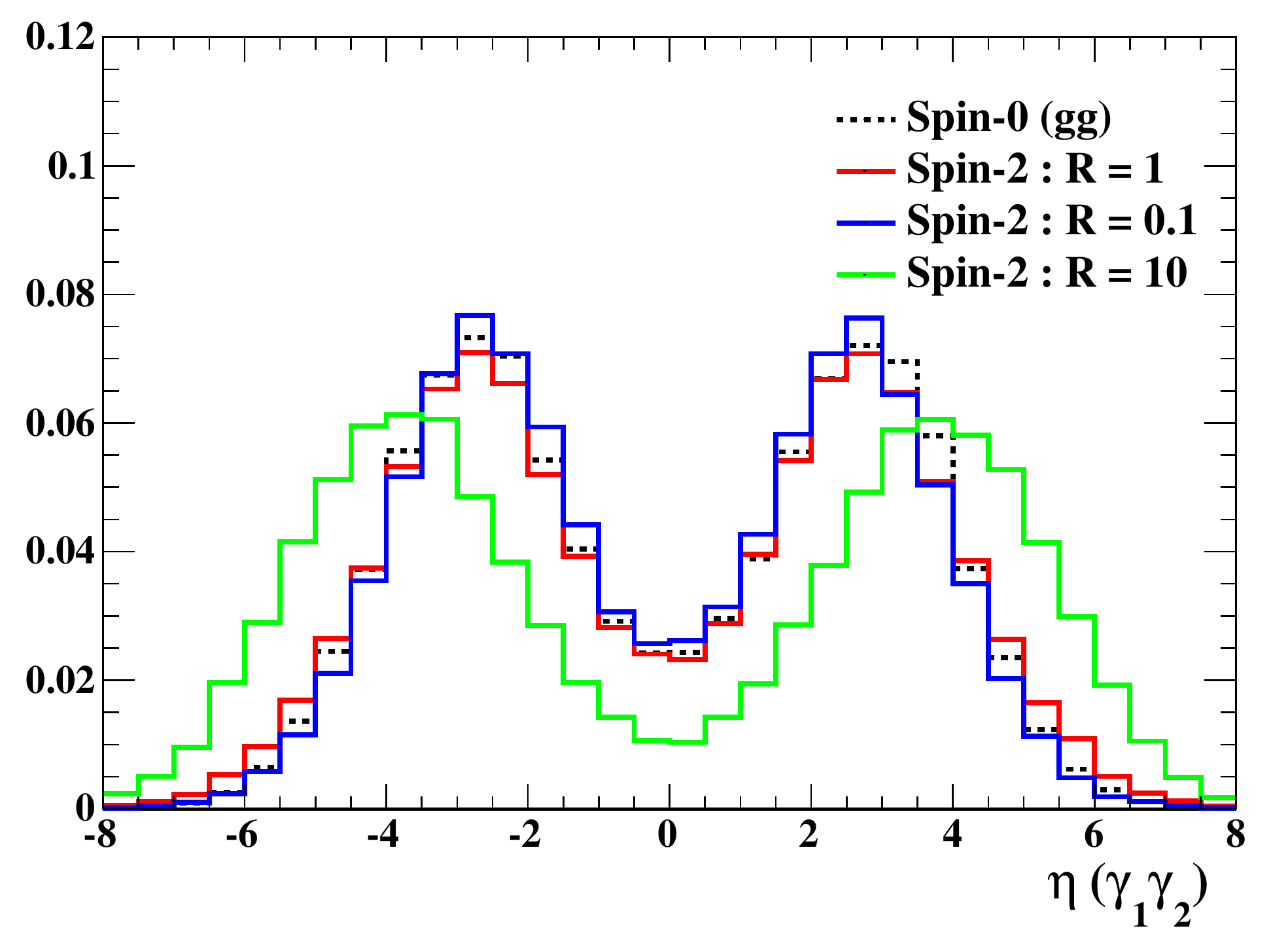}%
\includegraphics[width=.33\textwidth]{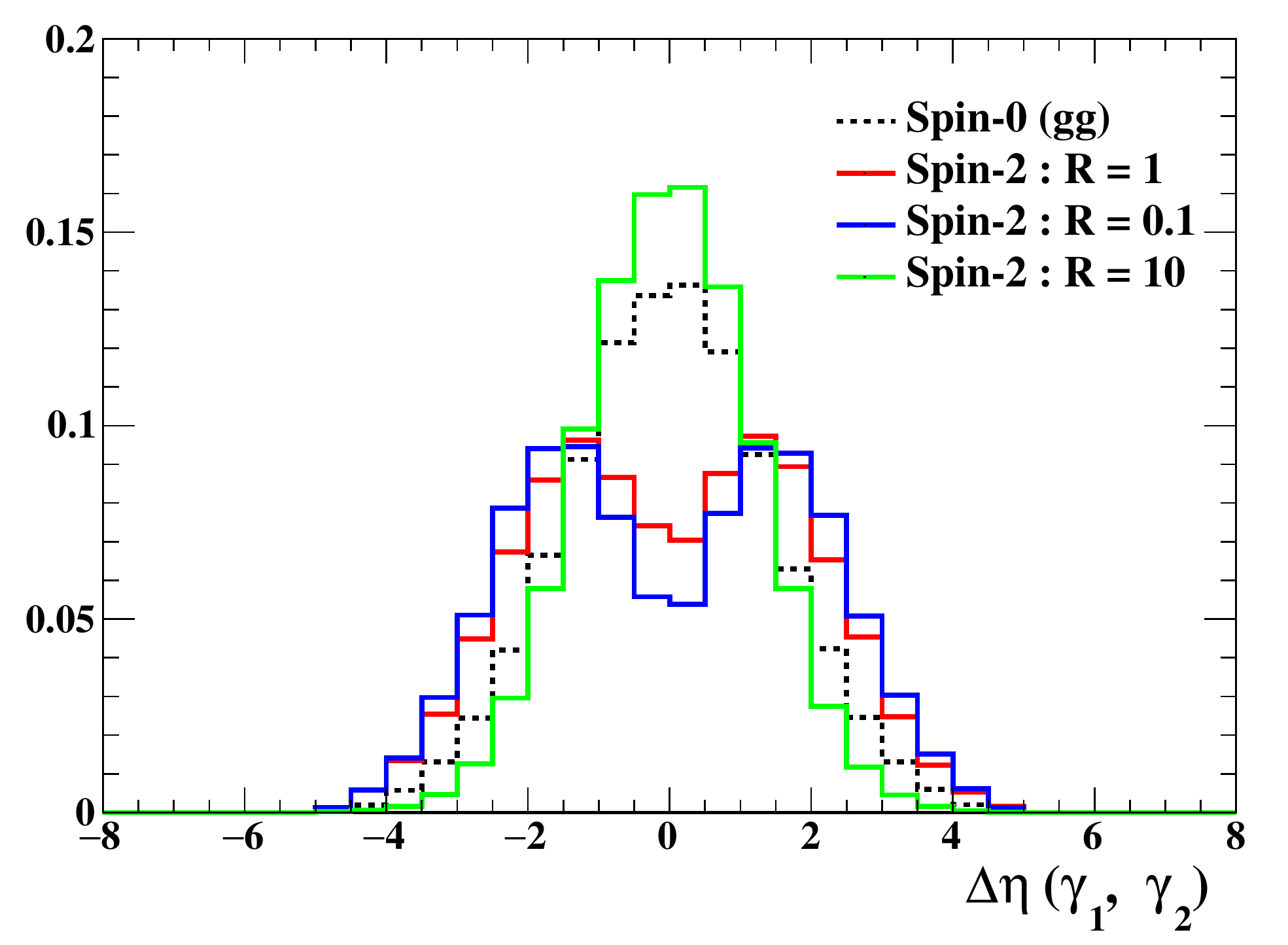}
\includegraphics[width=.33\textwidth]{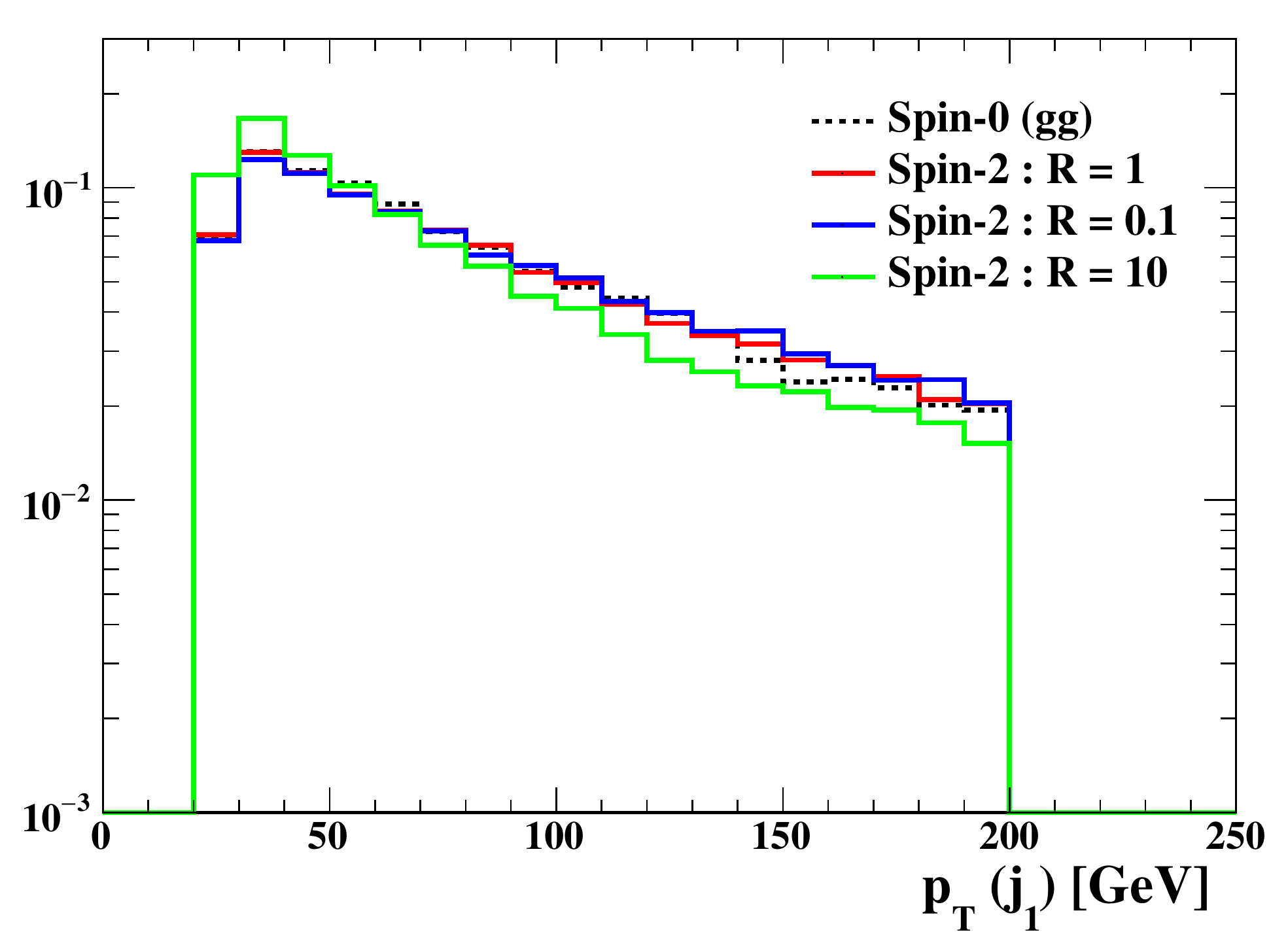}%
\includegraphics[width=.33\textwidth]{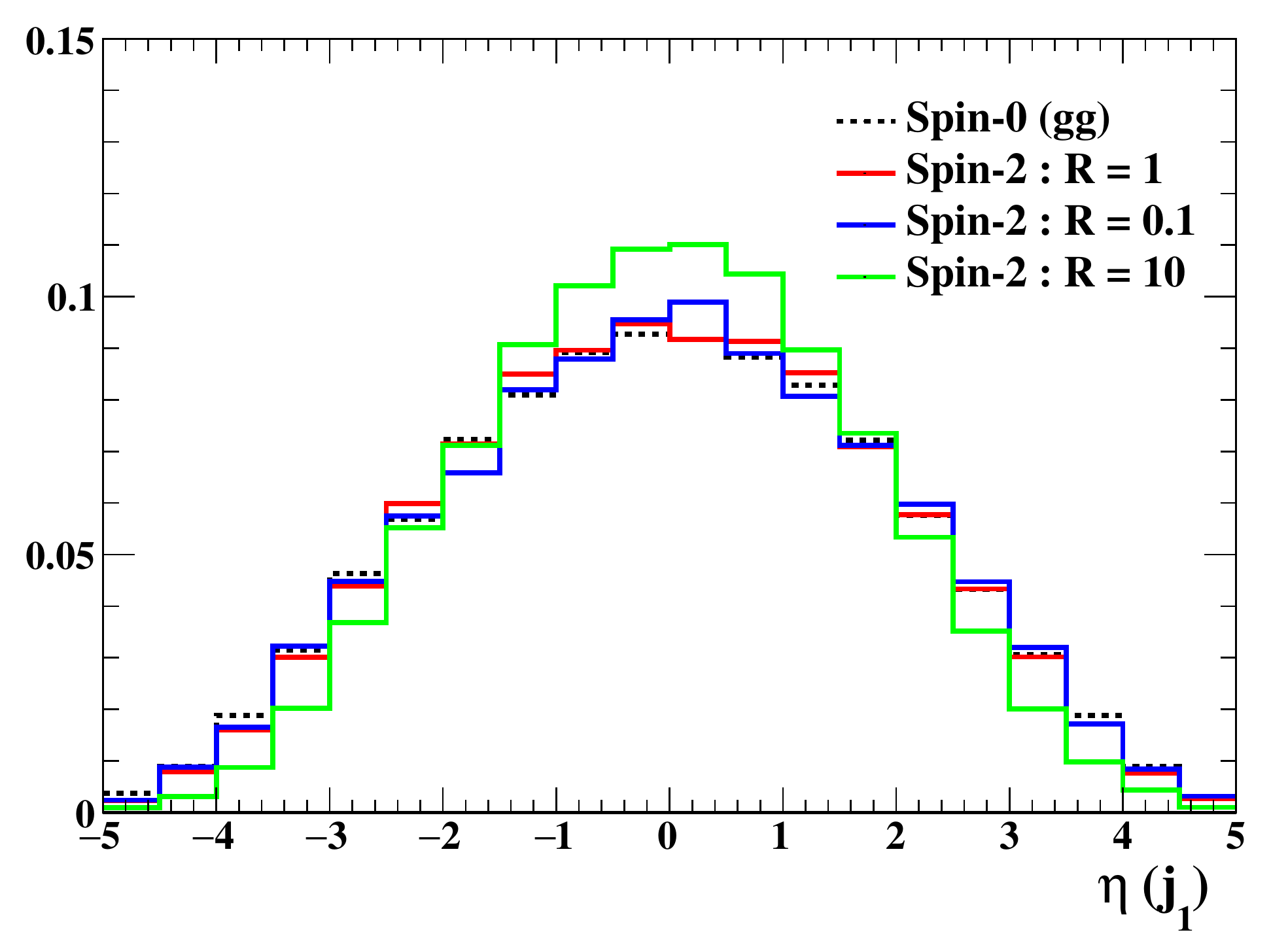}%
\includegraphics[width=.33\textwidth]{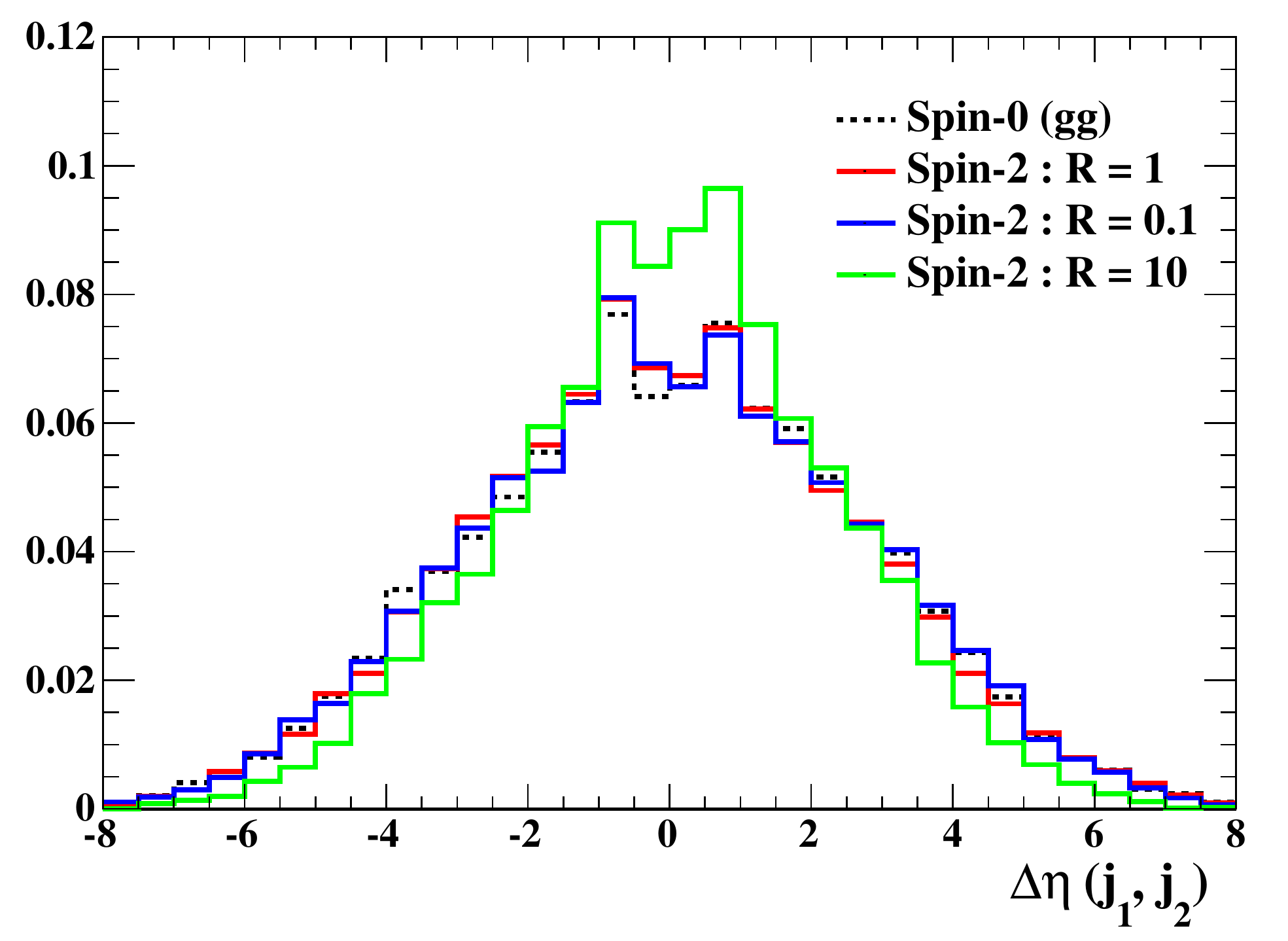}
\includegraphics[width=.33\textwidth]{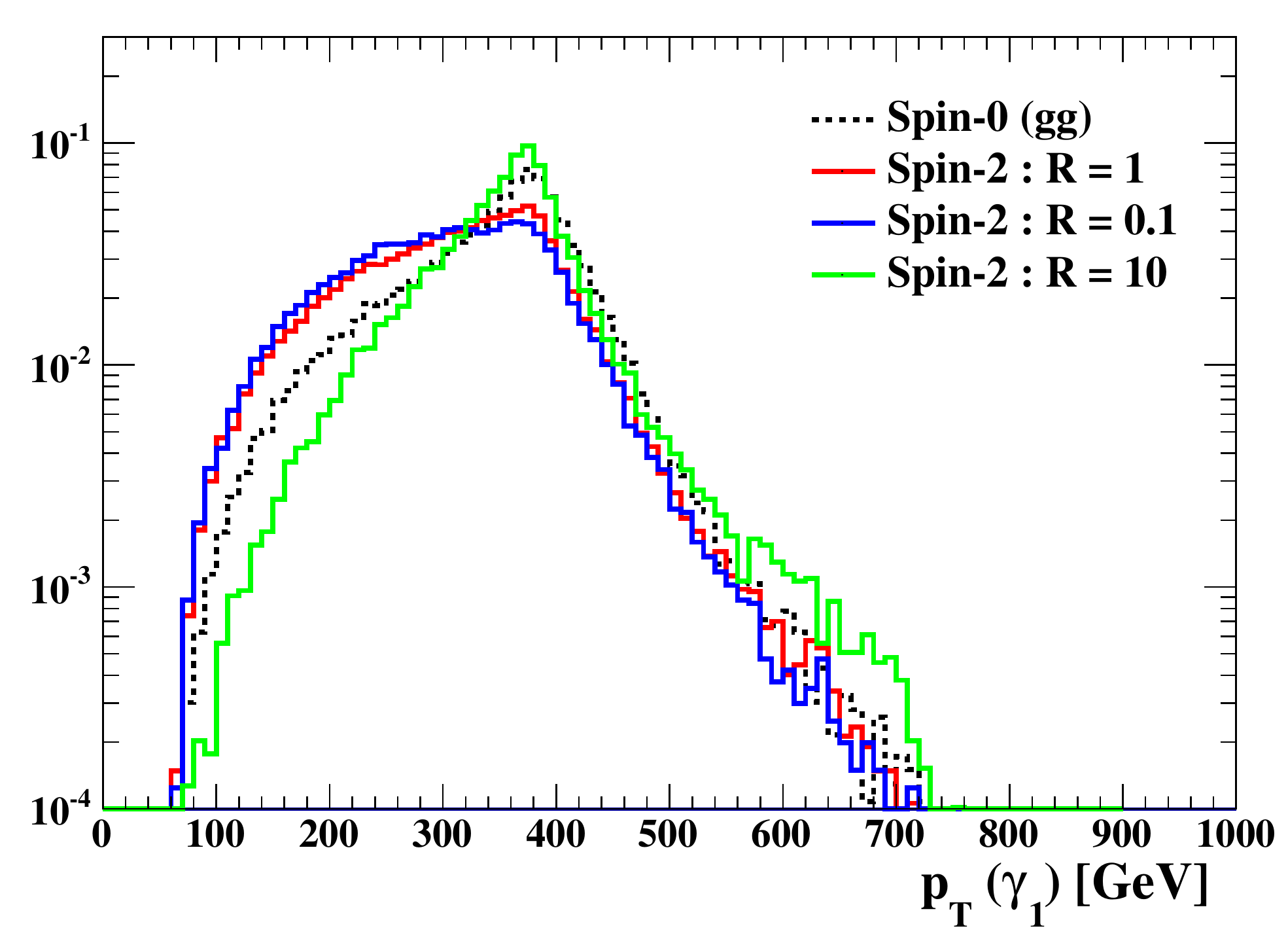}%
\includegraphics[width=.33\textwidth]{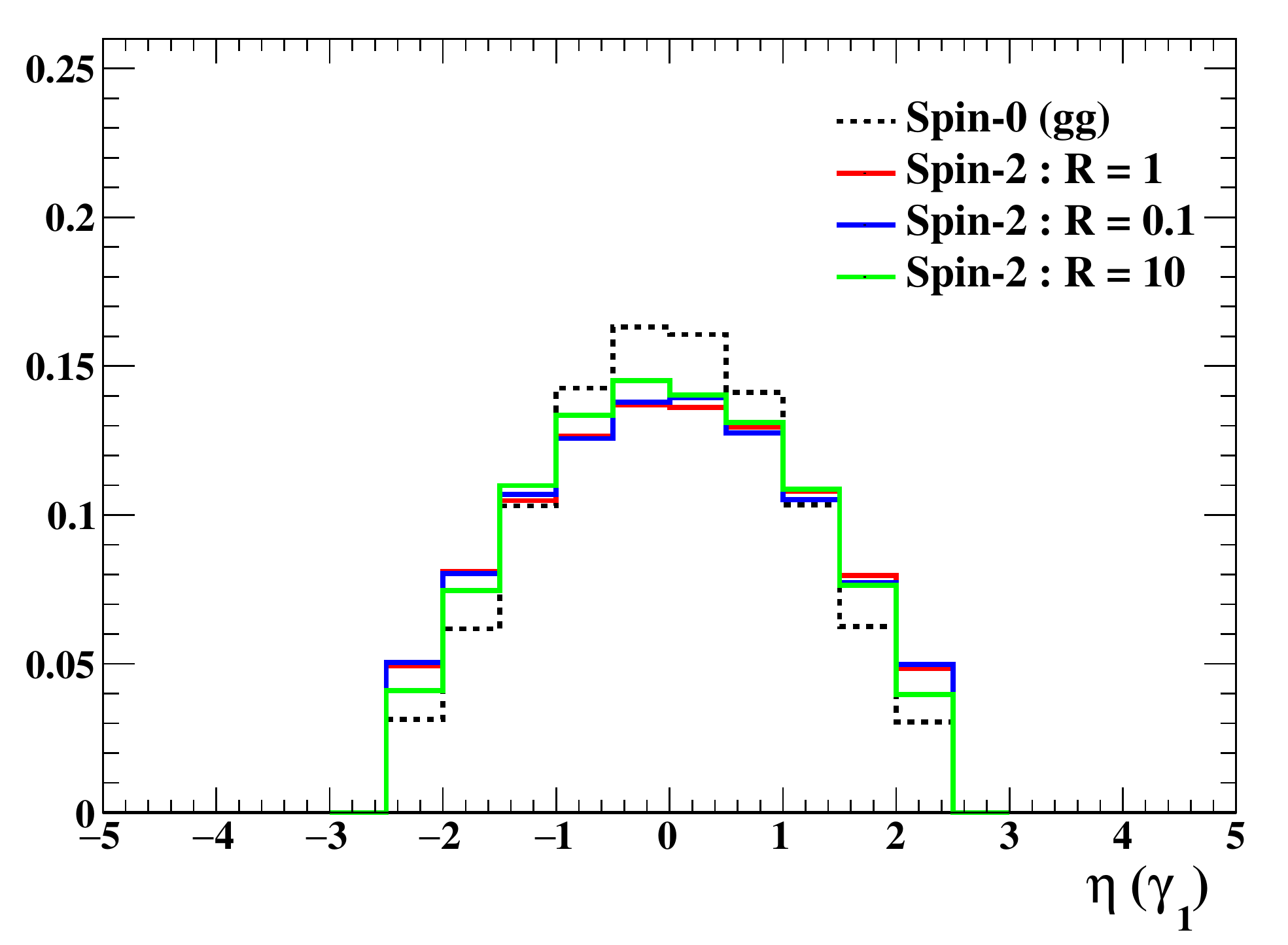}%
\includegraphics[width=.33\textwidth]{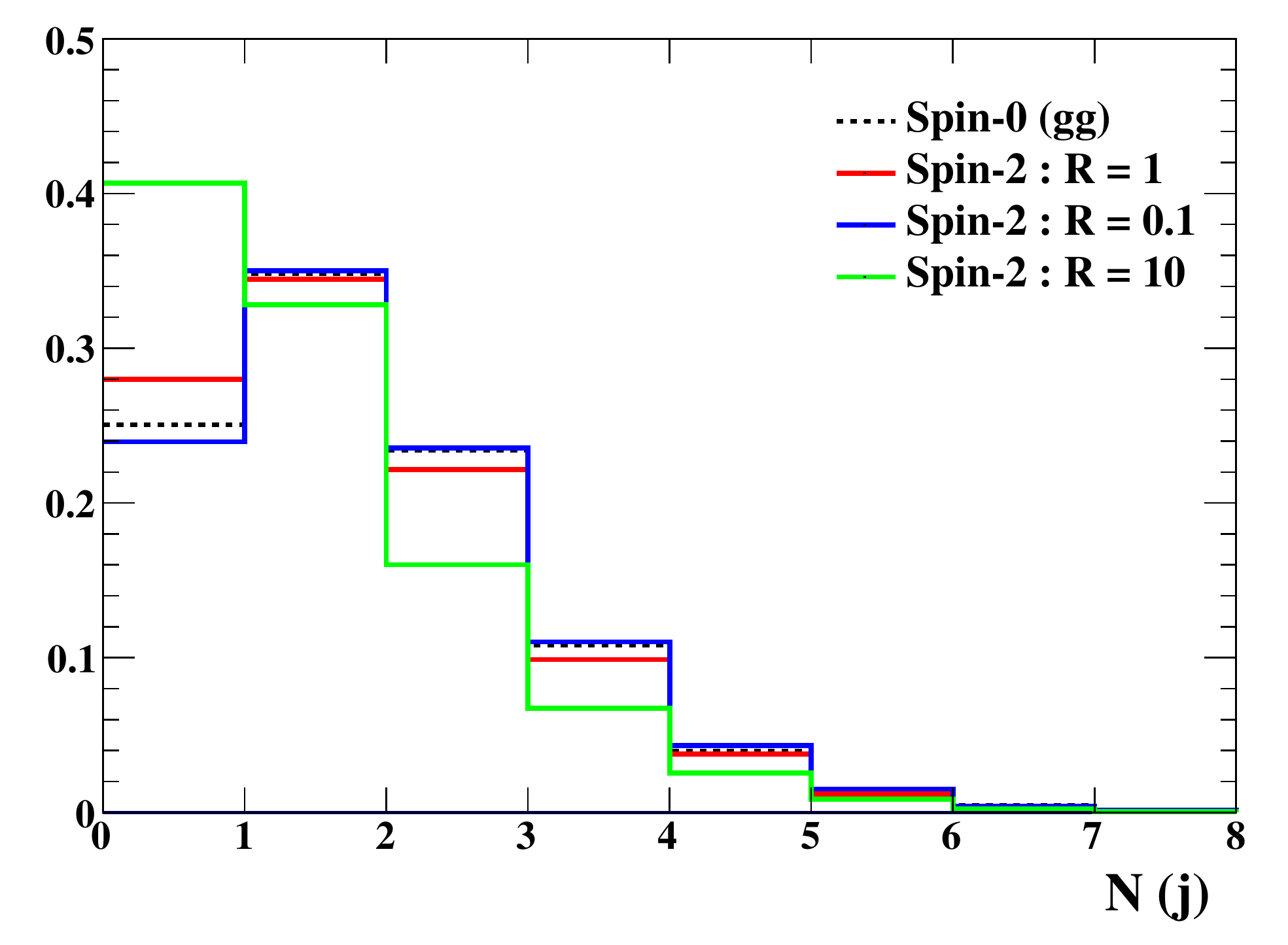}
\caption{
Normalised distributions in $p_T$ and $\eta$ of the diphoton system and the leading jet and photon as well as in $\Delta\eta$ of the photons and jets for the 750~GeV spin-2 scenario at the 13~TeV LHC.
The red, blue and green lines show the $R=1$, $0.1$ and $10$ cases, respectively.
For reference, the gluon-initiated spin-0 case is shown as black dashed line. 
A cut of $p_T(j)<200$~GeV is imposed as explained in the text.}
\label{fig:spin2}
\end{figure}

Let us now turn to the spin-2 case. 
As pointed out in~\cite{Artoisenet:2013puc}, and depicted in Fig.~\ref{fig:spin2_nocut}, 
the non-universal coupling scenario ($R\equiv\kappa_q/\kappa_g\ne1$) gives rise to a unitarity-violating behaviour at higher order in QCD.%
\footnote{We note that, unlike the $R=10$ (quark-dominant) case, the distributions for $R=0.1$ (gluon-dominant) is very similar to the $R=1$ case since the gluon-initiated process is dominant for the universal coupling case, as mentioned before.}
To avoid such behaviours, or to show results with respect to any UV completion of the spin-2 model, in the following spin-2 analysis we require the tagging jets to fulfill~\cite{Englert:2012xt}
\begin{align}
	p_T(j) < p_T^{\rm max}(j) = 200~{\rm GeV}\,.
\end{align}
In Fig.~\ref{fig:spin2} we present the same set of distributions as in Fig.~\ref{fig:spin0} for the case of a spin-2 resonance with mass of 750~GeV (and width of 45~GeV). Here, we compare $R=1$ (red), $0.1$ (blue) and $10$ (green). 
The gluon-initiated spin-0 case is also shown by black dotted lines as a reference, for easier comparison with Fig.~\ref{fig:spin0}. 
Note that here we require the above $p_T^{\rm max}(j)$ cut even for the spin-0 case in order to perform a meaningful comparison. 
We see that while $q\bar q$ dominated production differs from $gg$ dominated production in several of the distributions, most notably the jet activity, distinction of the $gg$ initiated spin-0 and spin-2 cases is less obvious. 
This was also observed for the case of the 125~GeV Higgs characterisation in~\cite{Artoisenet:2013puc}. 
There are, however, some differences in the heavy resonance decay; 
indeed most promising for differentiating spin-0 from spin-2 are the rapidity separation between the two photons, $\Delta\eta(\gamma_1,\gamma_2)$,
and to some extent the leading photon $p_T$ and $\eta$ distributions~\cite{Han:2015cty,Martini:2016ahj}, simply due to the different decay distributions between the spin-0 and spin-2 resonances.

\subsection{Heavier parent resonance}\label{sec:disheavier}

\begin{figure}
\center
\includegraphics[width=.48\textwidth]{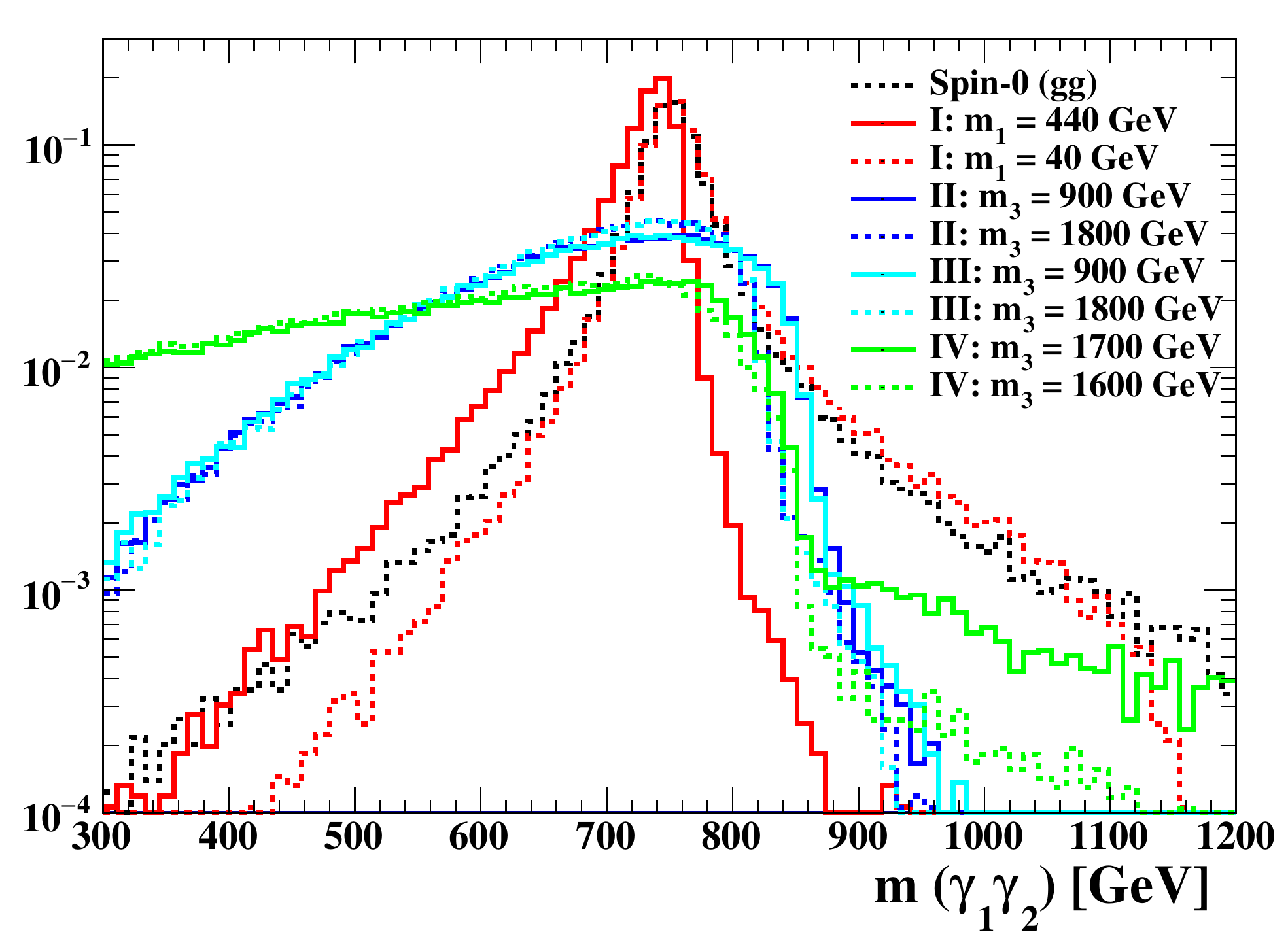}\quad
\includegraphics[width=.48\textwidth]{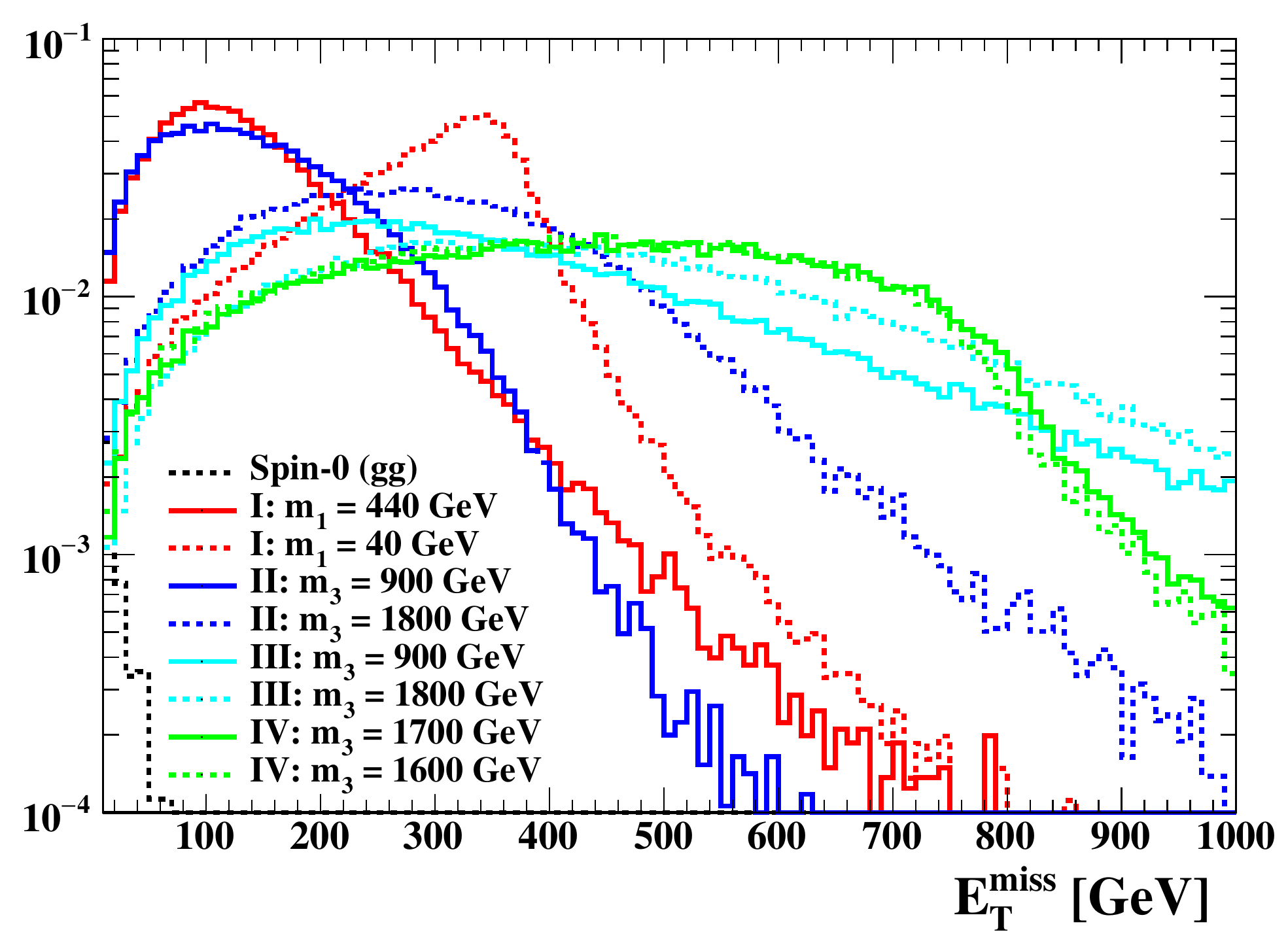}
\caption{ 
Normalised distributions of diphoton invariant mass (left) and missing transverse energy (right) for the heavier parent scenarios I--IV. For reference, the gluon-initiated spin-0 case is shown as black dashed line. 
}\label{fig:MET}
\end{figure}

We next contrast the above results to the different cases of a heavier parent resonance described in Section~\ref{sec:scenariosParent}. To begin with, we show in Fig.~\ref{fig:MET} (left) the diphoton invariant mass 
distributions for the different benchmark scenarios from Table~\ref{tab:scenarios14}.  
We see that with a precise lineshape analysis one should be able to discriminate the three-body decay and antler scenarios (II--IV) from the cases where the two photons originate from a two-body decay of a 750~GeV particle. 
Even for the diphoton coming from the 750~GeV resonance (with a 45~GeV width) in scenario~I, the lineshape can differ from the one for
direct resonance production (black dashed) depending on $m_1$. 

\begin{figure}
\center
 \includegraphics[width=.33\textwidth]{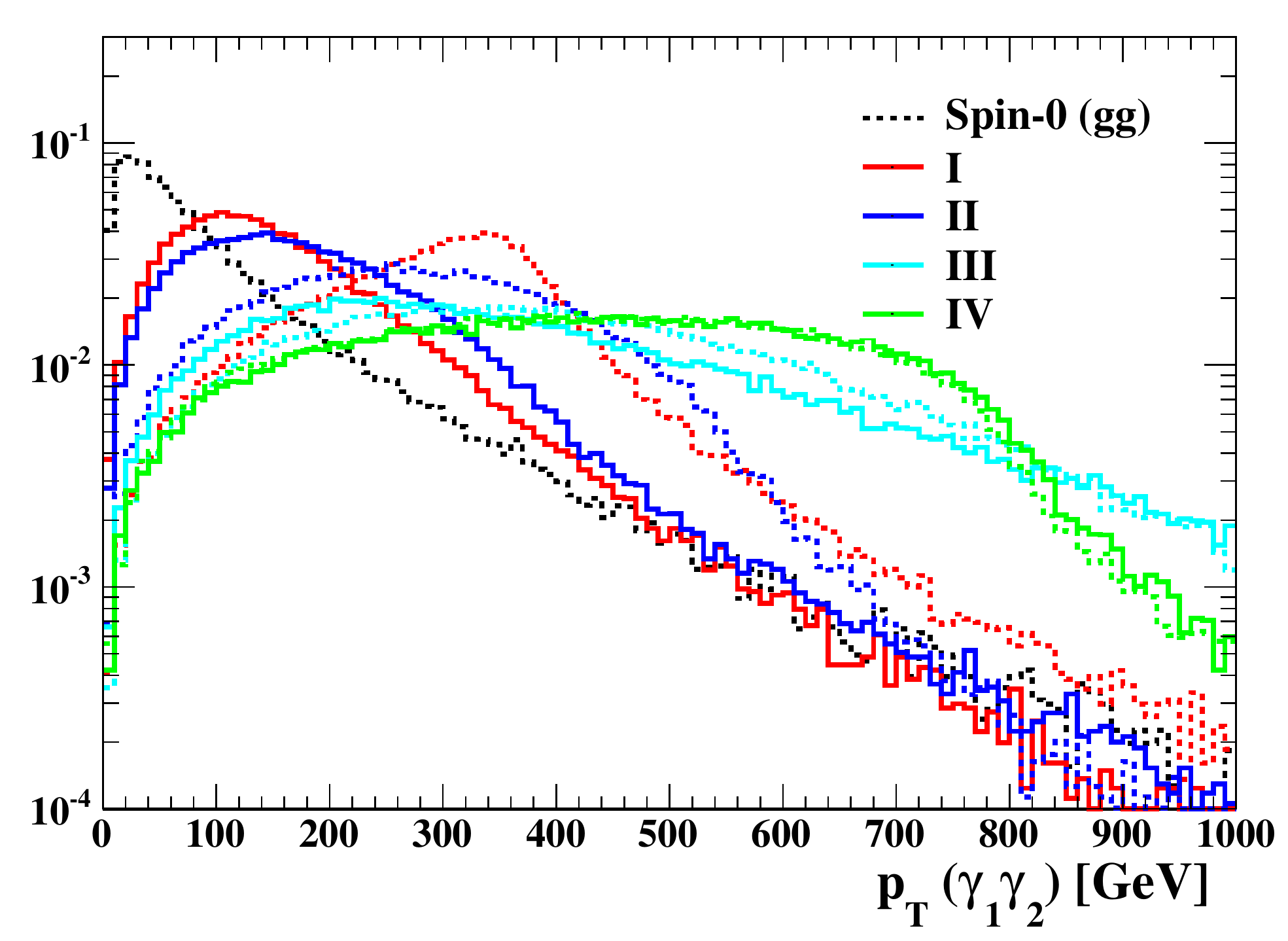}%
 \includegraphics[width=.33\textwidth]{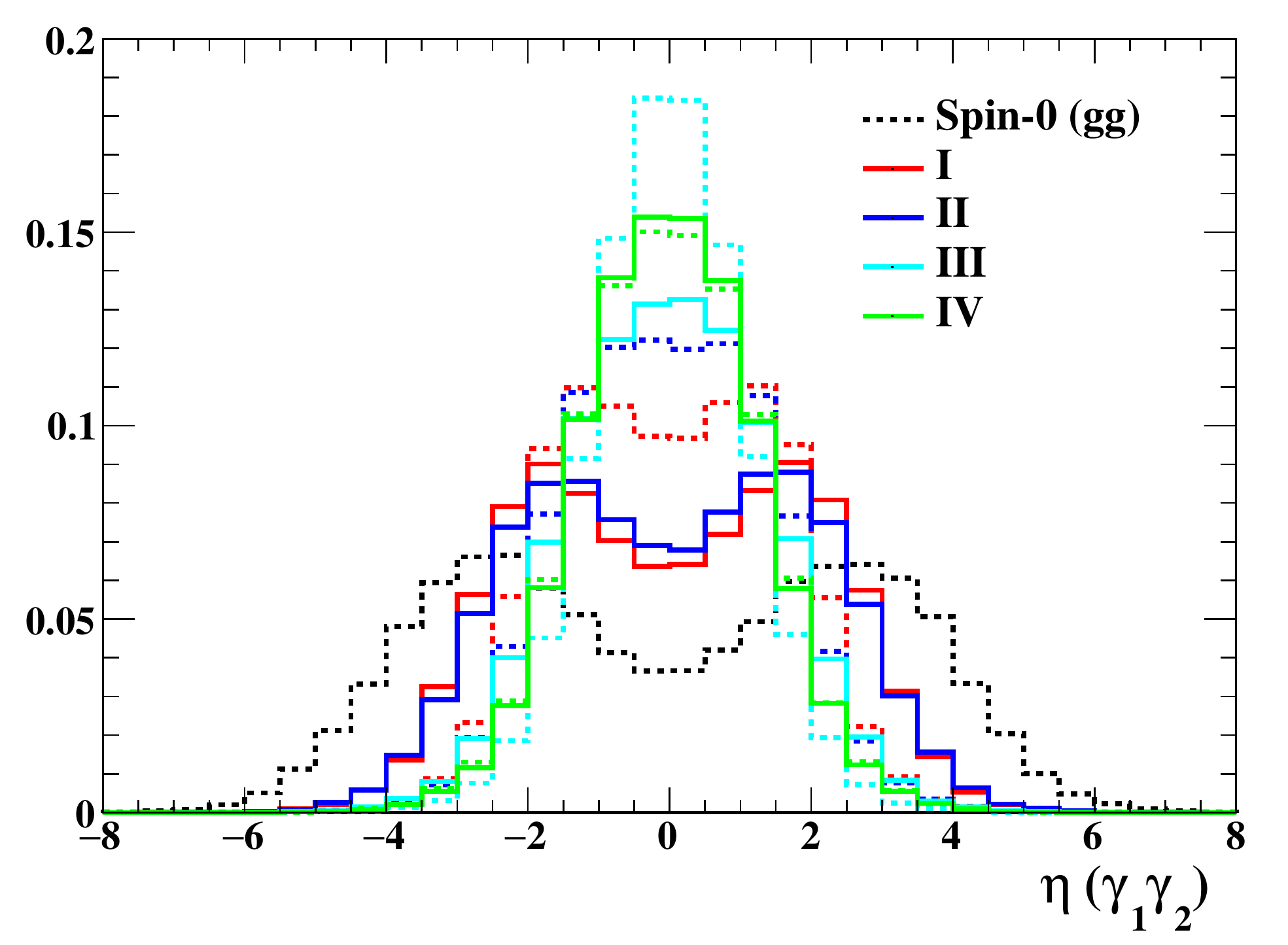}%
 \includegraphics[width=.33\textwidth]{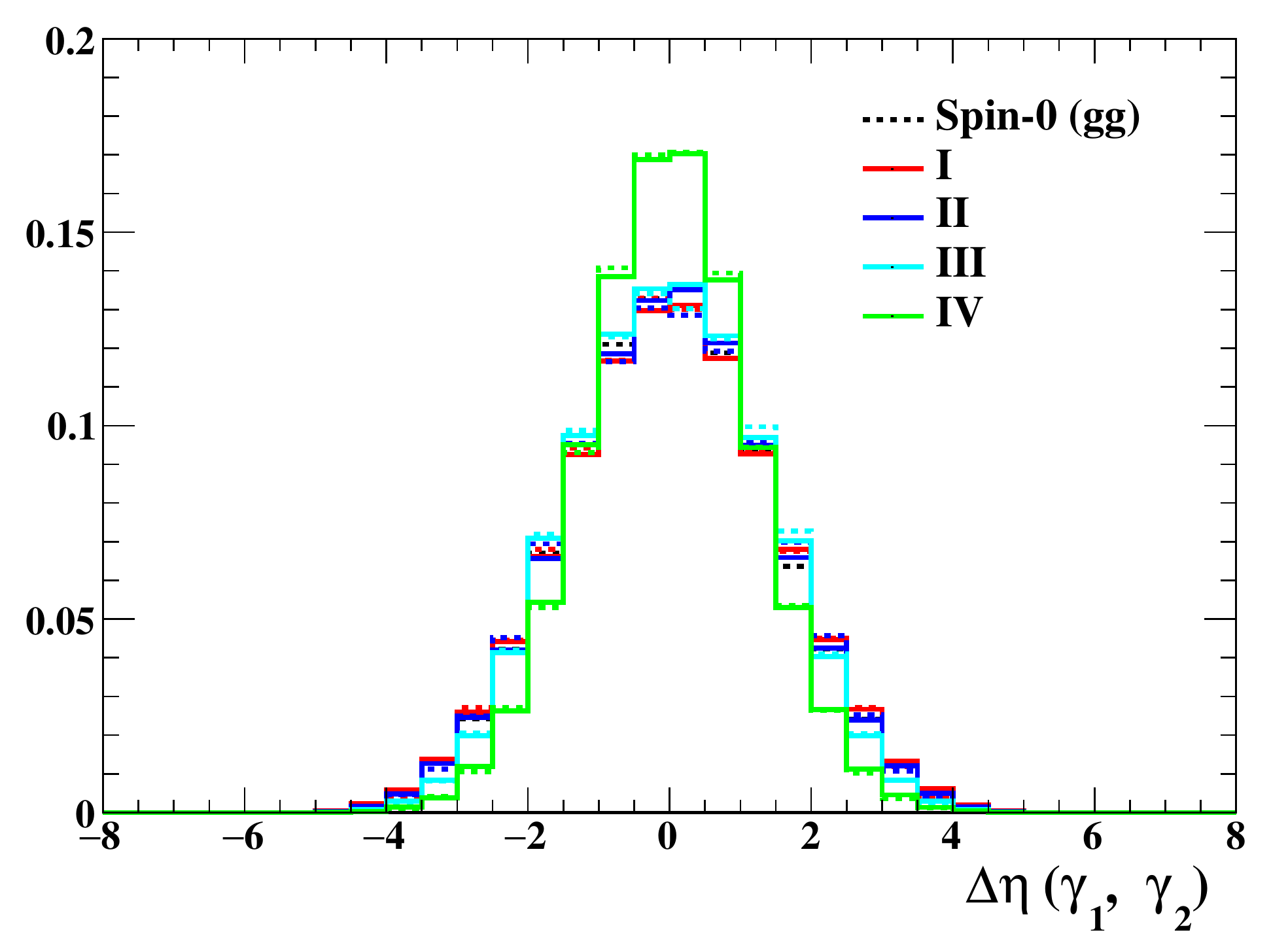}
 \includegraphics[width=.33\textwidth]{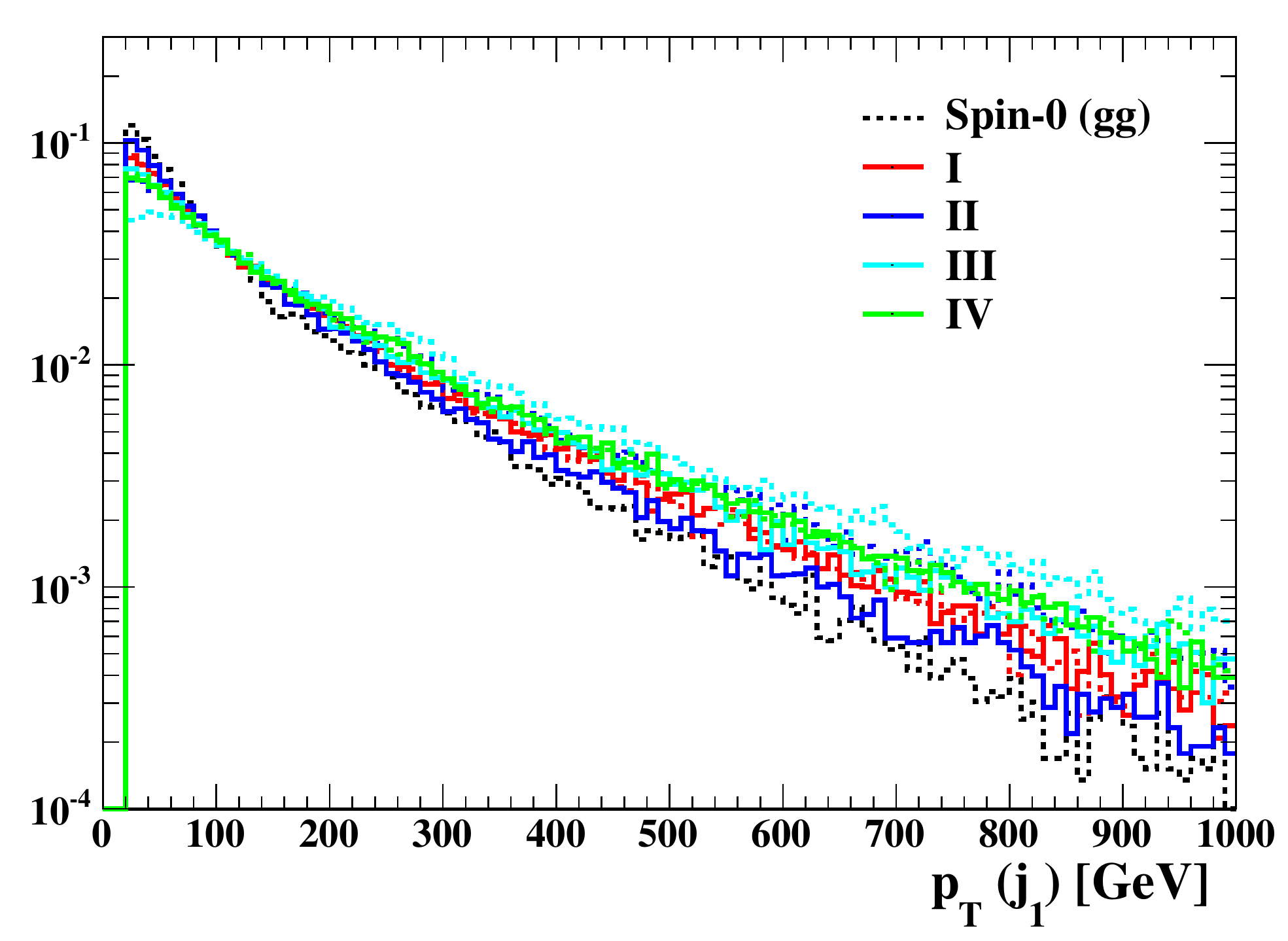}%
 \includegraphics[width=.33\textwidth]{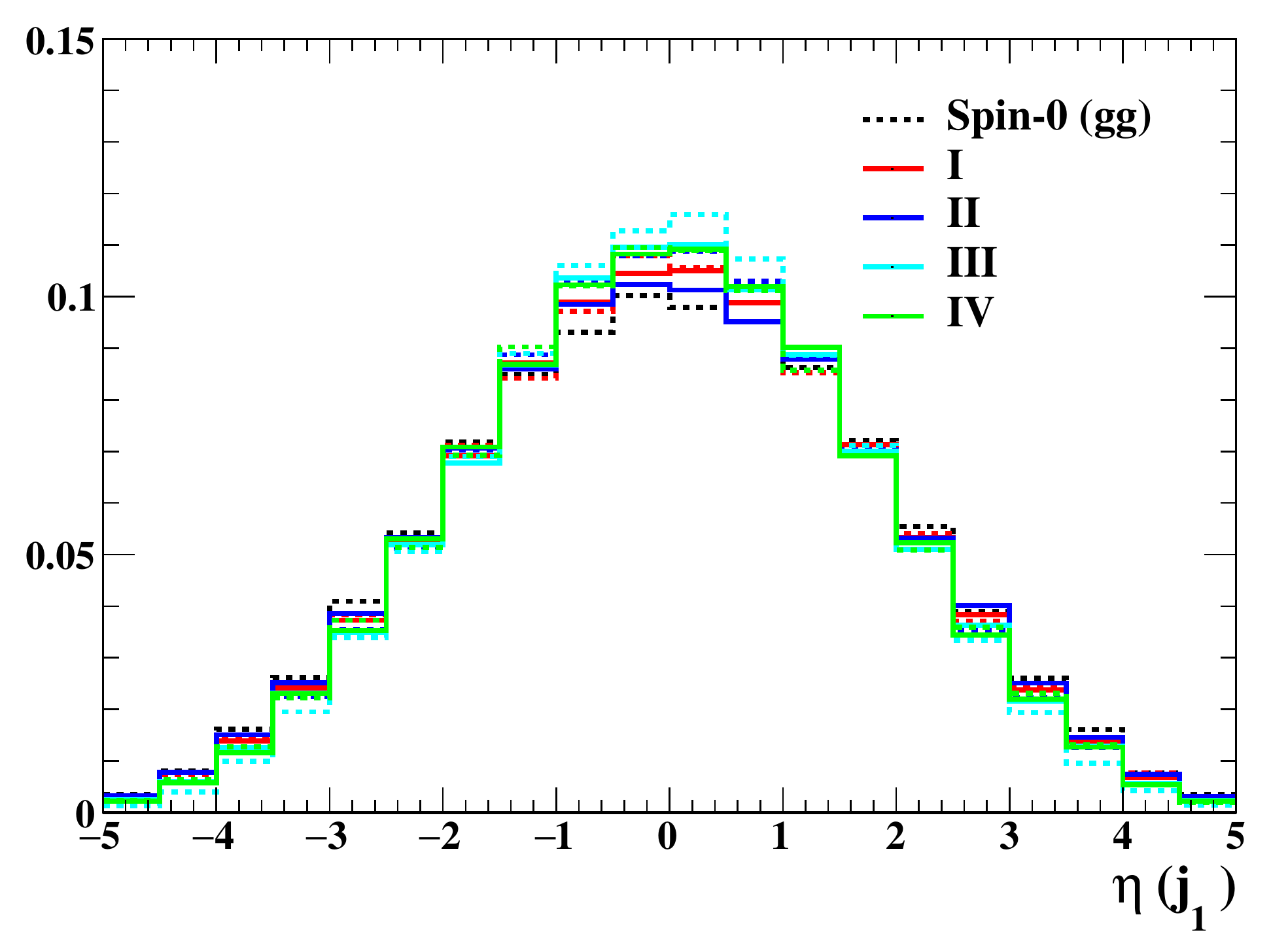}%
 \includegraphics[width=.33\textwidth]{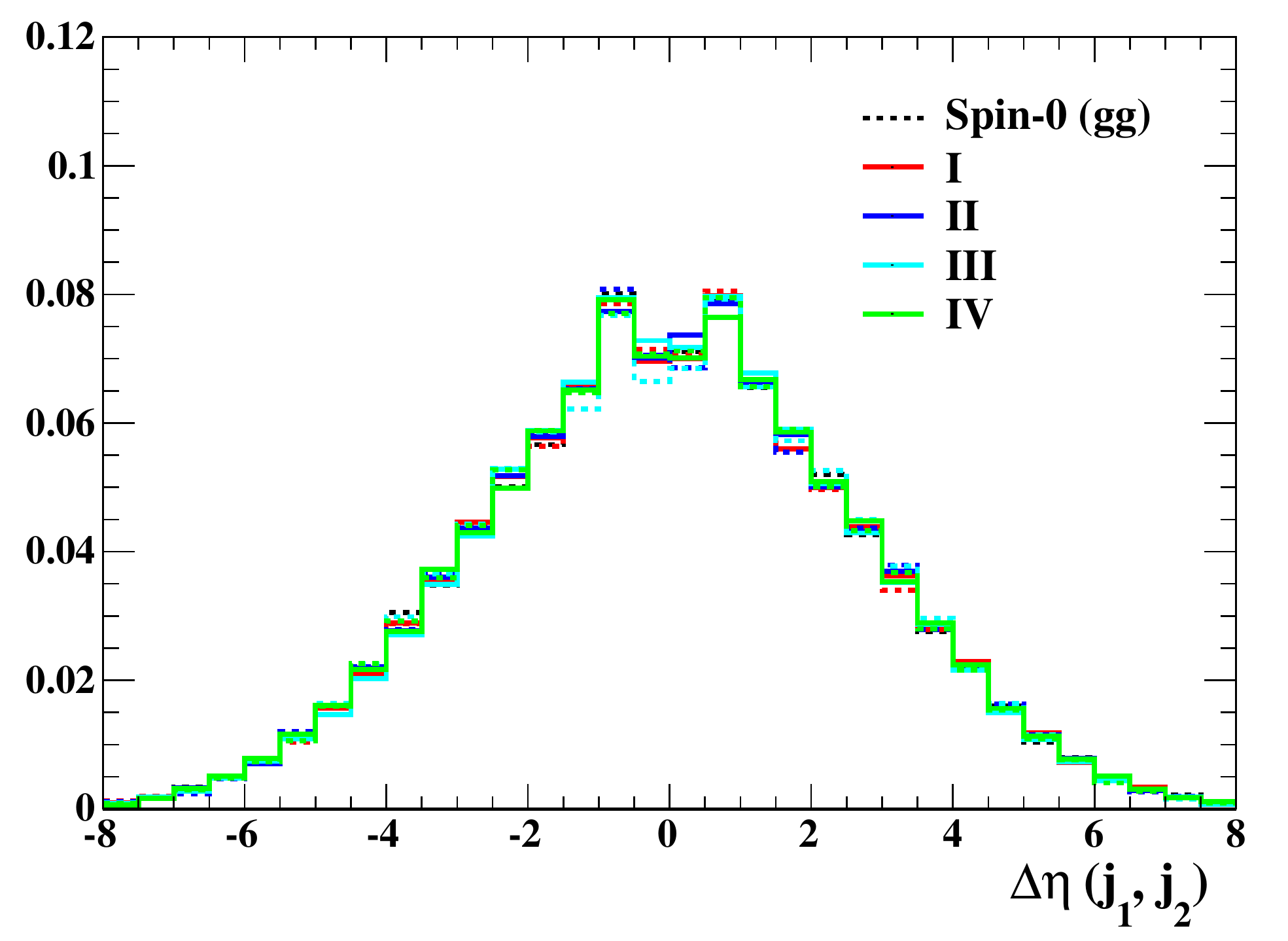}
 \includegraphics[width=.33\textwidth]{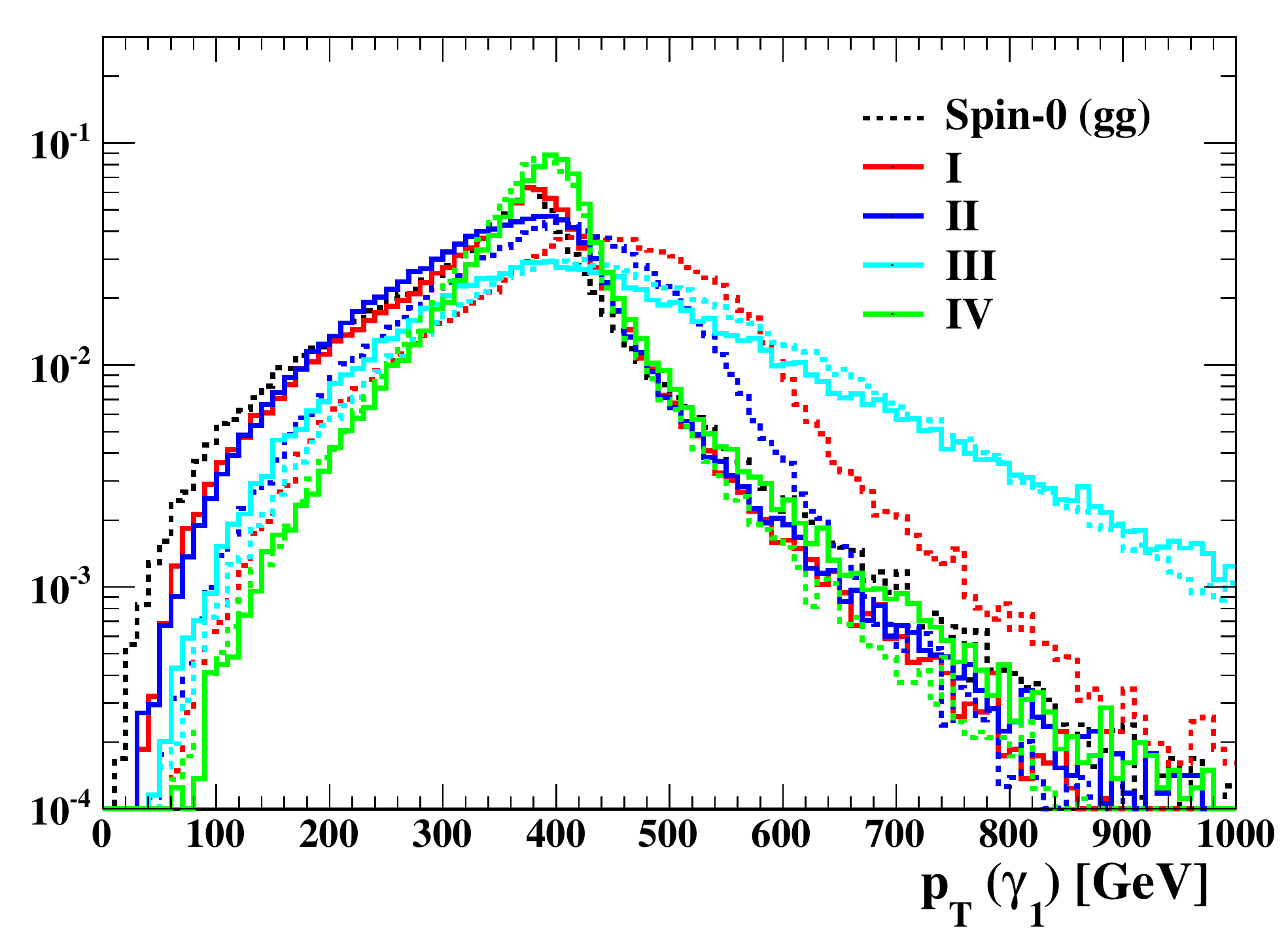}%
 \includegraphics[width=.33\textwidth]{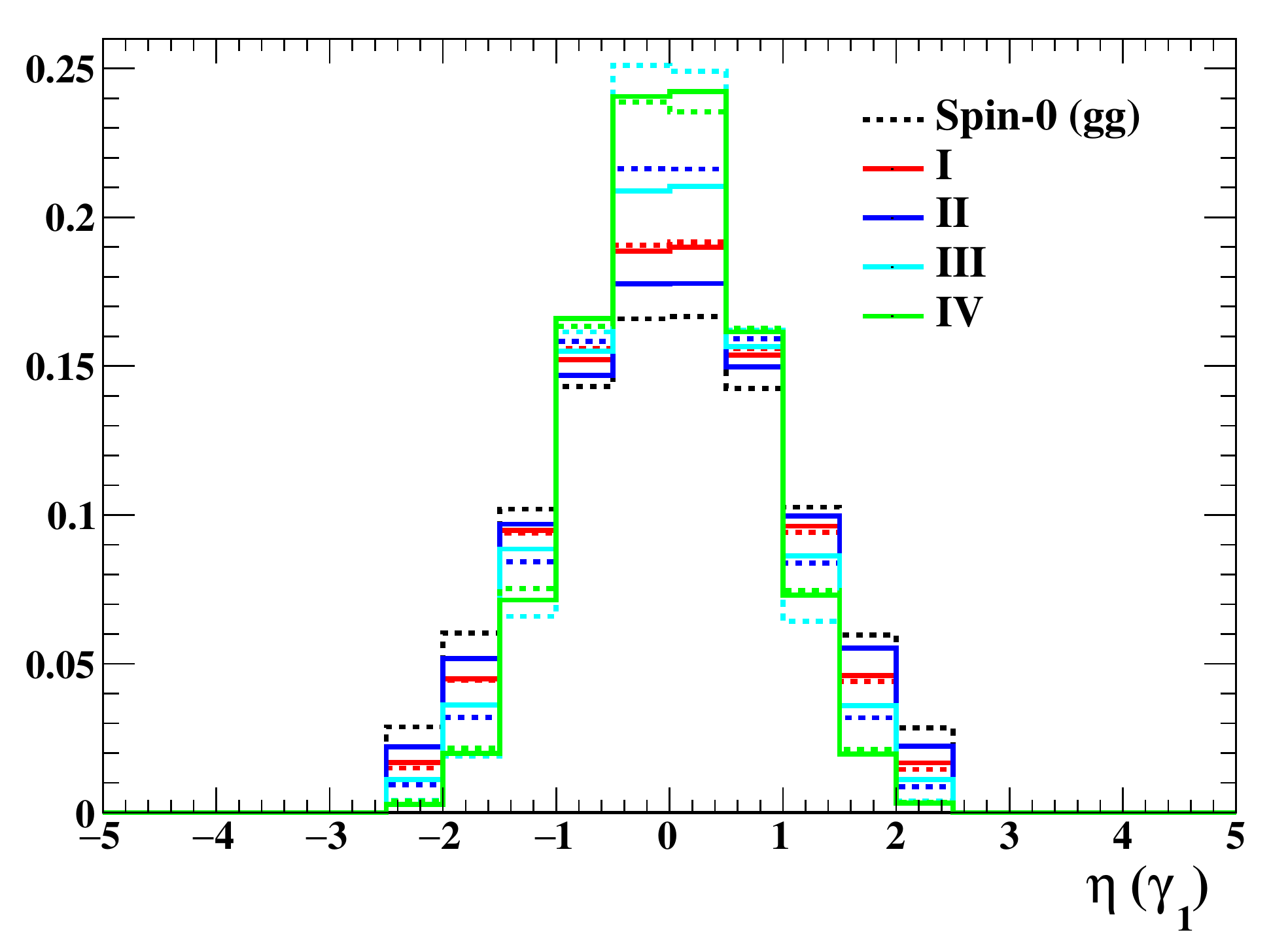}%
 \includegraphics[width=.33\textwidth]{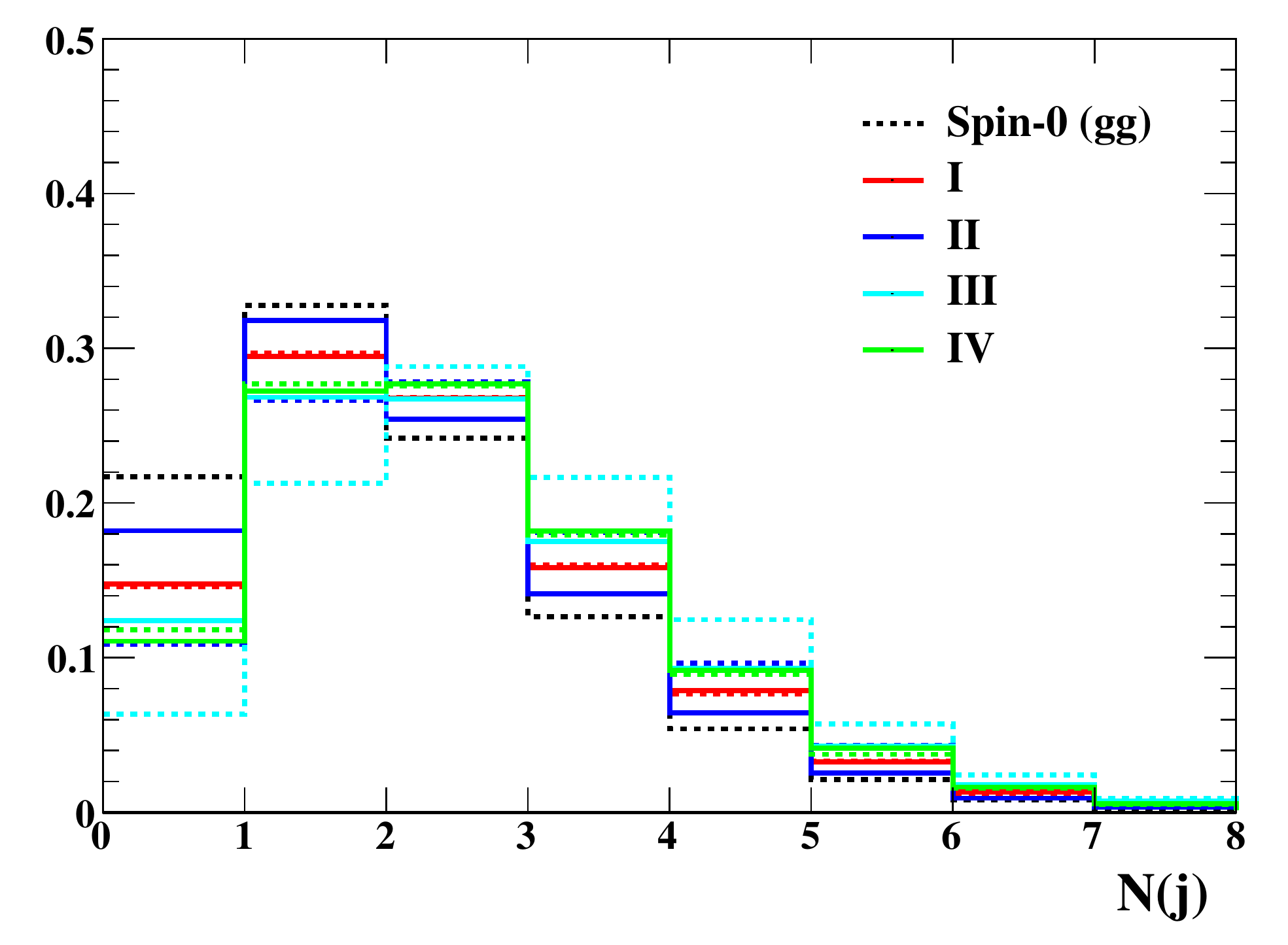}
 \caption{
 Normalised distributions for the heavier parent scenarios I--IV. For each scenario, the full (dashed) lines are for the first (second) mass combination, cf.\ Table~\ref{tab:scenarios14} and Fig.~\ref{fig:MET}. 
 For reference, the gluon-initiated spin-0 case is shown as black dashed line.}
\label{fig:cascade}
\end{figure}

If $S_1$ or $\chi_1$ are invisible, an observable that can be more readily exploited with less data 
to discriminate the heavier parent from the direct 750~GeV resonance case 
is the amount of missing transverse energy, $E_T^{\rm miss}$, shown in the right panel in Fig.~\ref{fig:MET}. 
Depending on the precise mass pattern, the $E_T^{\rm miss}$ distribution may also help to discriminate between scenarios I--IV. We note that scenarios III and IV as well as scenarios I with a light $S_1$ and the 3-body decay scenario II with a very heavy parent all lead to very high $E_T^{\rm miss}$. While so far ATLAS and CMS have not provided any details on the event structure of the diphoton excess, such high $E_T^{\rm miss}$ would have been a striking feature and difficult to miss. 
In this respect scenarios I and II with masses that minimize the amount of $E_T^{\rm miss}$ (red and blue solid lines) 
seem most interesting. Nonetheless for these cases the missing energy still peaks around 100~GeV, which would be a powerful discriminator against the direct 750~GeV resonance production discussed in Sec.~\ref{sec:dis750}. 

To obtain complementary information to the above and/or 
if $S_1$ or $\chi_1$ are not invisible but lead to soft decay products because of, e.g., a hidden valley cascade, 
one can make use of the `conventional' kinematic distributions that we already considered for the 750~GeV spin-0 and spin-2 resonance cases. These are shown in Fig.~\ref{fig:cascade} for the heavier parent scenarios. 
While the distributions involving jets offer little discriminating power, the diphoton $p_T$ and $\eta$ distributions are rather distinct. Concretely the diphoton system is harder and more central depending on the scenario. 
Additional information can be obtained from the $p_T(\gamma_1)$ spectrum. 
Putting everything together it seems feasible to distinguish not only between the 750~GeV and heavier resonance cases but also among the heavier parent scenarios I--IV, although distinguishing between scenarios I and II is somewhat more involved. Here note that in the sequential resonance case there are two free parameters, $m_3$ and $m_1$, while in the 3-body decay case $m_3$ and $m_1$ are tightly related once the $m_{\gamma\gamma}$ spectrum is fixed. 
Comparing scenario I with $(m_3, m_1)=(1200,440)$~GeV to scenario II with $m_3=900$~GeV we see that the former leads to a somewhat softer $p_T(\gamma_1\gamma_2)$ spectrum, and $p_T(\gamma_1)$ exhibits a notch (on log scale) around 375~GeV. 

\section{Conclusions}\label{sec:conclusions}

Should the observed excess in diphoton events at 750~GeV turn into a discovery with the accumulation of more data, the next step will be to elucidate its precise nature. An immediate question in this context will be whether we are dealing with the direct production of a new 750~GeV spin-0 or spin-2 particle that decays into a pair of photons, or with a heavier particle that follows a more complicated decay pattern with the masses of the involved particles conspiring to give two photons with an invariant mass spectrum peaking around 750 GeV. 
The characteristics of additional activity present in the events, such as 
the amount of missing energy, the jet multiplicity, or the presence of $b$-jets or other particles accompanying the two photons 
constitute an important piece of information to this end.  
On a longer timescale, one can envisage a detailed characterisation of the diphoton signal in terms of kinematic distributions. 

As a preliminary step towards such a program, in this work we studied kinematic distributions that may help determine the nature of the putative 750 GeV excess. Using a simple parametrisation of the underlying interactions, we analysed the $p_T$, $\eta$ and $\Delta\eta$ distributions of photons and jets and the overall jet activity expected for a 750 GeV spin-0 resonance produced through $gg$ or $b\bar b$ fusion, a 750 GeV spin-2 resonance produced in $gg$ or $q\bar q$ fusion, and four different scenarios for a heavier spin-0 parent produced in $gg$ fusion and undergoing 3-body or cascade decays.  We found that combinations of the distributions of the diphoton system and the leading photon can help distinguish the topology and mass spectra of the different scenarios, while patterns of QCD radiation can help differentiate the production mechanisms. 
Moreover, the presence of missing energy can help disentangle the direct resonance scenario from the heavy parent one if the latter involves (effectively) invisible particles. In this spirit, the study of such distributions constitutes a powerful complementary approach to both, the search for other decay modes of the new state(s) responsible for the diphoton signal and  
standard direct searches for additional particles that could accompany the new state(s). 

While our conclusions about the prospects of distinguishing between different topologies are generally optimistic, one has to bear in mind the limitations of the EFT approach\footnote{See e.g.\ \cite{Brehmer:2015rna} for a recent discussion in the context of Higgs EFTs.} that we employed in our analysis. 
For example, the presence of relatively light new particles in loops (for weakly coupled models) or form factors (for strongly coupled ones) could bring about some momentum-dependence of the underlying interactions, which would distort some of the distributions we have considered. In this case there should however also be other observable effects, most notably the eventual detection of additional new particles as the sensitivity of the LHC searches improves with more data.

The future of the 750 GeV excess remains, of course, unknown. Our results can, however, also be of relevance for other potential excesses that might be observed during the LHC Run 2. In any case, we are eagerly looking forward to the next round of data-taking that might (hopefully) turn the present excess into a discovery and thus open the door to a plethora of exciting new physics explorations.

\section*{Acknowledgements}

This work was supported in part by the ``Investissements d'avenir, Labex ENIGMASS'', the ANR project DMASTROLHC grant no.\ ANR-12-BS05-0006, the Theory-LHC-France initiative of the CNRS (INP/IN2P3) and the Research Executive Agency (REA) of the European Union under the Grant Agreement PITN-GA2012-316704 (HiggsTools). 
AG is supported by the ``New Frontiers'' program of the Austrian Academy of Sciences.

\newpage
\appendix
\section{Sequential resonance with $S_1=S_2$ \label{app:s1special}}

\begin{figure}
\center
\includegraphics[width=.48\textwidth]{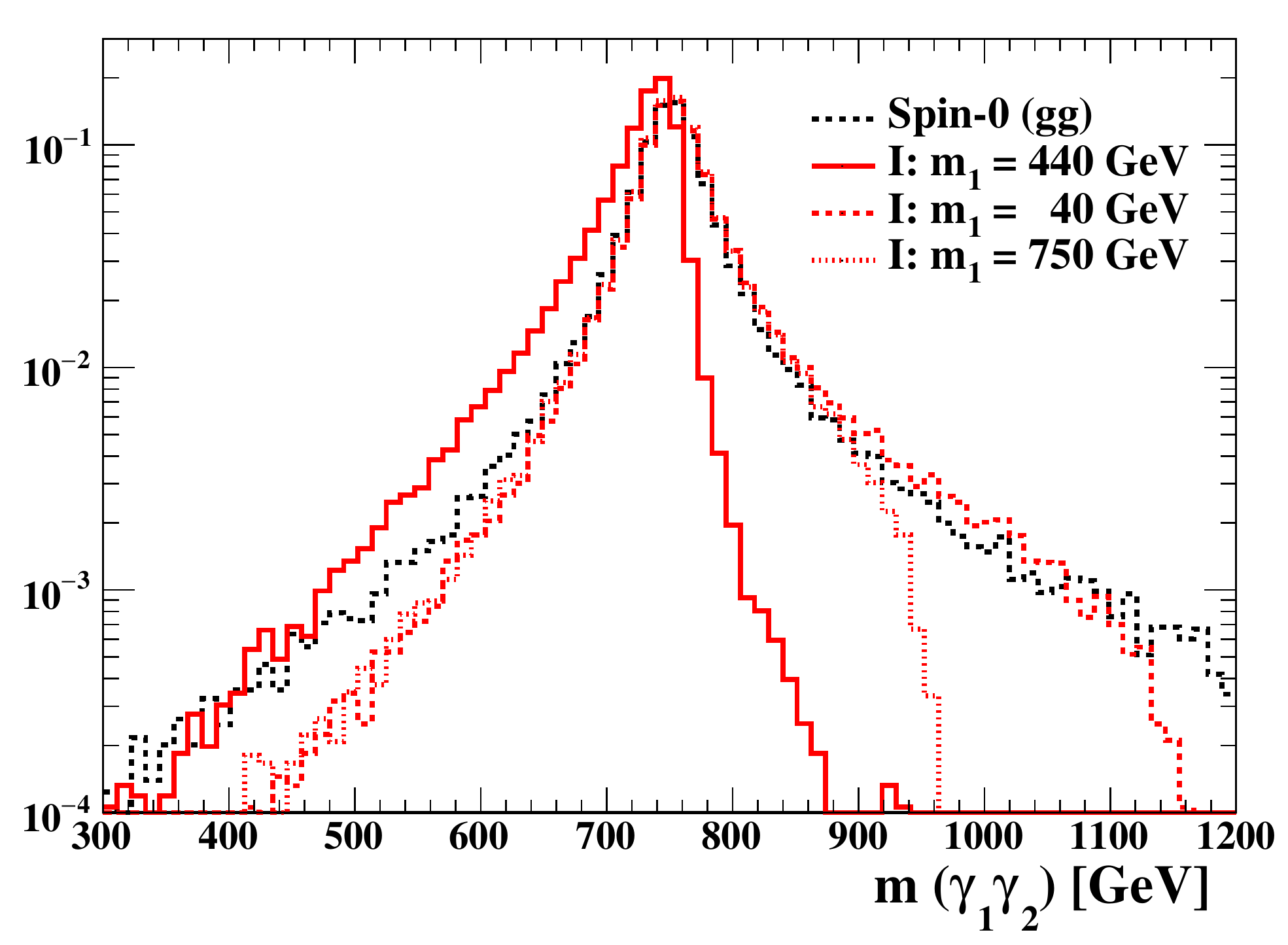}\quad
\includegraphics[width=.48\textwidth]{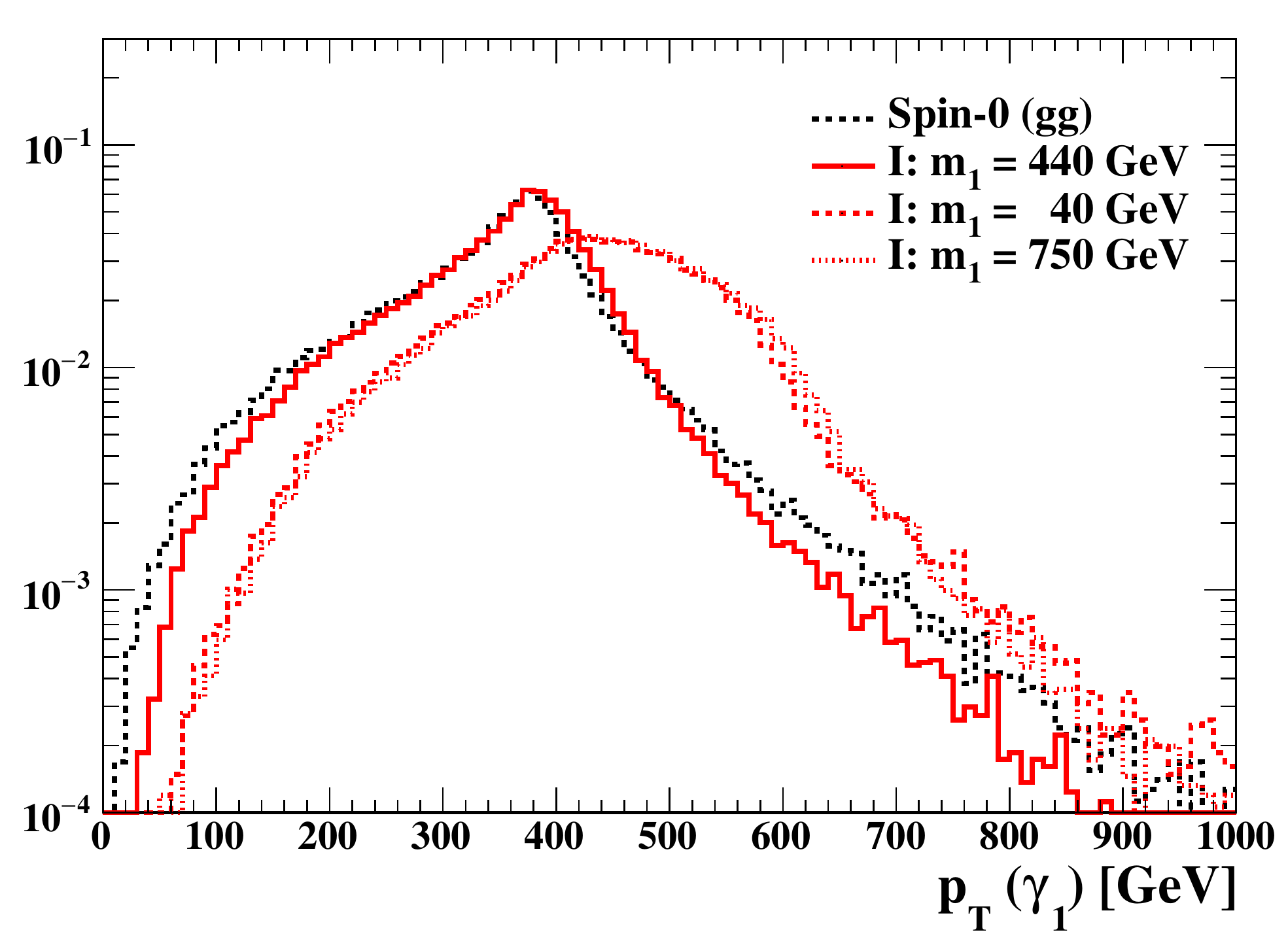}
\includegraphics[width=.48\textwidth]{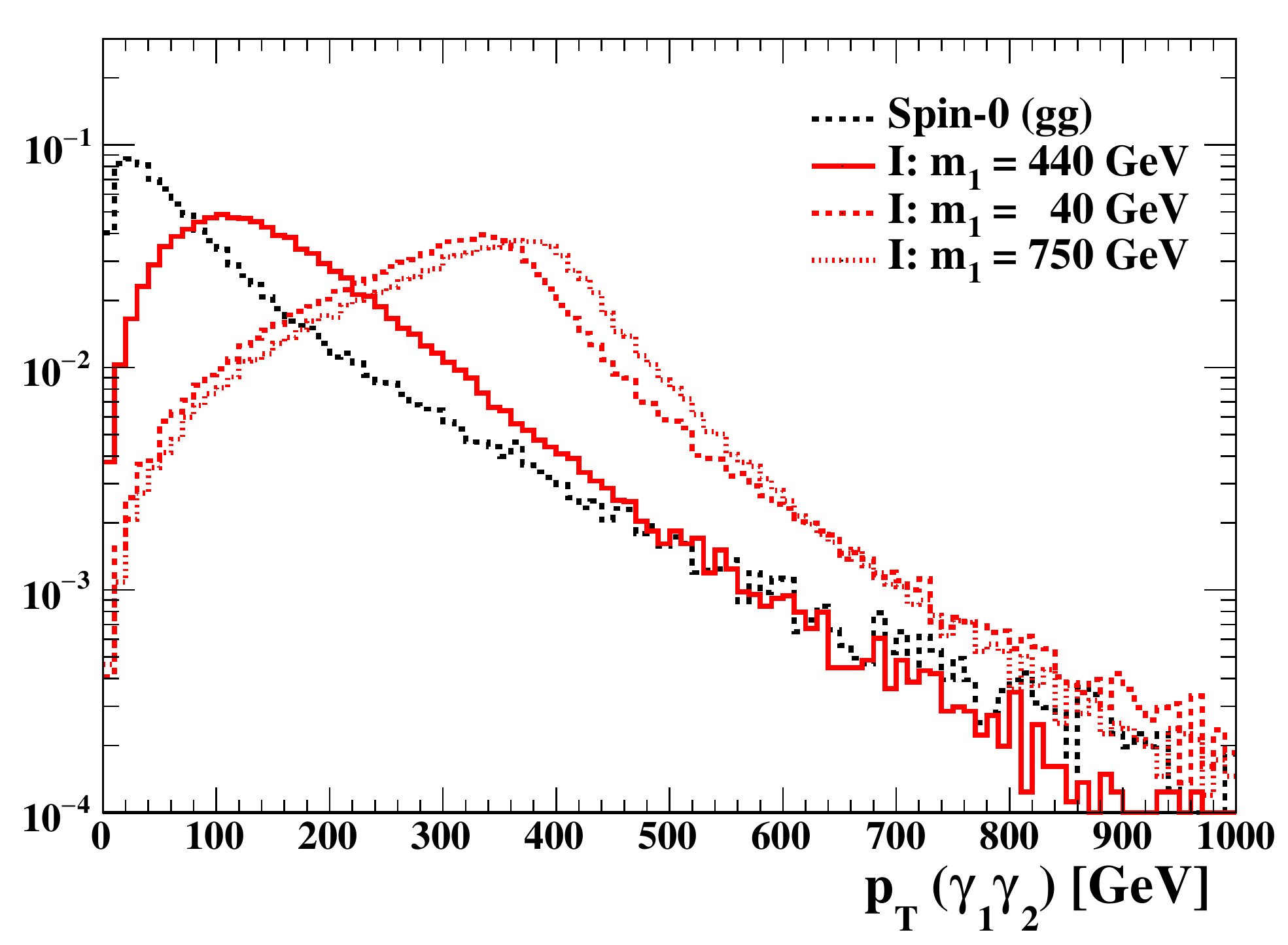}\quad
\includegraphics[width=.48\textwidth]{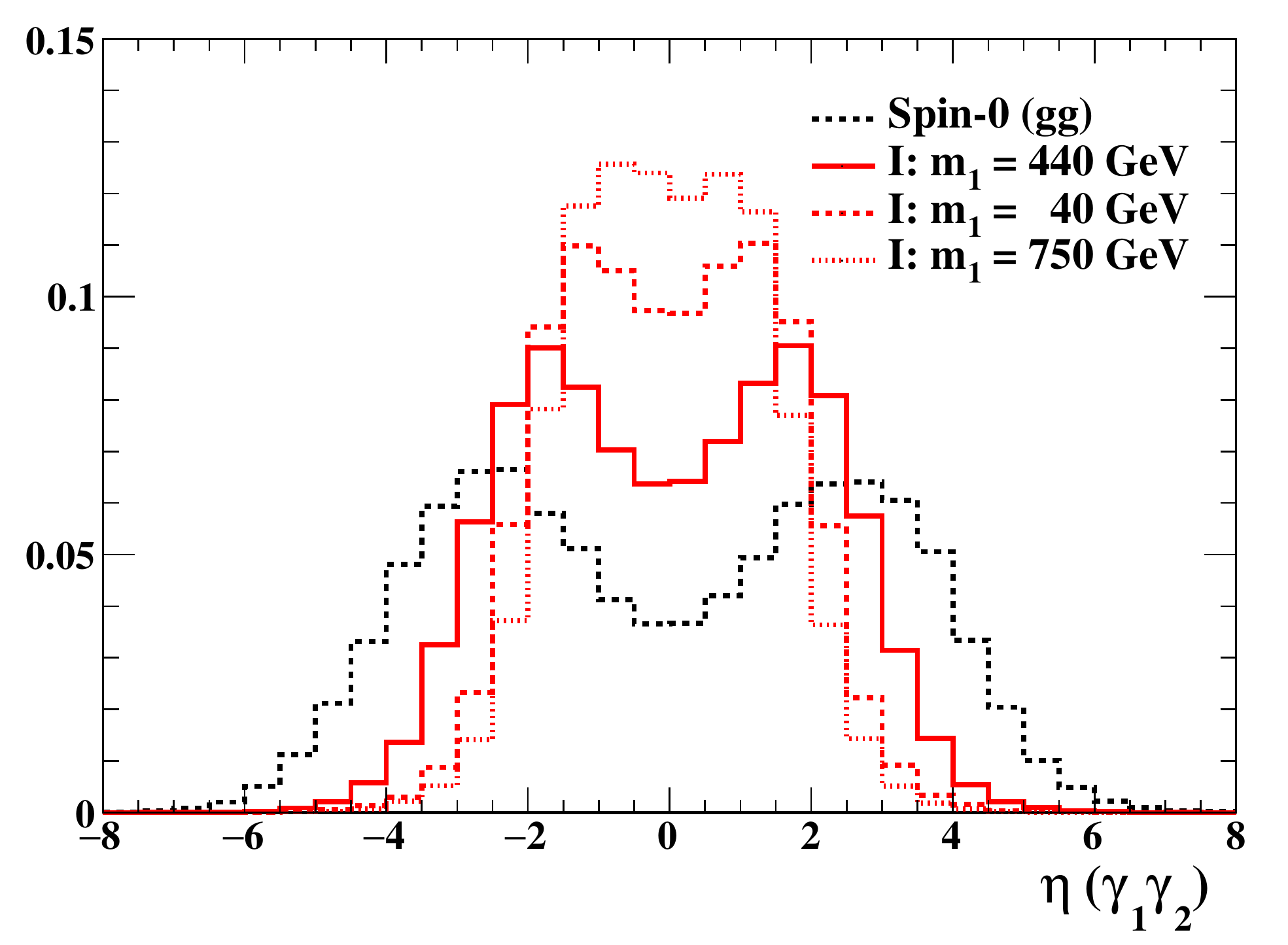}
\caption{ 
Normalised distributions for different mass combinations for the heavier parent resonance scenario I. For reference, the gluon-initiated spin-0 case is shown as black dashed lines. 
}\label{fig:case1}
\end{figure}

A special case of the `sequential resonance', i.e.\ our scenario~I given by the leftmost diagram in Fig.~\ref{fig:diagrams2}, is when $S_3$ decays into two identical particles $S_2=S_1$. This was considered in~\cite{Kim:2015ron}, where the authors studied the production of a heavier (pseudo)scalar resonance decaying into a pair of lighter pseudoscalars with mass of 750 GeV, which decay further into electroweak SM gauge bosons via the Wess-Zumino-Witten anomaly. While this case is distinct from the other `heavier parent' cases we considered --- the events would contain additional hard particles from the decay of the second $S_2$ (e.g., $pp\to S_3\to S_2S_2\to \gamma \gamma +ZZ$) which should be observable --- it is  interesting to compare the resulting  differential distributions of the diphoton signal to those of our scenario~I  benchmark points. This is exemplified in Fig.~\ref{fig:case1}, where the dotted red line shows the case $m_3=1700$~GeV,  $m_2=m_1=750$~GeV. 
Note that we assume exactly two photons; the possibility of the second $S_2$ also decaying into $\gamma\gamma$ is ignored.
Interestingly, most of the distributions look very similar to those of the $m_3=1200$~GeV, $m_2=750$~GeV, $m_1=40$~GeV case. The exception is $\eta(\gamma_1\gamma_2)$, which is more central and does not exhibit any dip at $\eta=0$. 

\section{Mass and width effects for the antler topology \label{app:antler}}

\begin{figure}
\center
\includegraphics[height=5.4cm]{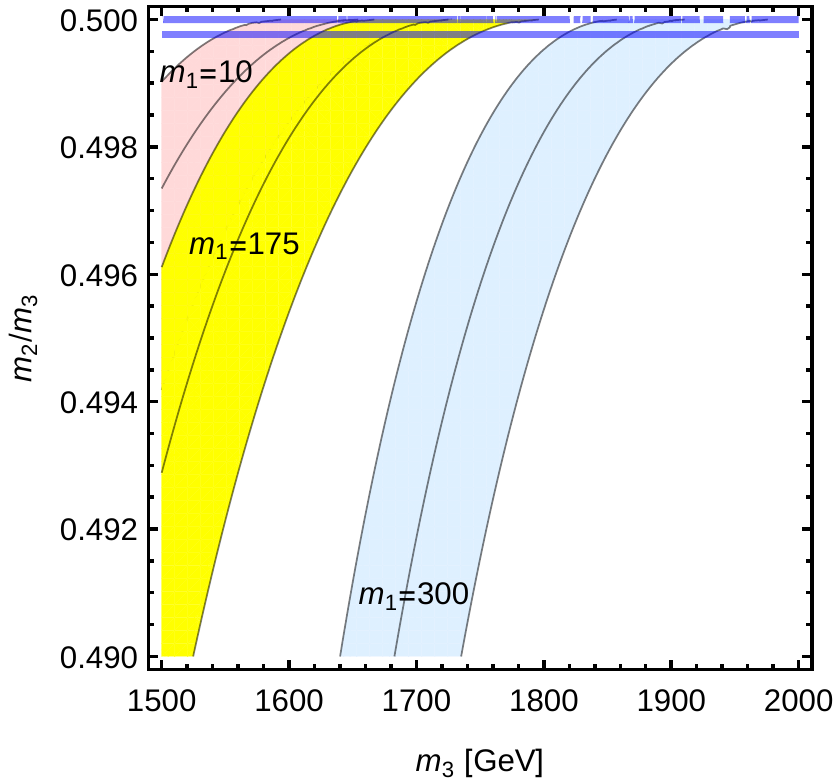}\qquad
\includegraphics[height=5.6cm]{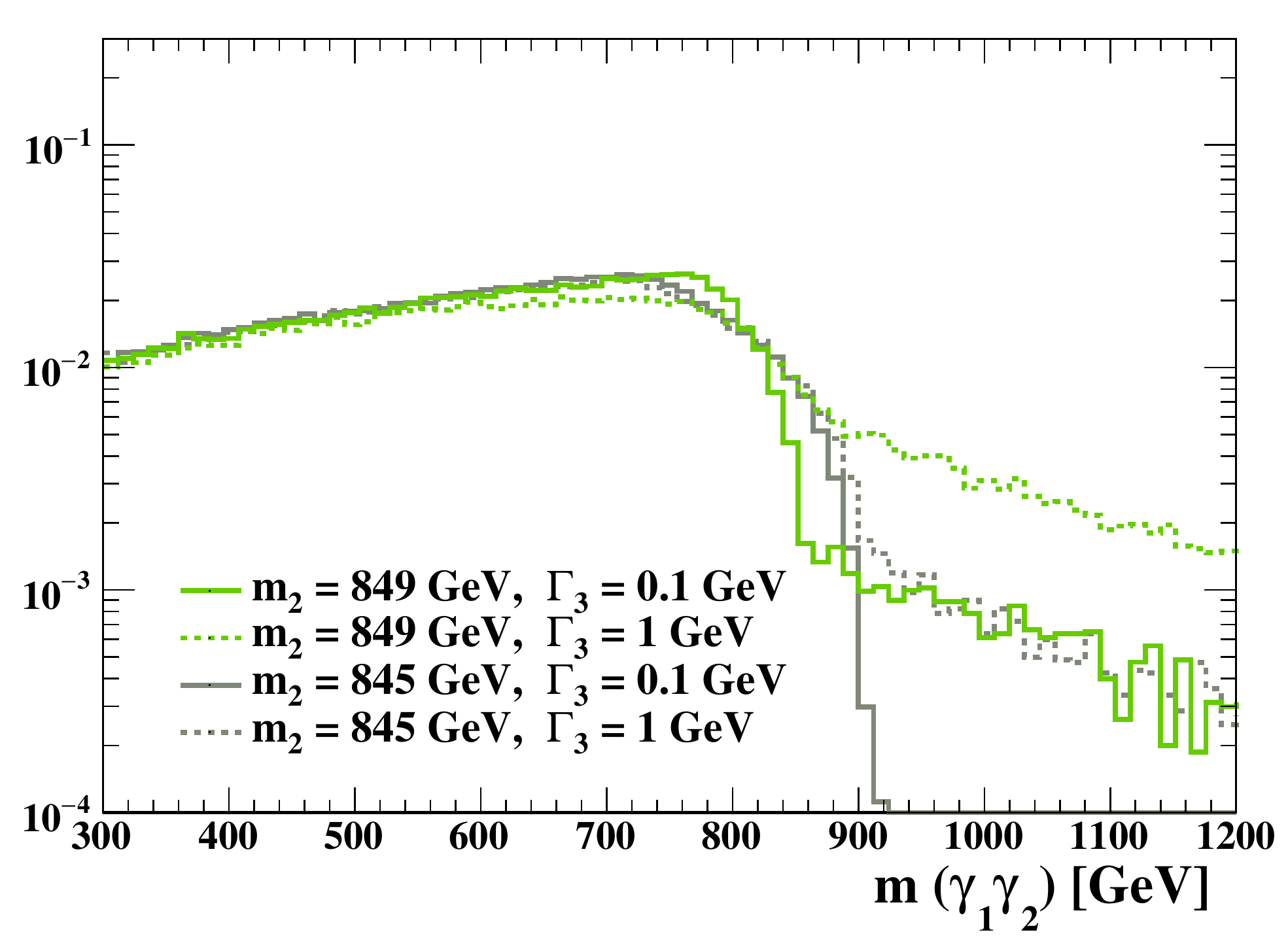}
\caption{ 
Left: Solutions for the antler topology that produce an endpoint in the diphoton invariant mass spectrum of $E=827^{+30.3}_{-36.9}$~GeV \cite{Cho:2015nxy} in the plane of $m_3$ versus $m_2/m_3$ for three different choices of $m_1=10$, 175 and 300~GeV (in light red, yellow and light blue, respectively). The $1\sigma$ limit $\eta=0.0322^{+0.0296}_{-0.0317}$ \cite{Cho:2015nxy} is satisfied between the two horizontal blue lines. Right: Sensitivity of the diphoton invariant mass spectrum to $m_2/m_3$ and the $S_3$ decay width, for $m_3=1700$~GeV and $m_1=75$~GeV.}
\label{fig:antler-masses}
\end{figure}

While the `heavier resonance' scenarios I--III can reproduce the observed diphoton excess in a rather generic manner, scenario IV (the so-called antler topology) is subject to some fine-tuning. 
First of all, as already noted in \cite{Cho:2015nxy}, obtaining the desired diphoton invariant mass spectrum requires a fine adjustment of $m_1$, $m_2$ and $m_3$. The interrelation between the three masses to obtain the correct endpoint is illustrated in Fig.~\ref{fig:antler-masses} (left). While a priori this does not look too constraining, the additional requirement that the cut-off in $m_{\gamma\gamma}$ be steep enough is a very severe constraint, pushing $m_2$ extremely close to $m_3/2$:  the $1\sigma$ range from the fit in \cite{Cho:2015nxy} is above the blue line in Fig.~\ref{fig:antler-masses} (left). 

Related to this, we observe moreover a strong sensitivity to the decay widths. This is illustrated in Fig.~\ref{fig:antler-masses} (right), where we compare the $m_{\gamma\gamma}$ spectrum of the benchmark point $(m_3,\,m_2,\,m_1)=(1700,\,849,\,175)$~GeV obtained with $\Gamma_3=2\Gamma_2=0.1$ GeV (as used in Figs.~\ref{fig:MET} and \ref{fig:cascade}) to that obtained with somewhat larger but still narrow widths of  $\Gamma_3=2\Gamma_2=1$ GeV. We see that $m_{\gamma\gamma}$ quickly flattens out. 
Also shown for comparison is the resulting $m_{\gamma\gamma}$ spectrum when changing $m_2$ from 849~GeV to 845~GeV (i.e.\ $m_2/m_3=0.497$ instead of $0.499$). Here the dependence on the width is less dramatic, as we are a bit further away from the threshold. However, $m_{\gamma\gamma}$ is already too flat to provide a good explanation for the observed excess.

\clearpage
\bibliographystyle{JHEP}
\providecommand{\href}[2]{#2}\begingroup\raggedright\endgroup

\end{document}